\documentclass[runningheads]{svmult}

\usepackage{makeidx}   
\usepackage{graphicx}  
\usepackage{subeqnar}  
\usepackage{multicol}  
\usepackage{physmubb}  
\makeindex             

\newcommand	\gtsim	{\lower.5ex\hbox{$\; \buildrel > \over \sim \;$}} 
\newcommand	\ltsim	{\lower.5ex\hbox{$\; \buildrel < \over \sim \;$}} 

\newcommand	\beq	{\begin{equation}}	
\newcommand	\eeq	{\end{equation}}	

\newcommand	\Angstrom {{\rm \AA}}
\newcommand	\cm	{{\rm \,cm}}
\newcommand	\erg	{{\rm \,erg}}
\newcommand	\nm	{{\rm \,nm}}
\newcommand	\g	{{\rm \,g}}

\newcommand	\K	{{\rm \,K}}
\newcommand	\kms	{{\rm km\,s^{-1}}}
\newcommand	\micron	{{\,\mu{\rm m}}}
\newcommand	\s	{{\rm \,s}}
\newcommand	\yr	{{\rm \,yr}}
\newcommand	\eff	{{\rm eff}}
\newcommand	\eV	{{\rm \,eV}}

\newcommand	\rmH	{{\rm H}}
\newcommand	\HH	{{\rm H}_2}

\newcommand{\abs}{{\rm abs}}
\newcommand{\bg}{{\rm bg}}
\newcommand{\crit}{{\rm crit}}
\newcommand{\drag}{{\rm drag}}
\newcommand{\drift}{{\rm drift}}
\newcommand{\ext}{{\rm ext}}
\newcommand{\gr}{{\rm gr}}
\newcommand{\ISRF}{{\rm ISRF}}
\newcommand{\pd}{{\rm pd}}
\newcommand{\pe}{{\rm pe}}
\newcommand{\pr}{{\rm pr}}
\newcommand{\rad}{{\rm rad}}
\newcommand{\sca}{{\rm sca}}
\newcommand{\stst}{{\rm ss}}

\begin{document}

\title*{Astrophysics of Dust in Cold Clouds}
\toctitle{Astrophysics of Dust in Cold Clouds}
%
%
\titlerunning{Astrophysics of Dust in Cold Clouds}
%
\author{B. T. Draine\\
	Princeton University}
\authorrunning{B. T. Draine}
%
%

\maketitle              

\vspace*{-13em}
\centerline{\small To appear in {\it The Cold Universe: 
		Saas-Fee Advanced Course 32}}
\centerline{ed. D. Pfenniger.
Berlin: Springer-Verlag, 2003, in press}
\vspace*{11em}

Dust plays an increasingly important role in astrophysics.
Historically, dust was first recognized for its obscuring effects, and
the need to correct observed intensities for attenuation by dust
continues today.
But with the increasing sensitivity of IR, FIR, and submm telescopes,
dust is increasingly important as a diagnostic,
with its emission spectrum providing an indicator of physical conditions,
and its radiated power bearing witness to star populations of which
we might otherwise be unaware.
Finally, and most fundamentally, dust is now understood to play many
critical roles in galactic evolution.
By sequestering selected elements in the solid grains, and by catalyzing
formation of the H$_2$ molecule, dust grains are central to the
chemistry of interstellar gas.
Photoelectrons from dust grains can dominate the heating of gas in
regions where ultraviolet starlight is present,
and in dense regions the infrared emission from dust can be an important
cooling mechanism.
Finally, dust grains can be important in interstellar gas dynamics,
communicating radiation pressure from starlight to the gas,
and providing coupling of the magnetic field to the gas in
regions of low fractional ionization.

We would like to understand these effects of dust in the Milky Way,
in other galaxies, and as a function of cosmic time.
These lectures are organized around topics in the astrophysics
of dust in the the Milky Way, as this is our best guide to understanding
and modelling dust long ago and far away.

\section{Introduction to Interstellar Dust}

We begin with a brief review of some of the observational evidence
which informs our study of interstellar dust.

\subsection{Interstellar Extinction}

Through study of open star clusters in the Galaxy, Trumpler (1930)
found that distant stars were dimmed by something in addition to the inverse
square law, and concluded that interstellar space in the galactic plane
contained
``fine cosmic dust particles of various sizes ... producing the
observed selective absorption''.
Over the past 7 decades we have built on Trumpler's pioneering study,
but many aspects of interstellar dust -- including its chemical composition! --
remain uncertain.  Let us therefore begin by reviewing the different
ways in which nature permits us to study interstellar dust.

\begin{figure}
\begin{center}
\includegraphics[width=.6\textwidth,angle=270]{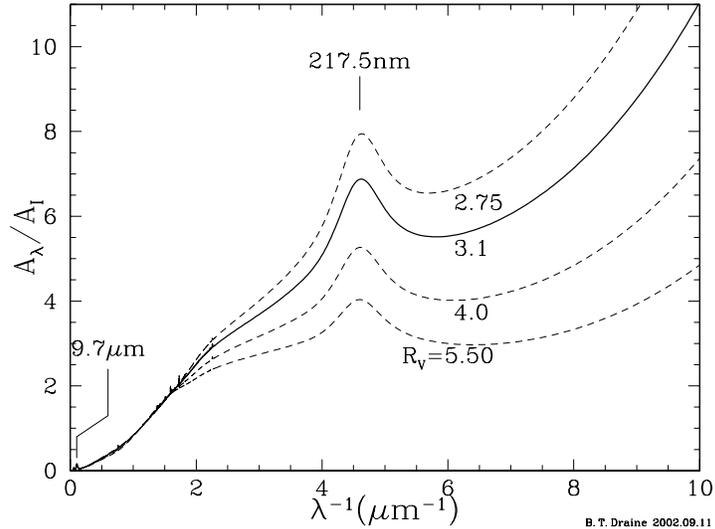}
\end{center}
\caption[]{
	\label{fig:extcurv}
	Extinction at wavelength $\lambda$, relative to the
	extinction at $I=900\nm$, as a function of
	inverse wavelength $\lambda^{-1}$, for Milky Way regions characterized
	by different values of $R_V\equiv A_V/E(B-V)$, where
	$A_B$ is the extinction at $B=4400\micron$, $A_V$ is the
	extinction at $V=5500\micron$, and the ``reddening''
	$E(B-V)\equiv A_B-A_V$.
	Note the rapid rise in extinction in the vacuum ultraviolet
	($\lambda \ltsim 0.2\micron$) for regions with $R_V\ltsim 4$.
	The normalization per H nucleon is approximately
	$A_{I}/N_{\rm H}\approx 2.6\times10^{-22}\cm^2/{\rm H}$.
	The silicate absorption feature (\S\ref{sec:silicate})
	at 9.7$\micron$ and the
	diffuse interstellar bands (\S\ref{sec:DIBs}) are barely visible.
	}
\end{figure}

Trumpler analyzed the interaction of light with interstellar dust, and
this remains our most direct way to study interstellar dust.
We use stars as ``standard candles'', and study the ``selective extinction''
-- or ``reddening'' -- of starlight by the dust.  
With the assumption that the extinction ($\equiv$ absorption + scattering) 
goes to zero at wavelengths $\lambda\rightarrow\infty$, 
and observing the star at sufficiently long wavelength (to, in effect,
determine its distance) one
can determine the attenuation of the starlight by dust as a function of
wavelength.  
Because atomic hydrogen absorbs strongly for $h\nu>13.6\eV$, it
it is only possible to measure the contribution of dust to the extinction
at $h\nu < 13.6\eV$, or $\lambda > 912\Angstrom$.
A typical ``extinction curve'' -- the extinction as a function of
wavelength or frequency --- is shown in Figure \ref{fig:extcurv},
showing the rapid rise in extinction in the vacuum ultraviolet.
Observed extinction curves vary in shape from one line-of-sight to another,
but appear to approximately form a one-parameter family 
(Cardelli et al.\ 1989); the parameter is often taken to be the
ratio $R_V\equiv (A_B-A_V)/A_V$, where $A_B$ and $A_V$ are the extinctions
measured in the B (4400\AA) and V (5500\AA) spectral bands.
A parametrization of the extinction curve was provided by Cardelli et al.;
the curves in Fig.\ \ref{fig:extcurv} were calculated using a more
recent prescription by Fitzpatrick (1999), with the extinction in
the infrared following Draine (1989b). 

We will discuss dust grain optics below, but it is clear that if the
dust grains were large compared to the wavelength, we would be in the
``geometric optics'' limit and the
extinction cross-section would be independent of wavelength.
Therefore the tendency for the extinction to rise even at the shortest
wavelengths where we can measure it tells us that grains
smaller than the wavelength must be making an appreciable contribution to
the extinction at all of observed wavelengths.   
As we will see below, ``small'' means
(approximately) that $2\pi a |m-1|/\lambda \ltsim 1$, where $m(\lambda)$ is
the complex refractive index.  
Thus if $|m-1|\approx 1$ at 
$\lambda=0.1\micron$,
\begin{itemize}
\item we must have large numbers of grains with $a\ltsim .015\micron$.
\end{itemize}

\subsection{Scattering of Starlight by Dust Grains}

When an interstellar cloud happens to be unusually near one or more
bright stars, we have a ``reflection nebula'', where we see the
starlight photons which have been scattered by the dust in the cloud.
The spectrum of the light coming from the cloud surface shows the
stellar absorption lines, showing that scattering rather than some
emission process is responsible.
By comparing the observed scattered intensity with the estimated intensity of
the starlight incident on the cloud, it is possible to infer the
albedo of the dust -- the ratio of scattering cross section to extinction
cross section.
The result is that in the optical the interstellar dust mixture has
an albedo $\omega\approx0.5$  -- scattering is about as important as
absorption -- and the grains are somewhat forward scattering, with
$\langle \cos\theta\rangle \approx 0.5$.
Rayleigh scattering by particles small compared to the wavelength has
$\langle\cos\theta\rangle\approx 0$, so this tells us that 

\begin{itemize}
\item the particles
dominating the scattering at $\lambda\approx 0.6\micron$ \\
have $a\gtsim \lambda/2\pi\approx 0.1\micron$.
\end{itemize}

\begin{figure}
\begin{center}
\includegraphics[width=.6\textwidth,angle=0]{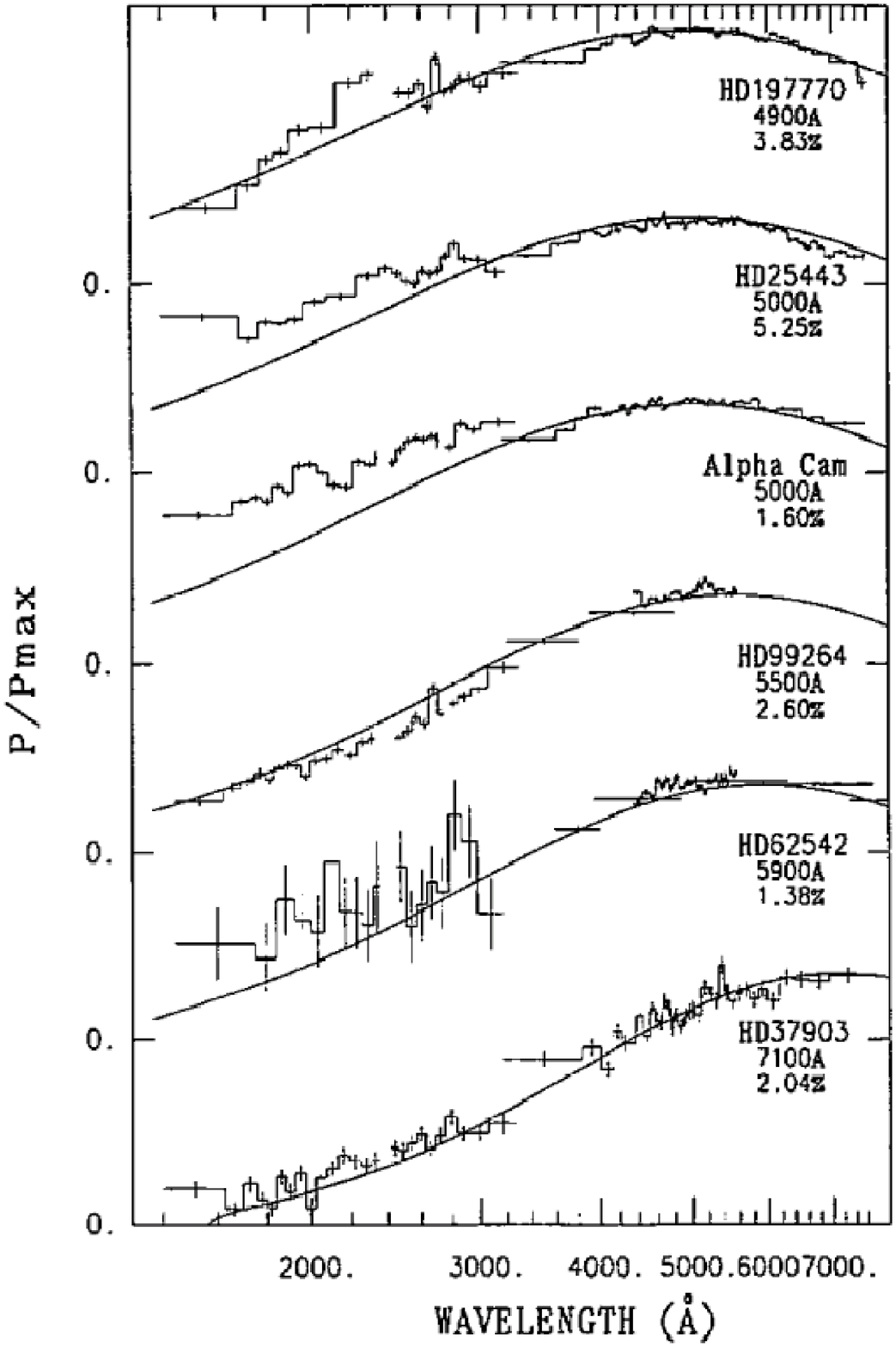}
\end{center}
\caption[]{
	Linear polarization in the ultraviolet measured by Clayton et al.
	(1992).
	The solid line is the ``Serkowski law'' fit to the data.
	Figure from Clayton et al (1992).
	}
\label{fig:uvpol}
\end{figure}

\begin{figure}
\begin{center}
\includegraphics[width=.6\textwidth,angle=270]{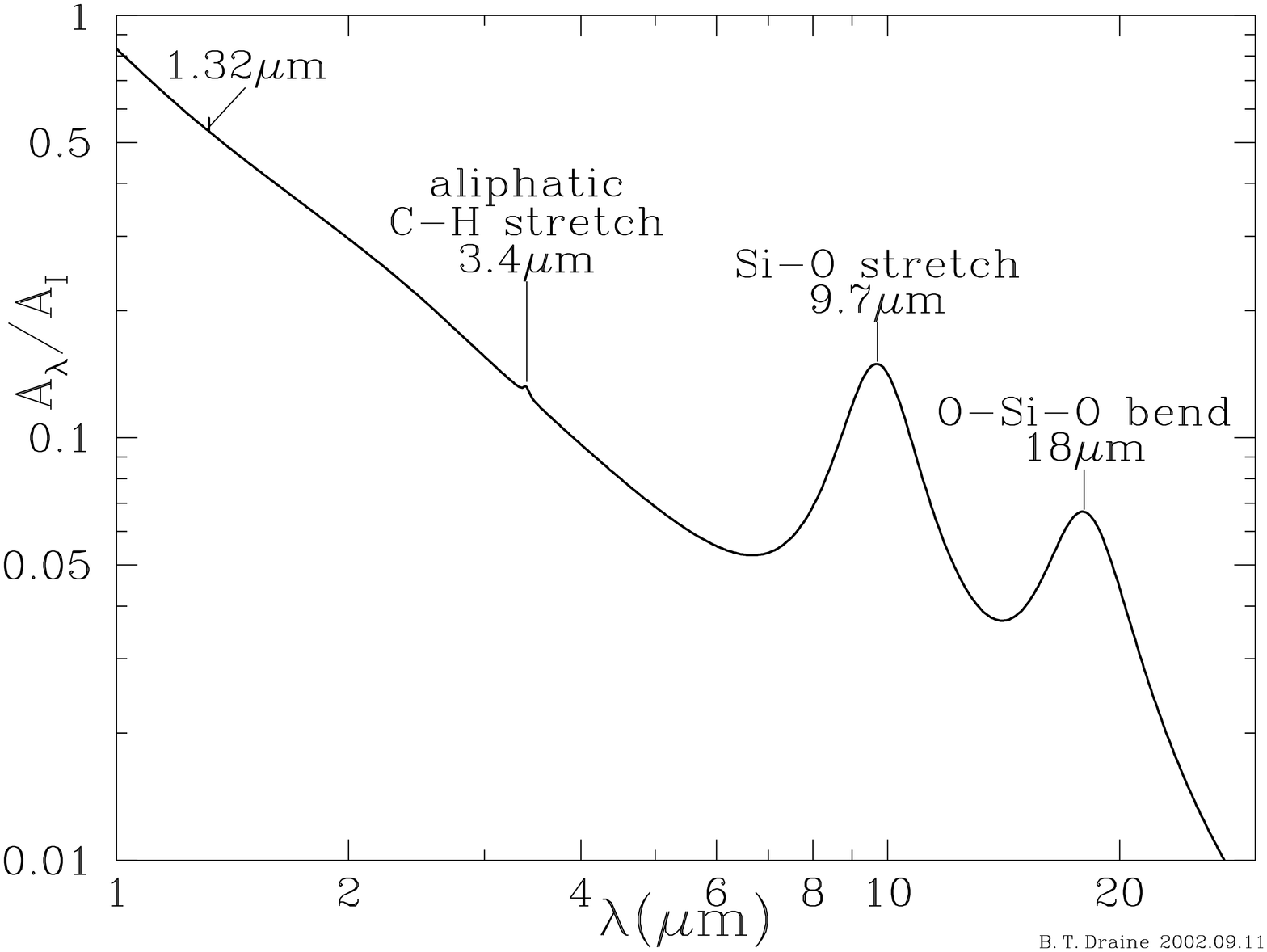}
\end{center}
\caption[]{Infrared extinction, relative to
	extinction at $I=900\nm$, showing the strong 9.7$\micron$
	and 18$\micron$ silicate features, the
	3.4$\micron$ aliphatic C-H stretch, and
	a weak unidentified DIB at 1.32$\micron$ (Joblin et al. 1990).
	}
\label{fig:IR_ext}
\end{figure}

\subsection{Polarization of Starlight}

The polarization of starlight was discovered in 1949
(Hall 1949; Hall \& Mikesell 1949; Hiltner 1949a,b).
When it was realized that the degree of polarization tended to be larger
for stars with greater reddening, and that stars in a given region of
the sky tended to have similar polarization directions, it was obvious that
the polarization is produced by the interstellar medium:
light propagating through the interstellar medium becomes linearly
polarized as a result
of preferential extinction of one linear
polarization mode relative to the other.
The polarization percentage typically peaks near the V band ($5500\Angstrom$),
and can be empirically described by the ``Serkowski law''
(Serkowski 1973):
\beq
p(\lambda) \approx p(\lambda_{\max})\exp[-K \ln^2(\lambda/\lambda_{\max})]
~~~,
\eeq
with $\lambda_{\max}\approx 5500\Angstrom$ and $K\approx 1.15$.
This ``linear dichroism'' of the interstellar medium is due to
dust grains which are partially aligned by the interstellar magnetic
field.
The peak polarization $p_{\max}$ is found to fall within an
envelope $0 < p_{\max} \leq 0.09 (E(B-V)/{\rm mag})$, 
or $0 < p_V \ltsim 0.03\tau_V$;
the maximum values are presumed to arise on sightlines where the
magnetic field is uniform and perpendicular to the line-of-sight.
While the ``Serkowski law'' was put forward as a fit to the observed
polarization at $0.3\micron \ltsim \lambda \ltsim 1\micron$, it turns out
to give a surprisingly good approximation to the measured linear 
polarization in the vacuum ultraviolet (Clayton et al 1992,
Wolff et al 1997) as seen in Figure \ref{fig:uvpol}

The mechanism responsible for the grain alignment remains a fascinating
puzzle, which we will discuss below.
Independent of the grain alignment mechanism, however, we can infer the
sizes of the interstellar grains responsible for this polarization by
noting that the extinction rises rapidly into the UV whereas the polarization
drops (Kim \& Martin 1995).
This can be understood if the grains responsible for the
polarization have
sizes $a$ such that $a\approx (\lambda_{\max}/2\pi) \approx 0.1\micron$:
then as one proceeds into the UV one moves toward the ``geometric optics''
limit where both polarization modes suffer the same extinction, so
the polarization goes to zero:
\begin{itemize}
\item The extinction at $\lambda\approx 0.55\micron$ has an appreciable
contribution from grains with sizes $a\approx 0.1\micron$ which are
nonspherical and substantially aligned.
\item The grains with $a\ltsim 0.05\micron$ which dominate the
extinction at $\lambda\ltsim 0.3\micron$ are either spherical (which seems
unlikely) or minimally aligned.
\end{itemize}

\subsection{Spectroscopy of Dust: The 2175\AA\ Feature}

Of what is interstellar dust composed?
One may look for spectroscopic clues in the extinction.
The extinction curves in Fig. \ref{fig:extcurv} show a conspicuous extinction
feature at $\lambda^{-1}=4.6\micron^{-1}$, or $\lambda = 2175\Angstrom$.
The feature is well-described by a Drude profile. 
The central wavelength
is nearly identical on all sightlines, but the width varies significantly
from one region to another (Fitzpatrick \& Massa 1986).

The strength of this feature implies that the responsible material must be
abundant (Draine 1989a):
it must be made from H, C, N, O, Mg, Si, S, or Fe.
Small graphite grains would have a strong absorption peak at about this
frequency (Stecher \& Donn 1965; Draine 1989a), 
due to $\pi\rightarrow\pi^*$ electronic excitations
in the $sp^2$-bonded carbon sheets.  Since the carbon skeleton of polycyclic
aromatic hydrocarbon (PAH) molecules resembles a portion of a graphite
sheet, such molecules also tend to have strong electronic transitions
at about this frequency.  It therefore seems likely that the $2175\Angstrom$
feature is due to some form of $sp^2$-bonded carbon material.

\subsection{Spectroscopy of Dust: The Silicate Features\label{sec:silicate}}
There is a conspicuous infrared absorption feature at 9.7$\micron$, shown
in Figure \ref{fig:IR_ext}.
Silicate minerals generally have strong absorption resonances due to
the Si-O stretching mode near $10\micron$, and it seems virtually certain
that the interstellar $9.7\micron$ feature is due to silicates.
This conclusion is strengthened by the fact that the $10\micron$ emission
feature is seen in the outflows from oxygen-rich stars (which would be
expected to condense silicate dust) but not in the outflows from
carbon-rich stars.
The interstellar $9.7\micron$ feature is seen both in emission
(e.g., in the Trapezium region in Orion [Gillett, Forrest, et al. 1975])
or in extinction in the interstellar medium
(Roche \& Aitken 1984).

Crystalline silicate minerals generally have sharp features in their
$10\micron$ absorption which are not seen in the broad interstellar 
$10\micron$ feature, leading to the conclusion that interstellar silicates
are probably amorphous.

Near 18$\micron$ warm interstellar dust shows another emission feature,
which is attributable to the Si-O-Si bending mode in amorphous silicates.

\begin{figure}
\begin{center}
\includegraphics[width=.6\textwidth,angle=270]{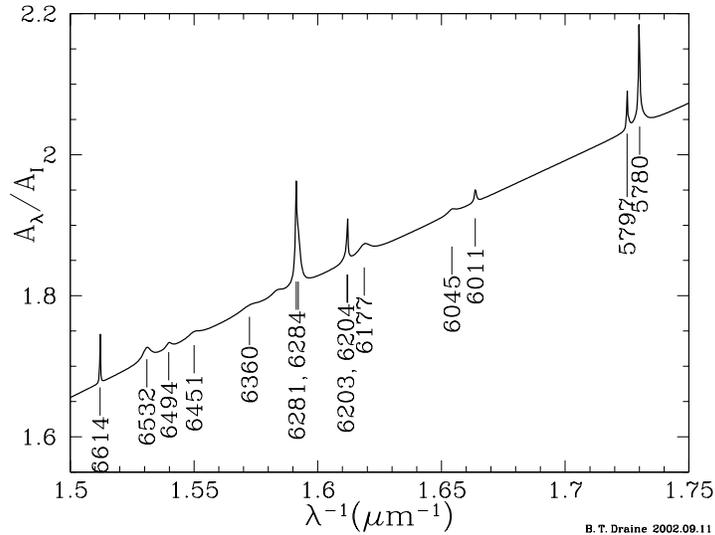}
\end{center}
\caption[]{Extinction at wavelength $\lambda$, relative to the
	extinction at $I=900\nm$, showing some of the diffuse
	interstellar bands.
	}
\label{fig:dibs}
\end{figure}

\subsection{Spectroscopy of Dust: Diffuse Interstellar Bands
	\label{sec:DIBs}}

The 3 features at $0.22\micron$, $9.7\micron$, and $18\micron$
are by far the strongest features seen in diffuse interstellar dust.
There are, in addition, numerous weaker features in the optical known
as the ``diffuse interstellar bands'', or DIBs.
Figure \ref{fig:dibs} 
shows the extinction for $1.5\micron^{-1}<\lambda^{-1}<1.75\micron^{-1}$, 
with several conspicuous DIBs present, most notably the
DIB at $0.5780\micron$.
The strongest DIB falls at 443.0nm.
Jenniskens \& Desert (1994) report a total of 154 ``certain'' DIBs in the
interval $0.38 - 0.868\micron$, plus another 52 ``probable'' detections.
DIBs were discovered 80 years ago (Heger 1922) and their interstellar
nature was established 68 years ago (Merrill 1934).

\begin{figure}
\begin{center}
\includegraphics[width=0.8\textwidth,angle=0]{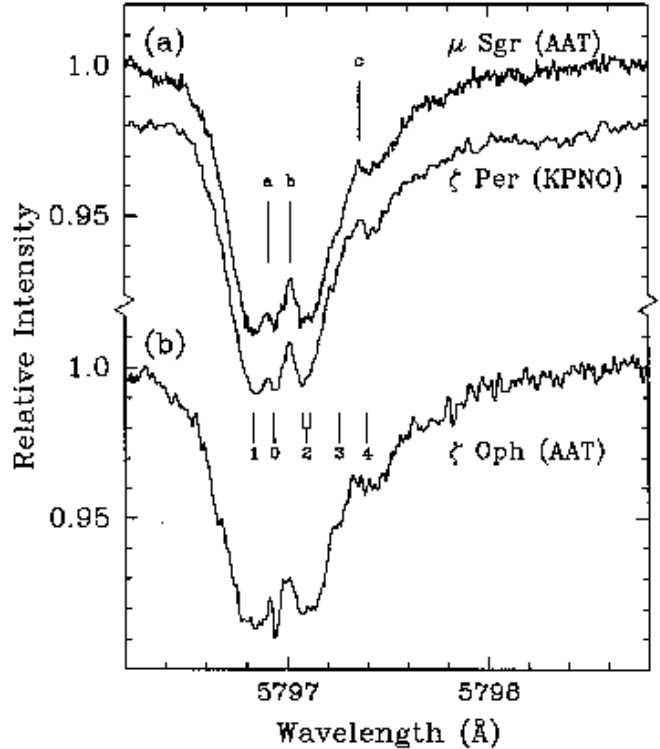}
\end{center}
\caption{Fine structure of the 5797$\Angstrom$ DIB toward $\zeta$ Oph,
$\zeta$ Per, and $\mu$ Sgr (from Kerr et al. 1998).
Note in particular the very narrow feature labelled 0.
	}
\label{fig:dib5797}
\end{figure}

It is embarassing that Nature has provided astrophysicists with this
wealth of spectroscopic
clues, yet as of this writing not a single one of the DIBs has been
convincingly identified!
It seems likely that some of the DIBs may be due to free-flying large
molecules; this hypothetis has received support from high resolution
spectra of the 5797$\Angstrom$ feature (see Figure \ref{fig:dib5797})
showing intrinsic ultrafine structure (Sarre et al. 1995; Kerr et al. 1998).

\subsection{Spectroscopy of Dust: The 3.4$\mu$m Feature}

There is a broad absorption feature at 3.4$\micron$ which is almost certainly
due to the C-H stretching mode in hydrocarbons.
A recent study by Pendleton \& Allamandola (2002) concluded that
hydrocarbons with a mixed aromatic (ring) and aliphatic (chain) character 
provided a good
fit to the observed interstellar absorption, including the $3.35-3.53\micron$
region.
This included hydrocarbon films deposited following laser-ablation of
amorphous carbon in Ar, followed by exposure to atomic H (Mennella et al. 1999)
or from a weakly-ionized plasma produced by laser-ablation 
of graphite in hydrogen (Scott \& Duley 1996; Duley et al. 1998).

\subsection{Spectroscopy of Dust: Ice Features\label{sec:spec_ice}}

In dark molecular clouds a number of additional absorption features are
seen, most notably a strong band at $3.1\micron$ which is attributed to
the O-H stretching mode in H$_2$O ice.
However, the $3.1\micron$ feature is {\it not} seen on sightlines which are
confined to diffuse interstellar clouds (Gillett, Jones et al. 1975).

When a strong $3.1\micron$ feature appears in absorption, a number of other
absorption features are also seen, including features due to
CO (4.67$\micron$), CH$_3$OH (3.53$\micron$),
and CO$_2$ (15.2$\micron$).
The shape of the $3.1\micron$ H$_2$O feature is indicative of the
type of ice and the impurities present in it.
The relative strengths of the various features indicate that H$_2$O 
is the dominant ``ice'' species,
with NH$_3$, CO, CH$_3$OH, and CO$_2$ as secondary constituents.

\begin{figure}
\begin{center}
\includegraphics[width=0.7\textwidth,angle=270]{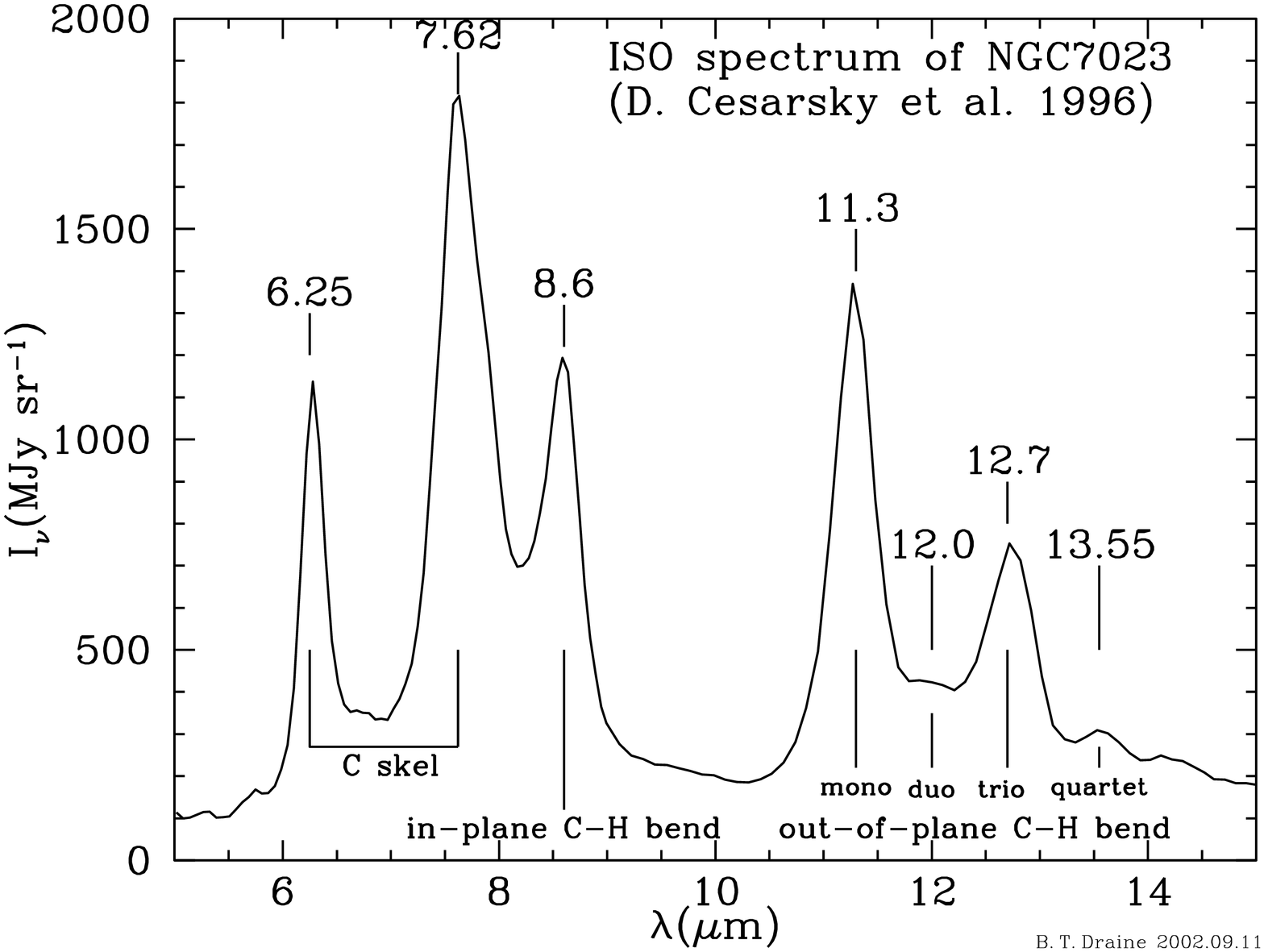}
\end{center}
\caption{Emission spectrum of the reflection nebula NGC 7023
measured by ISOCAM (Cesarsky et al. 1996).}
\label{fig:7023}
\end{figure}

\begin{figure}
\begin{center}
\includegraphics[width=0.7\textwidth,angle=0]{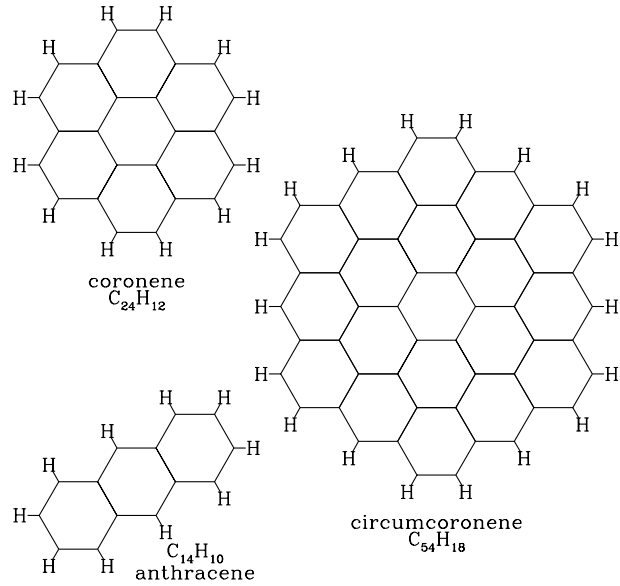}
\end{center}
\caption{Three examples of PAH molecules.}
\label{fig:pah}
\end{figure}

\begin{figure}
\begin{center}
\includegraphics[width=0.7\textwidth,angle=270]{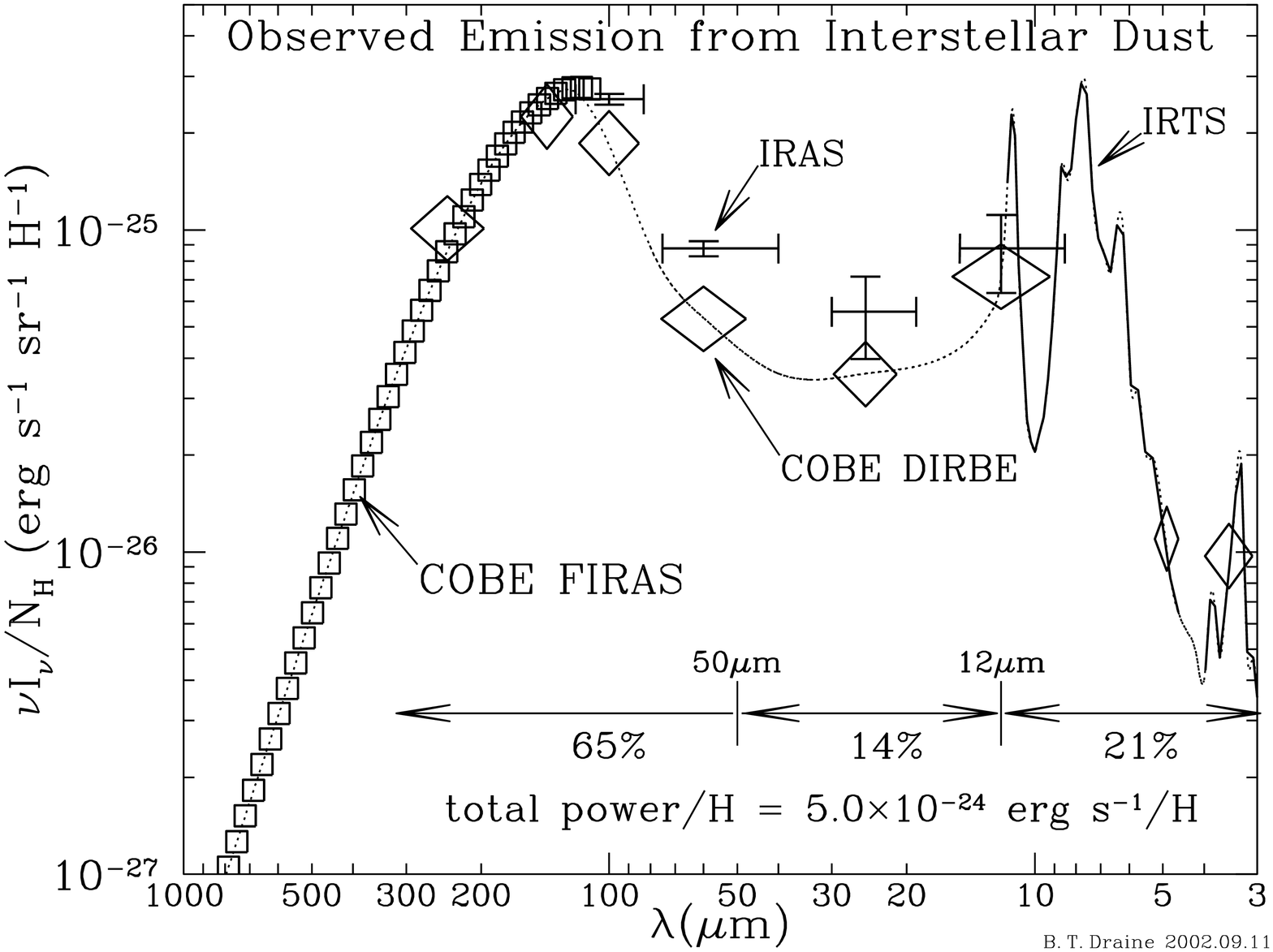}
\end{center}
\caption{Observed infrared emission per H nucleon from dust heated
by the average starlight background (see text).
Crosses: IRAS (Boulanger \& Perault 1988);
Squares: COBE-FIRAS (Wright et al. 1991);
Diamonds: COBE-DIRBE (Arendt et al. 1998);
Heavy Curve: IRTS (Onaka et al. 1996, Tanaka et al. 1996).
The interpolated dotted line is used to estimate the total power.
}
\label{fig:fir}
\end{figure}

\subsection{Spectroscopy of Dust: PAH Emission Features}

A wide variety of galactic objects, including planetary nebulae,
HII regions, photodissociation fronts, and reflection nebulae, have
strong infrared emission in the 3-13$\micron$ region.
Most of the radiated power emerges in 5 broad infrared bands,
at 3.3, 6.2, 7.7, 8.6, and 11.3$ \micron$.
In Figure \ref{fig:7023} we show the 4-15$\micron$ emission
observed from the reflection nebula NGC 7023 (Cesarsky et al. 1996).
These features are seen in many other objects as well (see, e.g.,
Boulanger et al. 1998).
The emission is quite strong: a significant
fraction of the starlight energy incident on the reflection nebula is 
reradiated in these infrared emission bands -- the fraction can be as
large as 10-20\% depending on the spectral type of the star -- so the
particles responsible for this emission must be quite abundant, as
we will discuss further below.

The emission is thought to result from vibrational modes of polycyclic
aromatic hydrocarbon (PAH) molecules.
PAHs can be thought of as a section of a graphite ``sheet'' with the C atoms
in a 2-dimensional hexagonal lattice, with the H atoms attached to C atoms
at the edges of the lattice.
Three examples of PAHs are shown in Figure \ref{fig:pah}.
When H atoms are attached to the edge of an aromatic ring,
\begin{itemize}
\item the C-H stretching mode tends to be at
3.3$\micron$,
\item the C-H in-plane bending mode is near 8.6$\micron$
\item the C-H out-of-plane bending mode tends to fall at 11.3, 11.9, 12.7, or
13.6$\micron$, depending on whether there are one, two, three, or four
adjacent C-H bonds (``solo-'', ``duo-'', ``trio-'', or ``quartet-'' C-H).
The observed spectra suggest approximately equal fractions of solo,
duo, or trio C-H, and little quartet C-H, consistent with what one would
expect for a mixture of large symmetric compact PAHs (Stein \& Brown 1991).
\item The carbon skeleton has C-C-C bending modes near 6.2 and
$7.7\micron$
\end{itemize}
The precise position of the bands, and their relative strengths,
vary from one individual PAH to another, and upon their state of ionization,
but the observed spectra
are in good agreement with what one might obtain from a mixture
of neutral and ionized PAHs (Allamandola, Hudgins, \& Sandford 1999).

\subsection{IR and FIR Emission}

The energy which grains absorb from starlight is reradiated at longer
wavelengths, mainly into the far-infrared.  We can estimate the emission
spectrum of the Galaxy using data from all-sky surveys by the IRAS
and COBE satellites, and pointed observations by the Japanese IRTS
satellite.
At high galactic latitudes we know that most of the dust is illuminated
by the local average starlight background.
Theoretical studies of dust grain heating by starlight suggested that
$a\gtsim 0.01\micron$ interstellar grains would be heated to temperatures
$15\K\ltsim T \ltsim 20\K$ by the diffuse starlight (see, e.g.,
Draine \& Lee 1984).

Because there is little dust at high galactic latitudes, the infrared
surface brightness is low, but from the 
correlation of infrared surface brightness with 21 cm emission
one can extract the infrared emission per H atom.
This has been done using photometry in the 4 IRAS photometric bands,
the 9 photometric bands of the DIRBE instrument on COBE,
and using the spectrophotometry by the FIRAS instrument on COBE.

The Japanese IRTS satellite obtained the 3-13$\micron$ spectrum of the
galactic plane ($l\approx45^\circ$, $b\approx 0$).
If we assume that the 3-13$\micron$ surface brightness is proportional
to the 100$\micron$ surface brightness, we can estimate the 3-13$\micron$
emission per H atom.
The results are shown in Figure \ref{fig:fir}.
The $3-13\micron$ emission appears to be dominated by the same infrared
emission features observed in reflection nebulae and planetary nebulae,
and widely attributed to polycyclic aromatic hydrocarbons.

If we interpolate to obtain the
dotted line in Figure \ref{fig:fir}, we can numerically
integrate to obtain the total infrared emission power per H nucleon
$=5.0\times10^{-24}\erg\s^{-1}\rmH^{-1}$.
About 2/3 of this power is radiated at wavelengths
$\lambda > 50\micron$, which is where interstellar dust at $T\ltsim20\K$
was expected to radiate.  However, about 1/3 of the power is radiated
at $\lambda < 50\micron$, where the emission from $T<20\K$ dust
should be negligible.
Furthermore, about 1/6 of the total power is radiated at $\lambda < 10\micron$,
primarily in 
6.6, 7.7, and 8.6$\micron$ PAH features.

\begin{figure}
\begin{center}
\includegraphics[width=0.7\textwidth,angle=270]{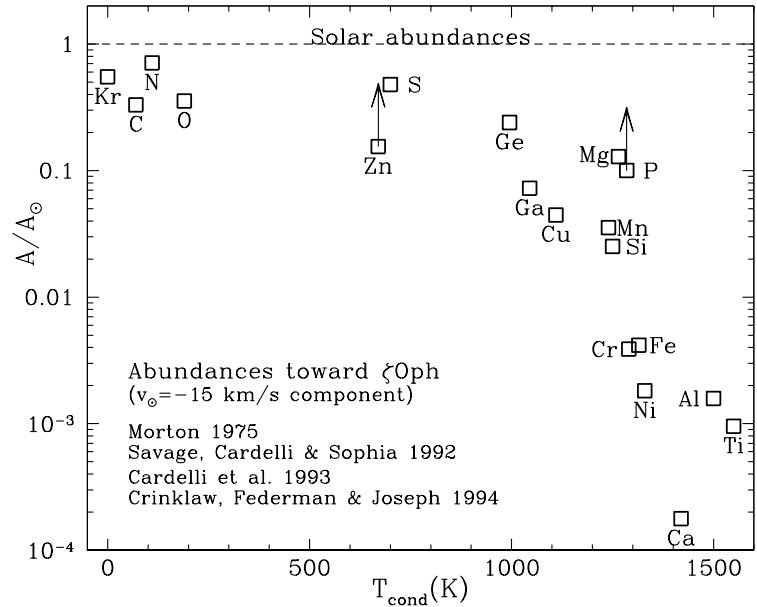}
\end{center}
\caption{Depletions in the diffuse molecular cloud toward $\zeta$Oph.
	}
\label{fig:depletions}
\end{figure}

\subsection{Interstellar Depletions: Atoms Missing from the Gas}

Narrow optical and ultraviolet absorption lines seen in absorption in
stellar spectra can be used to determine interstellar gas phase
abundances of many elements.
While some species, such as N and S, have gas phase abundances
(relative to H) which are approxiamtely solar, certain others, such
as Mg, Al, Si, Ti, Ca, Fe, Ni, Cr show abundances which are far below
solar.
Since we presume that the interstellar abundances are approximately solar
(Sofia \& Meyer 2001), the atoms missing from the gas must be locked up
in dust grains.
In Fig.\ \ref{fig:depletions} the abundances have been plotted as a function
of ``condensation temperature'' -- the temperature at which cooling,
solar abundance gas at LTE would begin to condense a particular element
into a solid or liquid phase.
Elements with $T_{\rm cond}\gtsim 1200 \K$ are able to form 
``refractory'' solids with large binding energies, and it is therefore
not surprising that the strongly depleted elements tend to be those
with high values of $T_{\rm cond}$.
The observed elemental depletions provide a clue to the composition of
interstellar dust.
Restricting ourselves to abundant elements, we see that the
``electron donor'' species in interstellar dust material must be 
predominantly C, Mg, Si, and Fe.
The grains could contain a significant amount of O, and perhaps H.

\subsection{A Provisional Grain Model}

\begin{figure}
\begin{center}
\includegraphics[width=0.7\textwidth,angle=270]{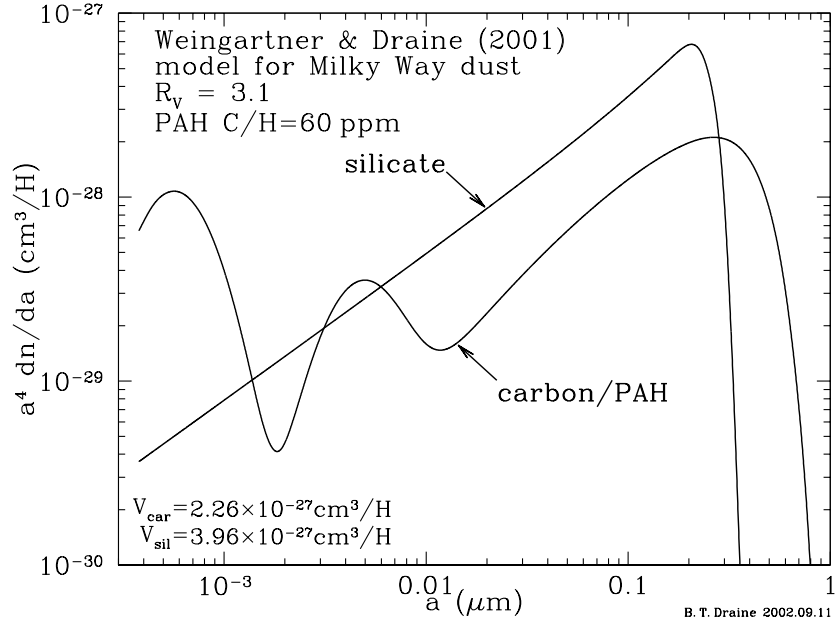}
\end{center}
\caption{WD01 size distribution for dust consistent with an
$R_V=3.1$ extinction curve for the local Milky Way.
The quantity plotted -- $a^4dn/da$ -- 
is proportional to the dust volume per logarithmic size interval.
}
\label{fig:sizedist}
\end{figure}
\begin{figure}
\begin{center}
\includegraphics[width=0.8\textwidth,angle=0]{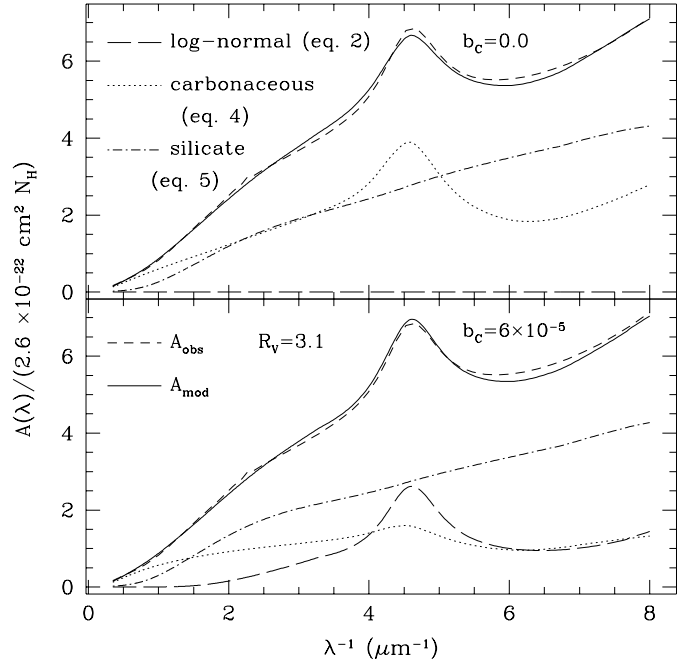}
\end{center}
\caption[]{Dashed line: average diffuse cloud extinction curve.
	Solid lines: model extinction curve for two extreme values of the
	ultrasmall carbonaceous grain abundance: $b_C=0$ (no PAHs) and
	$b_C=60$ppm C in PAHs.
	For each case the size distribution of carbonaceous and silicate grains
	was adjusted to obtain good overall agreeement with the 
	observed extinction.
	From Weingartner \& Draine (2001a).}
\label{fig:extcurv_mod}
\end{figure}
\begin{figure}
\begin{center}
\includegraphics[width=0.7\textwidth,angle=270]{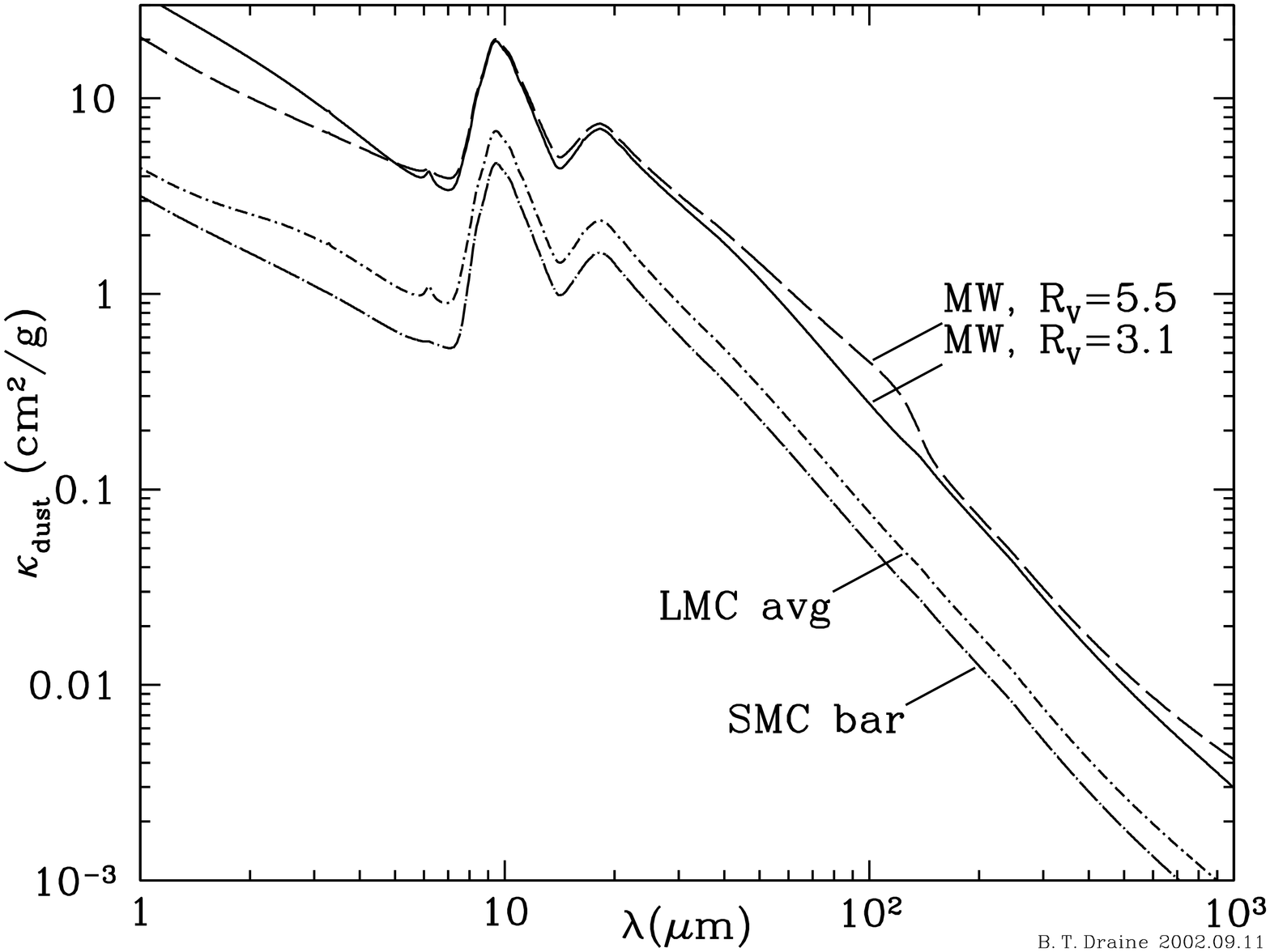}
\end{center}
\caption[]{Opacities $\kappa(\lambda)=$ 
(dust absorption cross section)/(mass of
gas + dust) for Milky Way dust with $R_V=3.1$ and $R_V=5.5$,
LMC average dust, and dust in the SMC bar, computed for the
carbonaceous/silicate dust model
(Weingartner \& Draine 2001a; Li \& Draine 2001).}
\label{fig:kappa}
\end{figure}

Below we will be discussing the astrophysics of interstellar dust.
It is helpful to have a specific grain model in mind so that one can
carry out calculations.
Accordingly, here we describe briefly a grain model which appears to
be consistent with observations.
The model consists of
a mixture of carbonaceous particles and silicate particles.
The carbonaceous particles have the physical and optical properties
of polycyclic aromatic hydrocarbon molecules when they are small
$a < 50\Angstrom$, or $N_{\rm C} \ltsim 6\times 10^4$ carbon atoms.
When they are larger, the carbonaceous grains are assumed to have
the optical properties of randomly-oriented graphite spheres.
The silicate grains are assumed to be amorphous silicates.

The emission in the $3-12\micron$ region from the diffuse ISM is attributed
to dust grains small enough that a single starlight photon can raise
the vibrational temperature high enough for thermal emission in the
observed vibrational modes.
In order for this emission to amount for $\sim21\%$ of the total radiated
power, very large numbers of very small dust grains are required.
Li \& Draine (2001) estimate that $\sim$15\% of the total carbon abundance
must be in carbonaceous particle containing $<10^3$ C atoms.
Since these particles must account for $\sim$20\% of the total absorption
of starlight, it is obvious that they must make a
signficant contribution to the interstellar extinction curve.

Weingartner \& Draine (2001a, hereafter WD01) 
have shown that there are dust size distributions
which include the required large numbers of ultrasmall carbonaceous grains
and which are compatible with the observed 
extinction curves for various
regions in the Milky Way, LMC, and SMC galaxies.
The size distributions are by no means unique -- the observed extinction
curve cannot be ``inverted'' -- 
but they are nevertheless strongly constrained,
so the WD01 size distributions are probably not too
far from the truth.  
Figure \ref{fig:sizedist} shows a grain size distribution appropriate for
diffuse clouds in the local Milky Way.
The total mass in grains with $a\ltsim 100\Angstrom$ is constrained by
the observed extinction in the ultraviolet, but since these grains
are in the ``Rayleigh limit'' the extinction is virtually independent of
the actual sizes and numbers of these particles, provided only that the
constraint on the total volume of these grains is satisfied.

The size distribution for $a\ltsim 15\Angstrom$ carbonaceous grains is
adjusted to make the predicted infrared emission agree with observations.
The bimodal nature of the carbonaceous size distribution for 
$a\ltsim 100\Angstrom$ is probably an artifact of the fitting procedure.

\subsection{Far-Infrared and Submm Opacities \label{sec:opacities}}

If dust grains in a cloud have a temperature $T_d$, the emergent intensity is
\beq
I_\nu = B_\nu(T_d) \left[1 - e^{-\tau_d}\right]
~~~,
\eeq
\beq
\tau_d(\lambda) = \kappa(\lambda) \int \rho ~ds
~~~,
\eeq
where the dust opacity
\beq
\kappa(\lambda) = 
\frac{1}{1.4n_{\rm H}m_{\rm H}}\int da \frac{dn_d}{da} C_\abs(a,\lambda)
~~~,
\eeq
$\rho$ is the total mass density (gas + dust),
and we have assumed $n_{\rm He}/n_{\rm H}=0.1$.
Dust opacities $\kappa$ are given in Fig.\ \ref{fig:kappa}; these opacities
are also available at {\tt http://www.astro.princeton.edu/$\sim$draine/dust/dust.html}.

At long wavelengths 
the optical depth $\tau_d \ll 1$.
If we can estimate the dust temperature,
the cloud mass per area can be obtained from the observed intensity
$I_\nu$ by
\beq
\int \rho~ds \approx \frac{1}{\kappa(\lambda)}
\frac{I_\nu}{B_\nu(T_d)}
~~~.
\eeq
The derived mass density depends sensitively on the value of $T_d$ unless
one is in the Rayleigh-Jeans limit, $\lambda \gtsim hc/kT_d$, in which
case $\int\!\rho\,ds \propto T_d^{-1}$.

\section{Optics of Interstellar Dust Grains}

Our knowledge of interstellar dust is based in large part on the
interaction of dust grains with electromagnetic radiation:
absorption, scattering, and emission.
The interaction of electromagnetic radiation with a ``target'' will
depend on the composition of the target (which determines the ``response''
of the target material to applied electric and magnetic fields) and the
geometry of the target -- both the shape and the size.
Lacking definite knowledge of both the composition and the grain geometry,
we must make some assumptions to proceed.

\subsection{Cross Sections, Scattering Matrices, and Efficiency Factors}

We are generally interested in calculating the following quantities
for unpolarized plane waves incident on the target:
\begin{itemize}
\item $C_\abs =$ the total absorption cross section.
\item $C_\sca =$ the total scattering cross section.
\item $dC_\sca/d\Omega =$ differential scattering cross section.
	This is related to the Muller matrix element $S_{11}$
	by $dC_\sca/d\Omega = S_{11}/k^2$, where $k=2\pi/\lambda$.
\item polarization $P$ for the radiation scattered in a particular
	direction.
\end{itemize}

In some cases we wish to explicitly consider polarized light.
For this case the Stokes vector
for the scattered radiation is obtained by multiplying the Stokes vector
of the incident radiation by the 4$\times$4 Mueller scattering matrix
$S_{ij}$.
The 4$\times$4 scattering matrix (which, for a fixed grain, 
is a function of direction of incidence and direction of scattering)
fully characterizes the scattering properties of a grain.
See Bohren \& Huffman (1983) or Mishchencko, Hovenier, \& Travis (2000)
for discussions of scattering concepts.
	
It is convenient to normalize scattering and absorption cross sections
by dividing them by a geometric cross section.
In the case of spherical target of radius $a$, the standard convention
is to simply divide by
the geometric cross section $\pi a^2$.
In the case of nonspherical targets, there is more than one convention in
use.  Some authors divide by the geometric cross section of the
target as seen from the direction of the illumination.
Other authors divide by the average geometric cross section for random
orientations.

My preference is to normalize by dividing by the geometric cross section
of an ``equal volume sphere''.
Thus for a target of mass $M$, and made of a material with density $\rho$,
the volume is $V=M/\rho$, and
we define ``efficiency factors'' $Q_\sca$, $Q_\abs$, 
$Q_\ext\equiv Q_\sca+Q_\abs$ by
\beq
Q = \frac{C}{\pi a_{\rm eff}^2}   ~~~~{\rm where}~~
a_{\rm eff} \equiv \left(\frac{3V}{4\pi}\right)^{1/3} ~~~.
\eeq

\subsection{Grain Geometry?}

Since interstellar dust is able to polarize starlight, we know that at least
some interstellar dust grains must be appreciably nonspherical.
Nevertheless, for many purposes we will approximate dust grains as
spheres for the simple reason that we can use ``Mie theory'' 
to calculate scattering
and absorption of light by spherically-symmetric targets with sizes
comparable to the wavelength.
Exact series solutions have been found for spheroids (Asano \& Yamamoto 1975,
Asano \& Sato 1980; Voshchinnikov \& Farafonov 1993) but numerical
calculations are much more challenging than for spheres.
Solutions for other shapes (other than the unphysical case of infinite
cylinders) are not available.

\subsection{Dielectric Functions\label{sec:dielectric}}

First, a few words about conventions:
We will follow c.g.s. electromagnetism, so that
$\epsilon=\mu=1$ for the vacuum.
We will represent plane waves as proportional to 
$e^{ikx-i\omega t}$; with this convention, absorption corresponds to
${\rm Im}(\epsilon) > 0$.

When an electric field is applied to a material, there are two
distinct types of response possible: the response of the {\it bound}
charge and the {\it free} charge.
The {\it bound} charge (e.g., electrons and nucleus within an
atom) undergoes a finite displacement when the electric field is
applied, with the individual atoms or molecules acquiring an electric dipole
moment in response to the applied electric field.
The response of the bound charge is characterized by a complex
dielectric function $\epsilon^{(bound)}(\omega)$.

The {\it free} charge responds in the form of an electric current density,
$J=\sigma E$, where $\sigma(\omega)$ is the electrical conductivity.

When dealing with a monochromatic electric field 
it is convenient to define a dielectric function $\epsilon$ describing
the response of both bound and free charge:
\beq
\epsilon = \epsilon^{(bound)}(\omega) + \frac{4\pi i \sigma(\omega)}{\omega}
~~~.
\eeq
With this definition, the free charge current $J$ is now absorbed into
the displacement current $(1/4\pi)\partial D/\partial t$,
so that 
Maxwell's equation becomes
\beq
\nabla\times B = \frac{1}{c}\frac{\partial D}{\partial t}
~~~,
\eeq
where $D=\epsilon E$.

{\bf\medskip\noindent\ref{sec:dielectric}.1 A Model Insulator\medskip}

It is instructive to consider a simple classical model for the
dielectric response of an insulator.  Suppose the insulator consists
of ``molecules'', with no permanent dipole moment, but which can
be polarized in response to the local applied electric field plus
the electric field from all of the other polarized molecules.
The dipole moment of a molecule is $qx$, where $x$ is the
displacement of the bound charge $q$.  Suppose the displacement $x$
behaves like a driven, damped harmonic oscillator,
\beq
m\ddot{x} = qE -m\omega_0^2 x -m\dot{x}/\tau_0
~~~,
\eeq
where $\omega_0$ is the resonant frequency, and $\tau_0$ is a damping time.
For a periodic electric field $E\propto e^{-i\omega t}$ we
can readily solve for the molecular polarizability $\alpha=qx/E$:
\beq
\alpha(\omega) = 
\frac{\alpha_0}{1-(\omega/\omega_0)^2-i\omega/(\omega_0^2\tau_0)}
~~~,
\label{eq:alpha}
\eeq
where
\beq
\alpha_0 \equiv \frac{q^2}{m\omega_0^2}
\eeq
is the zero-frequency polarizability of the molecule.
We now consider a medium consisting of a number density $n$ of such 
molecules.
Each now responds to the externally applied electric field
plus the electric field due to all of the other polarized molecules.
In the low frequency limit, the effective dielectric function $\epsilon$ of
the medium is related to the molecular polarizability through the
famous Clausius-Mossotti relation (see, e.g., Jackson 1975)
\beq
\epsilon = 1 + \frac{4\pi n\alpha}{1-(4\pi/3)n\alpha}
~~~.
\label{eq:clausius-mossotti}
\eeq
For molecules on a cubic lattice, the Clausius-Mossotti relation 
(\ref{eq:clausius-mossotti}) is exact in the limit $|m|kd\ll 1$,
where $d\equiv n^{-1/3}$ is the intermolecular spacing and 
$k\equiv\omega/c$.
For finite $kd$, corrections to $O[(kd)^3]$ have been derived by
Draine \& Goodman (1993).
Substituting (\ref{eq:alpha}) into (\ref{eq:clausius-mossotti}) we
obtain
\beq
\epsilon = 1 + 
\frac{4\pi n\alpha_0}
{1-(4\pi/3)n\alpha_0 - (\omega/\omega_0)^2 - i\omega/(\omega_0^2\tau_0)}
~~~.
\eeq
In the low frequency limit $\omega\rightarrow 0$, we have
\beq
{\rm Re}(\epsilon^{(bound)}) \rightarrow \epsilon_0^{(bound)} \equiv
1 + \frac{4\pi\alpha_0}{1-(4\pi/3)\alpha_0}
~~~,
\eeq
from which we note that
\beq
4\pi\alpha_0 = 3\left[\frac{\epsilon_0^{(bound)}-1}{\epsilon_0^{(bound)}+2}
\right]
~~~.
\eeq
At low frequencies, the imaginary part of $\epsilon^{(bound)}$ varies
linearly with frequency:
\beq
{\rm Im}(\epsilon^{(bound)}) \approx A\omega
\label{eq:Imomega_ins}
~~~,
\eeq
\beq
A =
\frac{3}{\omega_0^2\tau}
\frac{\epsilon_0^{(bound)}-1}{\epsilon_0^{(bound)}+2}
~~~.
\eeq
An insulator has no mobile charges ($\sigma=0$), and for our model
at low frequencies we have
\begin{eqnarray}
{\rm Re}\left(\epsilon\right)&\rightarrow& \epsilon_0^{(bound)} 
= const
~~~,
\\
{\rm Im}\left(\epsilon\right)&\rightarrow& A\omega
~~~.
\label{eq:Imeps_ins}
\end{eqnarray}

{\bf\medskip\noindent\ref{sec:dielectric}.2 A Model Conductor\medskip}

For a material with a density $n_e$ of free electrons, 
a simple classical model for the
electron dynamics would be 
\beq
m_e \ddot{x} = eE - m_e \dot{x}/ \tau_e
~~~,
\eeq
where $\tau_e$ is the electron collision time.
With this equation of motion, the 
conductivity $\sigma(\omega) \equiv n_e e \dot{x}/E$ 
is just
\beq
\sigma(\omega) = \frac{\sigma_0}{1-i\omega\tau_e}
~~~,
\eeq
where 
\beq
\sigma_0=n_e e^2\tau_e/m_e
=
\frac{\omega_p^2\tau_e}{4\pi}
~~~,
\eeq
where $\sigma_0$
is the d.c. conductivity, and $\omega_p\equiv(4\pi n_e e^2/m_e)^{1/2}$
is the plasma frequency.
For this case, the low-frequency behavior of $\epsilon$ is 
\begin{eqnarray}
{\rm Re}\left(\epsilon\right)&\rightarrow& 
\epsilon_0^{(bound)}-4\pi\sigma_0\tau_e = const
\label{eq:Reomega_cond}
~~~,
\\
{\rm Im}\left(\epsilon\right)&\rightarrow&
A\omega + \frac{4\pi\sigma_0}{\omega}
~~~.
\label{eq:Imomega_cond}
\end{eqnarray}

Comparing (\ref{eq:Imeps_ins}) and (\ref{eq:Imomega_cond}), 
we see that the low-frequency 
behavior of ${\rm Im}\epsilon$ is qualitatively
different for an insulator or for a conductor.
Below we will see what this implies for the frequency-dependence
of the absorption by small particles.

\subsection{Calculational Techniques}

In many astrophysical applications we are concerned with particles
which are neither very small nor very large compared to the wavelength
of the incident radiation.
There are several different
methods which can be used to 
calculate scattering and absorption cross-sections by targets which 
are neither very large nor very small compared to the wavelength:
\begin{itemize}
\item Mie theory solution for spherical targets (Bohren \& Huffman 1983).
	Mie theory codes are highly-developed and readily available.
	They break down due to numerical roundoff error when the target
	becomes too large compared to the wavelength.
	A modified version of the Bohren \& Huffman code is available
	at\\
	{\tt http://www.astro.princeton.edu/$\sim$draine/scattering.html}
\item Series solution for homogeneous spheroid
	or layered spheroid (boundaries must be confocal spheroids)
	(Asano \& Yamamoto 1975; Asano \& Sato 1980; Voshchinnikov \&
	Farafonov 1993).
	Codes for spheroids have been developed but have not been widely used.
	Codes based on the Asano \& Yamamoto treatment 
	had a reputation for being somewhat
	numerically delicate (Rogers \& Martin 1979).  
	Voshchinnikov \& Farafanov (1993) have developed a treatment based upon
	separation of variables which is reported to be more robust.
	They have recently generalized this to core-mantle spheroidal targets
	(Farafanov, Voshchinnikov, \& Somsikov 1996).

\item Extended Boundary Condition Method (EBCM), often referred to
	as the ``T-matrix method'' (Mishchenko, Travis \& Macke 2000).
	The EBCM method (Mischchenko, Travis \& Macke 2000) is used to
	construct the ``T matrix'' which gives the coupling between
	vector spherical harmonic components of the incoming wave and
	vector spherical harmonic components of the scattered wave.
	Once the T-matrix has been obtained, orientational averages can
	be efficiently calculated.
	The EBCM method appears to be
	well-suited to targets with rotational symmetry which are not extremely
	elongated.
	EBCM T-matrix codes have been made available by M. Mishchenko at
	{\tt http://www.giss.nasa.gov/$\sim$crmim/t\_matrix.html}
\item Discrete Dipole Approximation (Draine \& Flatau 1994; Draine 2000).
	The Discrete Dipole Approximation can be quite readily applied to
	complex geometries.  It has been used to calculate the optical
	properties of graphite particles, with a highly anisotropic dielectric
	tensor (Draine 1988; Draine \& Malhotra 1993).  In recent years it
	has been accelerated by use of FFT techniques (Goodman, Draine,
	\& Flatau 1991), and its accuracy has been improved by refinements
	in the assignment of dipole polarizabilities (Draine \& Goodman 1993).
	The Fortran code DDSCAT is available from
	{\tt http://www.astro.princeton.edu/$\sim$draine/DDSCAT.html} ,
	and a detailed User Guide is available (Draine \& Flatau 2000).

\end{itemize}

\subsection{Scattering by Homogeneous Isotropic Spheres}

The solution to Maxwell's equations for a plane wave incident on a
homogeneous and isotropic sphere was independently obtained by
Mie (1908) and Debye (1909).  The solution -- now generally referred to as
``Mie theory'' -- is given in terms of a series
expansion in powers of the ``size parameter'' $x\equiv 2\pi a/\lambda$.
The expansion converges, but the number of terms which must be retained is
of $O(x)$.
The details of the Mie theory solution are nicely described in the
excellent monograph by
Bohren \& Huffman (1983).
There are a number of computer codes to evaluate the absorption cross
section and differential scattering cross sections given by Mie scattering
theory.
When $x$ is large, care must be taken to avoid errors due to
finite precision arithmetic in numerical evaluation
of the series solution.

For an incident monochromatic 
plane wave, the full Mie theory solution
depends on the dielectric function $\epsilon(\omega)$
-- describing
the electric polarization of the material in response to an applied
electric field oscillating at angular frequency $\omega$ --
and
the magnetic permeability function $\mu(\omega)$ --
characterizing the magnetization
of the material in response to an applied oscillating magnetic fields.
Because the magnetic response is generally negligible 
(i.e., $|\mu-1| \ll |\epsilon -1|$) at frequencies
above 100~GHz) it is customary to neglect the magnetization of the
grain material except insofar as it is due to eddy currents.

\begin{figure}
\begin{center}
\includegraphics[width=.8\textwidth,angle=0]{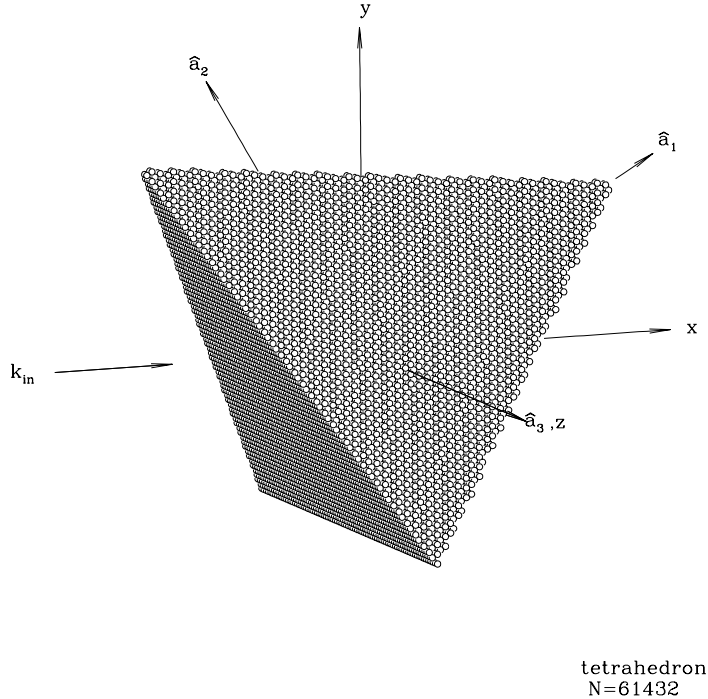}
\end{center}
\caption[]{Representation of a tetrahedron with a dipole array.
	The dipoles are actually pointlike -- here they are
	represented by small spheres for purposes of visualization.
	We consider incident radiation propagating along the 
	direction of the vector ${\bf k}_{\rm in}$ illustrated here.
	Taken from Draine (2000).
	}
\label{fig:tetratarg}
\end{figure}

\begin{figure}
\begin{center}
\includegraphics[width=.8\textwidth,angle=0]{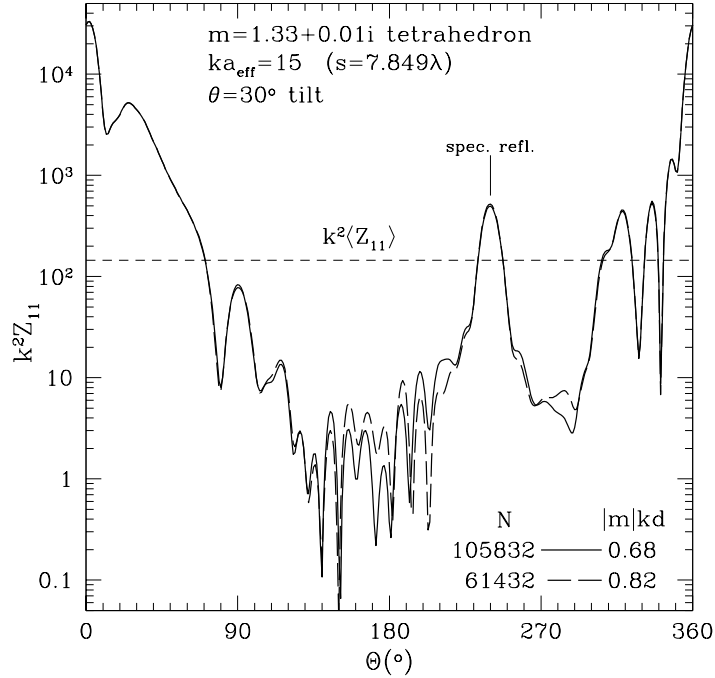}
\end{center}
\caption[]{$S_{11}=k^2 dC_\sca/d\Omega$ for scattering of unpolarized
	light bya tetrahedron,
	for scattering 
	directions in the x-y plane (see Fig.\ \ref{fig:tetratarg}),
	as a function of scattering angle $\Theta$.
	The tetrahedron has refractive index $m=1.33+0.01i$,
	has edges of length $s=7.849\lambda$, where $\lambda$ is
	the wavelength of the incident light in vacuo, and is
	oriented with an angle $\theta=30^\circ$ between $\bf{k}_{\rm in}$
	and $\hat{\bf a}_1$.
	The peak at $\Theta=240^\circ$ corresponds
	to the direction of specular reflection for geometric
	optics.
	Results are shown for a tetrahedron represented by
	$N=61432$ and $N=105832$ dipoles.
	The scattered intensities are in excellent agreement in all directions
	where the scattering is at all strong.
	The dashed line shows $k^2\langle Z_{11}\rangle=C_\sca/4\pi$.
	Taken from Draine (2000).
	}
\label{fig:tetra_15}
\end{figure}

\subsection{Discrete Dipole Approximation}

The discrete dipole approximation consists in replacing the (continuum)
target of specified dielectric function $\epsilon$ 
with an array of polarizable points (referred to as ``dipoles'').  
The array geometry is
chosen to mimic the target geometry.
The polarizabilities $\alpha$ of the points are chosen so that
an infinite lattice of such polarizable points would have the
same dispersion relation as the material of dielectric function $\epsilon$
(Draine \& Goodman 1993).
With FFT techniques employed to speed the calculation, it is now
feasible to calculate scattering and absorption by targets represented
by more than 100,000 polarizable dipoles on a workstation with 256 MB of
RAM.
There are great efficiencies if the dipoles are situated on a cubic lattice,
so DDSCAT requires that this be the case.

The DDA can be applied to inhomogeneous targets and targets with
complex geometries.
As an example of the type of problem which can be solved, in 
Fig.\ \ref{fig:tetratarg} we show a discrete-dipole array of 61432
dipoles intended to approximate a tetrahedral target.
In Fig.\ \ref{fig:tetra_15} we show the calculated scattered intensity
in the x-y plane as a function of scattering angle $\Theta$.
Notice the pronounced scattering peak at $\Theta=240^\circ$ -- this
is the direction where one would have a specular reflection peak
in the geometric optics limit.

For a fixed target size $D$, complex refractive index $m$,
and wavelength $\lambda$,
the DDA converges to the exact answer in the limit where the interdipole
separation $d\rightarrow0$: the target structure is well-resolved and
the dipole separation is small compared to the wavelength in the target
material.  
The criterion for separation small compared to $\lambda$ can be
written $|m|kd\ltsim 1$.
If $|m|$ is not too large, we find that
the overall scattering and absorption cross section is calculated fairly
accurately provided $|m|kd\ltsim 1$; the detailed scattering pattern will
be calculated accurately if $|m|kd\ltsim 0.5$.
For example, the two calculations in Fig. \ref{fig:tetra_15},
with $|m|kd=0.82$ and 0.68, are already
in fairly good agreement.

\subsection{Infrared and Far-Infrared}

When the wavelength is long compared to the target size, we may
use the ``dipole approximation'' (Draine \& Lee 1984) and
characterize the response of the target to the incident electromagnetic
field purely in terms of the (oscillating) electric and magnetic dipole 
moments of the target.
We may write
\beq
C_\abs = \frac{4\pi\omega}{c} 
\left[ {\rm Im}(\alpha_E^e) + {\rm Im}(\alpha_H^m)\right]
~~~,
\eeq
\beq
C_\sca = \frac{8\pi}{3}\left(\frac{\omega}{c}\right)^4 
\left[ |\alpha_E^e|^2 + |\alpha_H^m|^2\right]
~~~,
\eeq
where $\alpha_E^e$ is the (complex) 
electric polarizability along the direction of
the incident electric field vector, and
$\alpha_H^m$ is the (complex) magnetic polarizability along the direction
of the incident magnetic field vector.

Consider a homogeneous ellipsoidal particle composed of an isotropic material,
with semiaxes $a,b,c$.
If the applied electric field is along one of its principal axes,
the electric polarizability is
\beq
\alpha_E^e = \frac{V}{4\pi}\frac{\epsilon-1}{(\epsilon-1)L_E+1}
~~~,
\eeq
where $L_E$ is the ``shape factor'' along the direction of the applied
E field, and
$V=4\pi abc/3$ is the volume of the ellipsoid.
The shape factors along the three principal axes satisfy the
sum rule $L_1+L_2+L_3=1$ and may be obtained by numerical quadrature
(Bohren \& Huffman 1983).

Consider now a spheroid with semiaxes $a,b,b$ (prolate if $a/b>1$, oblate
if $a/b<1$).
For this case we have analytic expressions for the shape factors $L_a$ and 
$L_b=(1-L_a)/2$.
For a prolate spheroid ($a<b$) we have (van de Hulst 1957)
\beq
L_a = \frac{1-e^2}{e^2}
\left[\frac{1}{2e}\ln\left(\frac{1+e}{1-e}\right)-1\right]
~~~,
\eeq
and for an oblate spheroid ($a<b$) we have
\beq
L_a = \frac{1+e^2}{e^2}\left[1-\frac{1}{e}\arctan(e)\right]
~~~,
\eeq
where
\beq
e^2\equiv |1-(b/a)^2| ~~~.
\eeq

Exact solutions for $\alpha_E^e$ in the limit $\lambda\gg a_\eff$
are also available for layered grains
provided the interfaces are confocal spheroids (Gilra 1972; 
Draine \& Lee 1984).

For spheres we have $L_a=L_b=1/3$, and
\beq
C_\abs = \frac{9\omega V}{c}\frac{\epsilon_2}{(\epsilon_1+2)^2+\epsilon_2^2}
~~~,
\label{eq:cabs_fir}
\eeq
where $\epsilon_1\equiv{\rm Re}(\epsilon)$ and
$\epsilon_2\equiv{\rm Im}(\epsilon)$.

Eq. (\ref{eq:cabs_fir}) contains an important result: 
the absorption cross section of a grain with size $\ll\lambda$ is
simply proportional to the grain volume!
The far-infrared opacity therefore
depends only on the total volume of grain material present, but not
on the sizes of the particles, provided only that they are small compared
to the wavelength $\lambda$.

How do we expect the absorption cross section to depend on frequency
$\omega$ at low frequencies?
For an insulator, from (\ref{eq:Imeps_ins}) we see that
\beq
C_\abs \rightarrow \frac{9V}{c} \frac{A}{(\epsilon_0+2)^2}~~\omega^2
~~~,
\eeq
while for a conductor, we see from (\ref{eq:Imomega_cond}) that
\beq
C_\abs \rightarrow \frac{9V}{c} \left(\frac{1}{4\pi\sigma_0}\right)
~~\omega^2
~~~.
\eeq
Thus we see that for both insulating grains and conducting grains,
a simple physical model leads to $C_\abs \propto \omega^2$ at low
frequencies!
This is the basis for the expectation that dust opacities should
vary as $\lambda^{-2}$ in the far-infrared, as in the dust models
of Draine \& Lee (1984).

Now the above discussion has been based on simple classical models
for the response of charge to applied electric fields.
Real materials obey quantum mechanics, so results may differ.
Furthermore, the above discussion assumed that $\omega \ll \omega_0$ in
discussion of the response of the bound charge, but an amorphous
material could have very low frequency vibrational modes, so that
the assumption that $\omega \ll \omega_0$ might not be valid in
the far-infrared.
This might explain why some laboratory studies of amorphous materials
find different behavior: for example,
Agladze et al.\, (1996) find that the absorption coefficient
for amorphous MgO$\cdot$2SiO$_2$ varyies as $\omega^{1.2}$
between $800\micron$ and 4 mm.

Nature must be our guide, but it is worth keeping in mind that
opacities varying as $\omega^2$ emerge naturally from simple models.

\subsection{Kramers-Kronig Relations}

Purcell (1969) pointed out that the Kramers-Kronig relations can provide
useful constraints in dust modelling.
The Kramers-Kronig relations are general relations which apply
to ``linear response functions'', such as a dielectric function, which
specifies the response (e.g., the electric polarization) to
an applied stress (e.g., an applied electric field).
The {\it only} assumptions are that (1) the response is linear, and
(2) the system is causal -- the response can depend on the stress applied
in the past, but cannot depend upon the future.
With these very simple assumptions, it is possible to derive
the Kramers-Kronig relations (see Landau \& Lifshitz 1960 for a nice
derivation).
In the case of the dielectric function, the Kramers-Kronig relations are
\begin{eqnarray}
\epsilon_1(\omega_0) &=& 1 + \frac{2}{\pi} P \int_0^\infty d\omega
\frac{\omega \epsilon_2(\omega)}{\omega^2-\omega_0^2}
~~~,
\label{eq:kkreal}
\\
\epsilon_2(\omega_0) &=& \frac{2}{\pi}\omega_0 P \int_0^\infty d\omega
\frac{\epsilon_1(\omega)}{\omega_0^2-\omega^2}
~~~,
\label{eq:kkimag}
\end{eqnarray}
where $P$ indicates that the principal value is to be taken.

Thus the real and imaginary parts of $\epsilon(\omega)$ are by no means
independent -- if either one is known at all frequencies, the other
is fully determined.
In order to have a physically acceptable dielectric function, one approach
(see, e.g., Draine \& Lee 1984) is to specify $\epsilon_2$ at all 
frequencies, and then construct $\epsilon_1(\omega)$ using
(\ref{eq:kkreal}).

\subsection{Kramers-Kronig Relations for the ISM \label{sec:KK_ISM}}

Purcell showed that one could apply the Kramers-Kronig relations directly
to the interstellar medium.  Plane waves propagate through the
interstellar medium but are attenuated by scattering and absorption.
One can describe this attenuation by an imaginary component of
the dielectric constant $\tilde{\epsilon}$
of the interstellar medium,
where $|\tilde{\epsilon}-1|\ll 1$.

Recall that for a electromagnetic 
plane wave of frequency $\omega$, the wave vector
is $k = m(\omega) \omega/c$, where $m(\omega) = \sqrt{\epsilon}$ 
is the complex refractive index.
The electric field intensity decays as 
\beq
E \propto \exp\left[-{\rm Im}(k)x\right] = 
\exp\left[-(1/2){\rm Im}(\tilde{\epsilon})\omega/c\right]
~~~.
\eeq
The energy in the wave is proportional to $|E|^2$ and therefore
decays twice as rapidly as $E$.
Thus
\beq
n_{gr} C_\ext(\lambda) = \frac{\omega}{c} \tilde{\epsilon}_2
~~~.
\eeq

We can now apply the Kramers-Kronig relation (\ref{eq:kkreal}) to
obtain the real part of the dielectric function $\tilde{\epsilon}$
of the ISM at zero frequency:
\beq
\tilde{\epsilon}_1(0) -1 = \frac{2}{\pi}
\int_0^\infty \frac{d\omega}{\omega} \frac{c}{\omega} n_{gr} C_\ext(\omega)
~~~.
\eeq

\begin{figure}
\begin{center}
\includegraphics[width=.7\textwidth,angle=270]{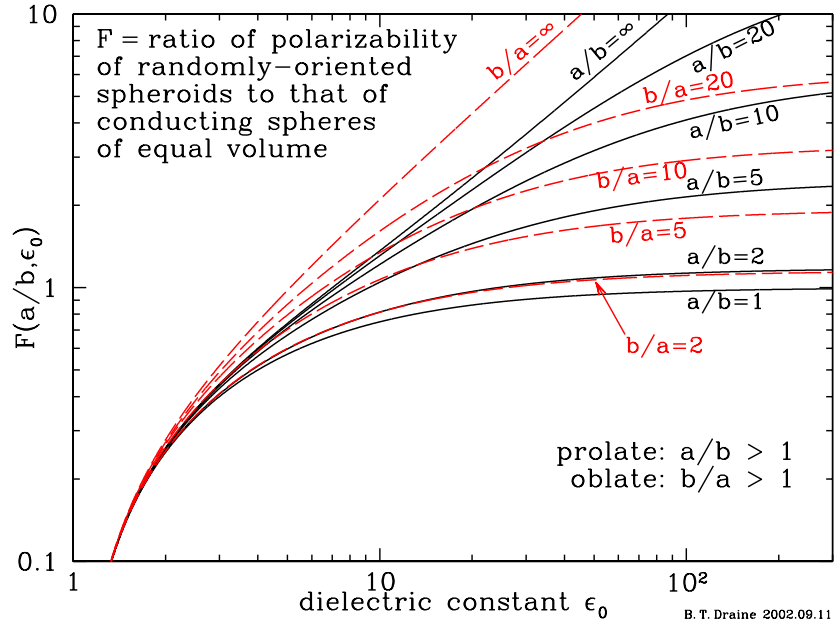}
\end{center}
\caption[]{Function $F(a/b,\epsilon_g)$ as a function of the static
	dielectric constant $\epsilon_0$, for selected values of
	axial ratio $a/b$.
	After Purcell (1969).
	}
\label{fig:purcellF}
\end{figure}

The static dielectric function of the ISM 
is directly related to the electric polarizability $\alpha_{gr}$
of the grains in it:
\beq
\tilde{\epsilon}_1(0) - 1 = 4\pi n_{gr} \alpha_{gr}(0)
~~~.
\eeq
Thus
\begin{eqnarray}
4\pi n_{gr} \alpha_{gr}(0) &=& \frac{2}{\pi} c n_{gr}
\int_0^\infty d\omega \frac{1}{\omega^2} C_\ext(\omega)
\\
&=& \frac{1}{\pi^2} n_{gr} \int_0^\infty d\lambda ~ C_\ext(\lambda)
~~~.
\end{eqnarray}
We have seen above how the static polarizability of a spheroidal 
grain depends on
its volume, shape, and dielectric function.
Averaged over random orientation, we have (for a dielectric grain)
\beq
4\pi\alpha_\gr(0) = 3 V F(a/b,\epsilon_0)
~~~,
\label{eq:alpha_vs_F}
\eeq
\beq
F(a/b,\epsilon_0) \equiv \frac{(\epsilon_0-1)}{3}
\left[
\frac{1}{(\epsilon_0-1)3L_a + 3} + \frac{2}{(\epsilon_0-1)3L_b + 3}
\right]
~~~.
\label{eq:Fdef}
\eeq
$F$ is just the orientationally-averaged polarizability relative to
the polarizability of an equal-volume conducting sphere.
Using (\ref{eq:alpha_vs_F}), we obtain the grain volume per H atom
in terms of an integral over the extinction per H atom:
\beq
\frac{n_{gr}V}{n_{\rm H}} 
=\frac{1}{3\pi^2 F}\int_0^\infty d\lambda ~ 
\frac{n_{gr}}{n_{\rm H}} C_\ext(\lambda)
\label{eq:grain_vol}
\eeq

Now consider a conducting material.
From (\ref{eq:Reomega_cond}-\ref{eq:Imomega_cond}) we see
that $\epsilon\rightarrow i\times\infty$, so that
\beq
F \rightarrow \frac{1}{9L_a} + \frac{2}{9L_b}
\eeq
so that a conducting sphere ($L_a=L_b=1/3$) has $F(a/b=1,\epsilon_0=\infty)=1$.

We do not know the grain shape, and we do not know the static dielectric
constant $\epsilon_0$.
However, from Figure \ref{fig:purcellF} we see that $F < 1.5$ unless the
grain is {\it extremely} elongated ($a/b > 20$ or $b/a > 20$)
{\it and} the static dielectric function is very large ($\epsilon_0>10$) --
as for a metal.
For a reasonable grain shape (e.g., $0.5<a/b<2$) and dielectric function
$\epsilon_0\approx 3$ we have $F\approx 0.4$.

Purcell's analysis is a delight, and has two important consequences:

{\bf\medskip\noindent\ref{sec:KK_ISM}.1 Grain volume per H \medskip}

It is of course not possible to measure the extinction per H atom
at wavelengths from 0 to $\infty$.
However, since $C_\ext>0$, measurements over a finite wavelength range
can be used to obtain a lower bound on $FV n_{gr}/n_{\rm H}$.
The extinction per H nucleon is fairly well-known from $0.1\micron$ to
30$\micron$; a numerical evaluation gives
\beq
\int_{0.1\micron}^{30\micron}\frac{\tau_\ext(\lambda)}{N_\rmH} d\lambda
\approx 1.1\times 10^{-25}\cm^3/\rmH
~~~.
\eeq
Approximately 50\% of the integral is contributed by wavelengths
$0.1 < \lambda < 1\micron$.
If we estimate $F\approx 0.4$ this gives 
\beq
\frac{n_{gr}V}{n_\rmH} \gtsim 9.3\times10^{-27}\cm^3/\rmH
~~~,
\eeq
or, for an assumed grain mass density $\rho=2.5\g\cm^{-3}$ (intermediate
between graphite and silicate) we have a lower bound on the ratio of
grain mass to H mass:
\beq
\frac{M_{gr}}{M_\rmH} \gtsim .014
~~~.
\eeq

{\bf\medskip\noindent\ref{sec:KK_ISM}.2 Asymptotic Behavior 
of $C_\ext$ at Long Wavelengths \medskip}

Eq. (\ref{eq:grain_vol}) tells us that 
$\int_0^\infty C_\ext(\lambda)d\lambda$
must be convergent, and therefore that $C_\ext(\lambda)$ {\it must} decline
more rapidly than $1/\lambda$:
grain models in which $C_\ext\propto 1/\lambda$
for $\lambda\rightarrow\infty$
are unphysical.
It is of course possible to have $d\ln C_\ext/d\ln\lambda \approx -1$
over a limited range of wavelengths, but this cannot be the asymptotic
behavior.

\begin{figure}
\begin{center}
\includegraphics[width=0.7\textwidth,angle=270]{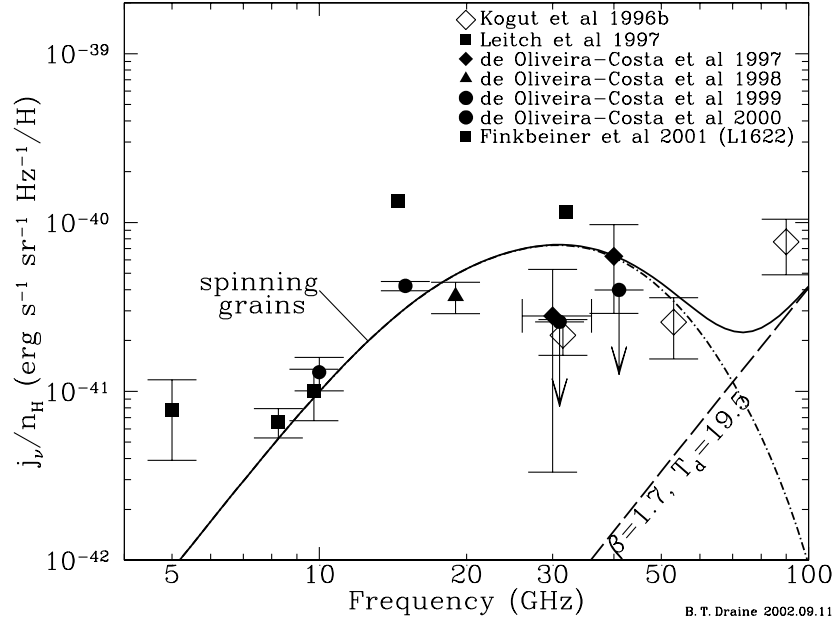}
\end{center}
\caption[]{
	  Microwave sky brightness correlated with 100$\micron$
	  emission.  Shown is the estimated microwave emissivity
	  per unit H nucleon.
	  }
\label{fig:microwave}
\end{figure}

\subsection{Microwave}

Dust grains would be expected to radiate thermally at microwave frequencies,
but this emission was expected to be quite weak, based on an extrapolation
from the thermal emission peak at $\sim100\micron$.
It was therefore a surprise when sensitive maps of the microwave
sky brightness (for the purpose of studying angular structure in the
cosmic background radiation) showed
relatively strong emission correlated with the Galactic $100\micron$
emission, and therefore with interstellar dust
(Kogut et al.\ 1996a,b).
Figure \ref{fig:microwave} shows the inferred microwave emissivity of
per H nucleon.
The dashed line labelled ``$\beta=1.7, T_d=19.5$'' is the contribution 
expected from thermal emission from dust grains if the dust opacity
$\propto \lambda^{-1.7}$ for $\lambda \gtsim 100\micron$.
We see that at 30 GHz ($\lambda = 1\cm$ 
the observed microwave emission is two orders
of magnitude above the value expected from classical dust grains.

The power radiated in the microwave is a tiny fraction of the total, so
this microwave emission is not an important source of cooling.
Nevertheless, we would like to know what process is responsible.
Possible mechanisms include synchrotron emission from
relativistic electrons, or free-free emission from a thermal plasma, but
these seem unable to account for the observed 15-50 GHz emission
(see Draine \& Lazarian 1999b).
We are left with interstellar dust.

The possibility of radio emission from rotating dust grains dates back
to Erickson (1957).
As discussed above, the strong infrared emission in the $3-12\micron$ range
appears to require a very large population of very small dust grains.
Ferrara \& Dettmar (1994) pointed out that these dust grains would produce
detectable radio emission if they were undergoing Brownian rotation
at the gas temperature.

Draine \& Lazarian (1998a,b) analyzed the grain rotational dynamics, and
showed that the population of ultrasmall grains required to understand
the infrared emission would be expected to produce microwave emission,
with a predicted spectrum (and intensity) as shown 
in Fig.\ \ref{fig:microwave}.  We will discuss the grain dynamics in
\S\ref{sec:rotation}, but this rotational emission still seems likely
to contribute a substantial fraction of the observed dust-correlated
microwave emission.
In principle, these rapidly spinning dust grains could be aligned,
leading to polarized emission, but the polarization fraction has been
estimated to be small (Lazarian \& Draine 2000).

However, if dust grains contain magnetic materials -- which does not
seem implausible, considering the large amount of Fe in interstellar grains --
there could also be {\it thermal} magnetic dipole emission from dust.
Consider a ferromagnetic domain in a dust grain.
This domain is spontaneously-magnetized: it is energetically favorable
to have alignment of electron orbital
angular momentum and electron spins.
The lowest energy state has some specific magnetization, but there are
nearby energy states where the domain has uniform magnetization with 
the same net magnetic
moment but in a slightly different direction.
In thermal equilibrium, it is possible for these higher energy states to
be excited, so that the magnetization direction will fluctuate.
This results in a time-varying magnetic dipole moment, and hence
magnetic dipole radiation.

The optics of magnetic dust grains with size small compared to the
wavelength has been developed by Draine \& Lazarian (1999a; hereafter
DL99a), and
the magnetic susceptibility $\mu(\omega)$ was estimated for several
magnetic materials, including metallic Fe and magnetite Fe$_3$O$_4$.
The spectrum and intensity of this radiation was estimated by DL99a, and 
it was found that {\it if} a substantial
fraction of the Fe in interstellar grains is incorporated into magnetic
materials (e.g., magnetite), the resulting thermal emission
could account for an appreciable fraction of the observed microwave emission!

How will we be able to distinguish between spinning dust grains and
magnetic dust grains as sources of microwave emission?  DL99a point out
that small grains are believed to be underabundant in dense
clouds (they presumably coagulate to form bigger grains), so that
the rotational emissivity per H nucleon would be reduced.  The
magnetic dipole emission from magnetic grain materials does not, however,
depend on the grain size.
Hence if
we see strong microwave emission from dust in dense clouds, this would
suggest that magnetic dust grains may be responsible.
Future pointed observations of dense clouds may answer this question.

\subsection{X-rays}

The scattering of X-rays by dust grains was first discussed by
Overbeck (1965), Slysh (1969), and Hayakawa (1970).
Because the refractive index of grain materials is close to unity at
X-ray energies, the scattering is through small angles.
As a result, images of X-ray point sources show a ``halo'' of
dust-scattered X-rays.  The observable halo can extend for tens of
arcminutes from the source.

For sufficiently small grains, this scattering can be calculated using
the Rayleigh-Gans approximation (see Bohren \& Huffman 1983), which is
valid if $|m-1|ka\ll 1$, where $a$ is the grain radius.
Because X-rays have large $k$, this condition may not be satisfied for
the larger interstellar grains, in which case one should resort to
full Mie theory.  This can be numerically challenging because the
number of terms which must be retained is of order 
$ka \approx 507 (a/0.1\micron)(E/{\rm keV})$, so high numerical
accuracy is required.
Smith \& Dwek (1998) have compared scattering halos estimated using the
Rayleigh-Gans approximation with Mie theory, and show that the Rayleigh-Gans
approximation fails for energies below 1 keV.
Scattering by grains with $|m-1|ka\ll 1$ and $ka\gg 1$ can be treated using
``anomalous diffraction theory'' (van de Hulst 1957).

Nova Cygni 1992, a bright X-ray nova, was observed by
ROSAT.
The observed X-ray halo has been compared to predictions for different
grain models by Mathis et al. (1995),
Smith \& Dwek (1998), and
Witt, Smith \& Dwek (2001), with differing conclusions.
The most recent study (Draine \& Tan 2002) concludes that our standard
dust model is in reasonable agreement with the observed X-ray halo.
Future observations of X-ray scattering halos
will be valuable to test and constrain grain models.

\section{IR and Far-IR Emission from Interstellar Dust}

\subsection{Heating of Interstellar Dust}

While dust grains can be heated by collisions with gas atoms and molecules,
starlight usually is the dominant heating process.
The rate of energy deposition, the rate of photon absorptions, and
the mean energy per absorbed photon are
\beq
\langle dE/dt\rangle_\abs =
\int_0^\infty d\nu ~ u_\nu ~ c ~ C_\abs(\nu)
~~~,
\eeq
\beq
\dot{N}_\abs = \int_0^\infty d\nu \frac{u_\nu}{h\nu} c C_\abs(\nu)
~~~,\eeq
\beq
\langle h\nu\rangle_\abs = \frac{\langle dE/dt\rangle_\abs}
{\dot{N}_\abs}
~~~,
\eeq
where $u_\nu d\nu$ is the photon energy density in $(\nu,\nu+d\nu)$,
and $C_\abs(\nu)$ is the photoabsorption cross section.
We assume that all of the energy of an absorbed photon is converted to
heat; this is not exactly correct, since energetic photons may eject
a photoelectron, or excite fluorescence, but these processes take away
only a small fraction of the total absorbed power.

Now suppose that the dust grain has energy $E$ distributed among its
vibrational degrees of freedom.  Except when the energy $E$ is
very small, the number of vibrational modes which can be excited is
very large.  The density of states has been discussed by Draine \& Li (2001):
even for a small molecule like coronene C$_{24}$H$_{12}$, there are 
$\sim10^{20}$
different states with total energy 
$E< 1\eV$; for C$_{4000}$H$_{1000}$ this number becomes $\sim10^{135}$.
As a result, the statistical notion of ``temperature'' can be used, even
for quite small grains: a grain with vibrational energy content $E$ is
assumed to be characterized by a temperature $T(E)$ such that if the
grain were in contact with a heat bath at temperature $T$, the expectation
value for its energy would be $E$.

At temperature $T$, the average power radiated by a grain is
\beq
\left(\frac{dE}{dt}\right)_{rad} = 
\int_0^\infty d\nu C_\abs(\nu)4\pi B_\nu(T)
~~~,
\eeq
where $C_\abs(\nu)$ is the angle-averaged absorption cross section for
photons of frequency $\nu$, and 
\beq
B_\nu(T) \equiv \frac{2 h\nu^3}{c^2}\frac{1}{e^{h\nu/kT}-1}
\label{eq:planck_func}
\eeq
is the Planck function.

It is now natural to determine the ``steady-state'' temperature $T_\stst$
at which the power radiated equals the power absorbed:
\beq
\int_0^\infty d\nu C_\abs(\nu)4\pi B_\nu(T_\stst) = 
\langle dE/dt\rangle_\abs
~~~.
\eeq
Once we solve for $T_\stst$, we
calculate the thermal energy content of the grain at this temperature,
$E(T_\stst)$.  There are two regimes:
\begin{itemize}
\item If $E(T_\stst)\gg \langle h\nu\rangle_\abs$, then individual
	photon absorption events do not substantially change the
	energy content of the grain, and we may assume that the
	grain temperature $T=T_\stst$.
	This is the ``steady heating'' regime, where we can approximate
	the discrete heating events as a continuous process.
\item If $E(T_\stst)\ll \langle h\nu\rangle_\abs$, then individual photon
	absorption events heat the grain up to peak temperatures 
	$T_{max} \gg T_\stst$, and the grain will usually cool to a temperature
	$T_{min} < T_\stst$ before the next photon absorption event.
	This is the ``stochastic heating'' regime, where the discrete
	nature of the heating has important consequences.
	As we will see below, stochastic heating is often important.
\end{itemize}
\begin{figure}[ht]
\begin{center}
\includegraphics[width=.75\textwidth,angle=270]{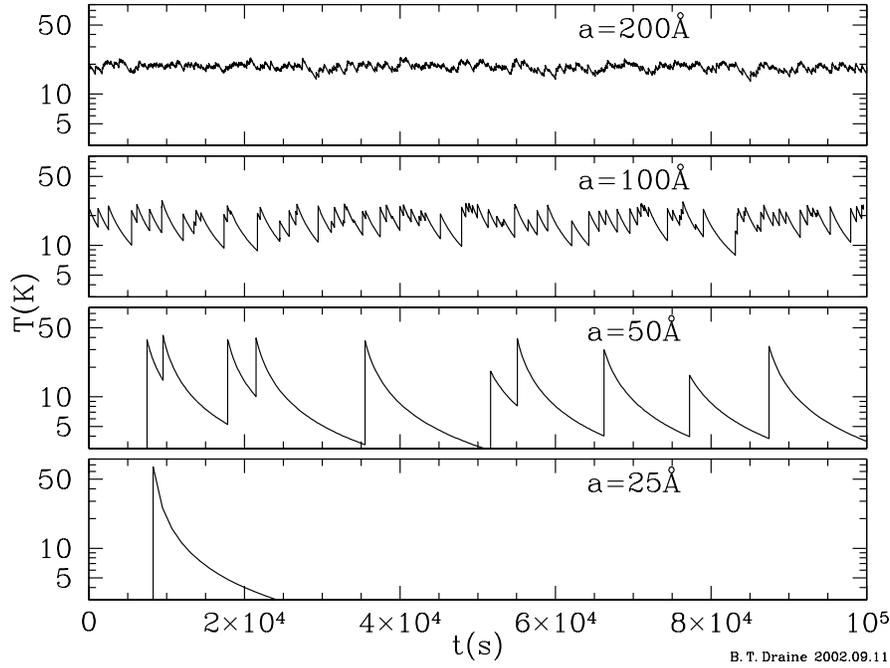}
\end{center}
\caption{A day in the life of an interstellar grain:
grain temperature vs. time for 4 grain sizes,
for grains heated by the average interstellar radiation field.
Grains with $a\gtsim200\Angstrom$ have a nearly constant temperature,
but $a\ltsim 100\Angstrom$ grains show conspicuous increases in
temperature following each photon absorption, with gradual cooling
between photon absorption events.}
\label{fig:T_vs_t}
\end{figure}
Figure \ref{fig:T_vs_t} shows the temperature history of 4 
grains of different sizes in the diffuse interstellar medium
over a $\sim10^5\s$ interval -- about one day.  The 
temperature of the $a=200\Angstrom$
grain fluctuates in a small range around $\sim20\K$.
However, as the grain size is reduced, the fluctuations become increasingly
extreme, with the peaks being higher and the low points being lower.
The smallest grain shown ($a=25\Angstrom$) absorbs about 1 starlight photon
per day; this grain spends most of its time quite cool, but immediately
following absorption of a photon the grain temperature reaches $\sim50\K$.
The grain cools by infrared emission; it is obvious that most of the infrared
emission must take place during the brief interval while the grain is ``hot''.
If we wish to calculate the time-averaged emission spectrum, we cannot use
the ``average'' grain temperature -- we need to integrate over a
distribution of grain temperatures.
This phenomenon is even more pronounced for smaller grains.
A grain with a radius of 5\AA\ has a volume only 1/125 that of the 25\AA\
shown in Figure \ref{fig:T_vs_t}; the mean time between photon absorptions
is of order $\sim10^7\s$, but a 10 eV starlight photon can raise the grain 
temperature to $\sim10^3\K$.

Let
$P(E)$ be the probability of the grain having vibrational energy 
$E^\prime \geq E$.
In the steady heating regime, we approximate $P(E)$ as a step function,
and $|dP/dE|$ as a delta function
$\delta(E-E_\stst)$.

\begin{figure}
\begin{center}
\includegraphics[width=.8\textwidth,angle=0]{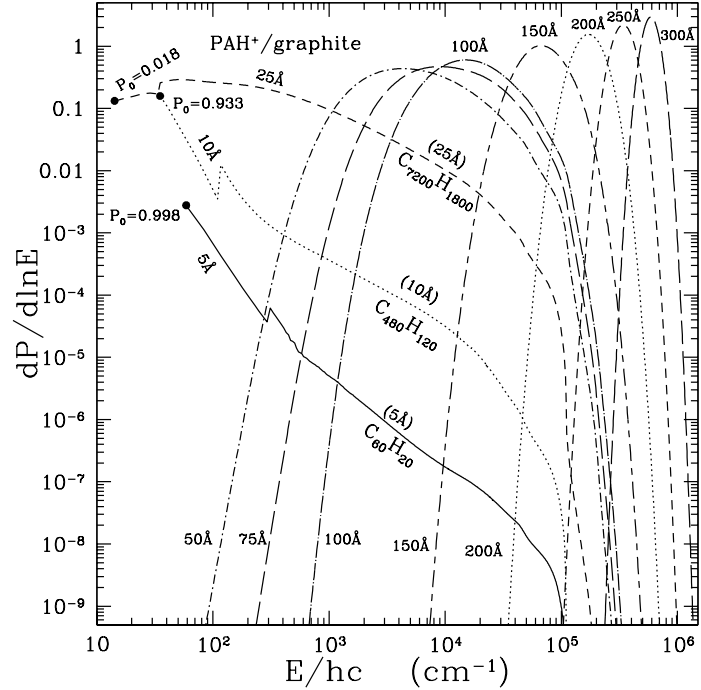}
\end{center}
\caption[]{Energy distribution functions for charged carbonaceous
	grains with radii $a=5,10,25,50,75,100,150,200,250,300\Angstrom$
	in the interstellar radiation field.
	The discontinuity in the 5, 10, and 25$\Angstrom$ curves is an
	artifact due to a change in the method of estimating the cooling
	when the energy is equal to the 20th vibrational mode.
	For 5, 10, and 25\AA, a dot indicates the first excited state,
	and the probability $P_0$ of being in the vibrational ground
	state is given.
	Taken from Li \& Draine (2001).
	}
\label{fig:dPdE}
\end{figure}

In the stochastic heating regime, we must solve for $P(E)$.  
Draine \& Li (2001a) show examples of energy
distribution functions.  If the energy states are grouped into bins $j=0,1,...$
(where $j=0$ is the ground state), then we can calculate the state-to-state
transition rates $T_{ji}$ for transitions $i\rightarrow j$ 
due to both photon absorptions and
photon emissions.  Once $T_{i\rightarrow j}$ is known, we define the diagonal
elements $T_{ii}\equiv-\sum_{j\neq i}T_{ji}$.
The steady state solution $P_j$ for the probability of being in state $j$
then satisfies the $N$ coupled linear equations
\beq
0 = \sum_{j=0}^M T_{ij} P_j ~~~{\rm for}~i=0,...,M
~~~.
\eeq
Using the normalization condition $\sum_{j=0}^M P_j=1$, we obtain a set of 
$M$ linear equation for the first $M$ elements of $P_j$:
\beq
\sum_{j=0}^{M-1}(T_{ij}-T_{iM})P_j = -T_{iM} ~~~{\rm for}~i=0,...,M-1
~~~,
\eeq
which we solve using standard techniques.  In practice, we take $M\approx 500$,
in which case iterative methods are required to efficiently solve for $P_j$.
Once the $P_j$ are determined for $j=0,...,M-1$ we could obtain $P_M$ by
subtraction $P_M=1-\sum_{j=0}^{M-1}$ but this is inaccurate; it is better to
instead use
\beq
P_M = \frac{-1}{T_{MM}}\sum_{j=0}^{M-1}T_{Mj}P_j   ~~~.
\eeq

Fig.\ \ref{fig:dPdE} shows $dP/d\ln E$ for carbonaceous grains heated by
the average intestellar radiation field (ISRF) due to starlight
in an H~I region,
as estimated by Mathis, Mezger, \& Panagia (1983, hereafter MMP).
For $a\ltsim 25\Angstrom$ grains $dP/d\ln E$ becomes very small for
$E/hc > (1/911\Angstrom)=1.1\times10^5\cm^{-1}$: this reflects the fact
that grains cool essentially completely between photon absorption events,
so that the energy content virtually never rises above the maximum energy
(13.6~eV) of the illuminating photons.  
As the grain size increases, the time between photoabsorptions goes down,
and the cooling time at fixed energy goes up;
when the grain size exceeds $\sim30\Angstrom$ there is a significant
probability of a photoabsorption taking place before the energy of the
previous photoabsorption has been radiated away.
However, stochastic heating effects are noticeable even for grains
as large as $300\Angstrom$ -- the energy distribution function has
narrowed considerably, but is still appreciably broad.

\begin{figure}
\begin{center}
\includegraphics[width=.8\textwidth,angle=0]{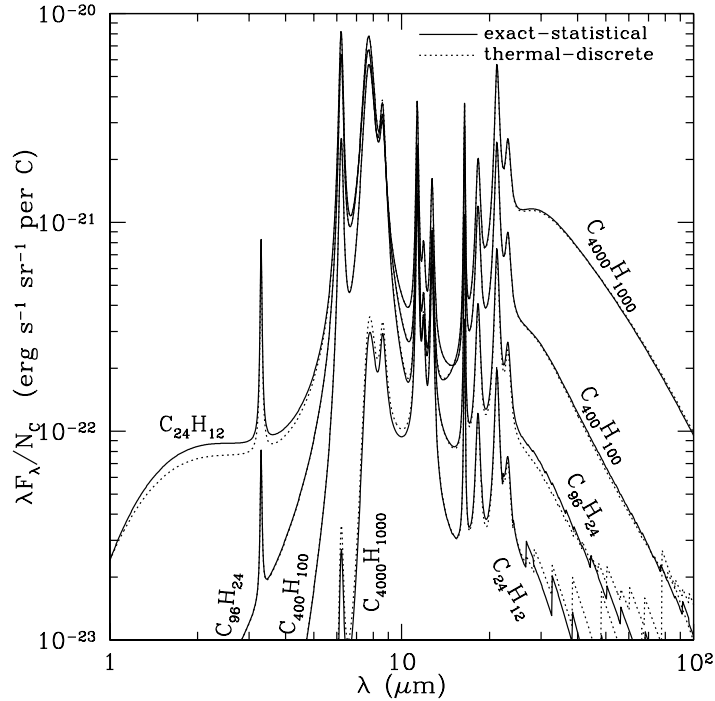}
\end{center}
\caption[]{IR emissivity per C atom for PAH$^+$ molecules of various
	sizes illuminated by the average ISRF.
	Solid line: ``exact-statistical'' calculation, which uses
	transition probabilities which do not involve the thermal
	approximation.
	Dotted line: ``thermal-discrete'' calculation where the
	spontaneous emission rates are calculated using a thermal
	approximation, as discussed in the text.
	The two methods are in excellent agreement, indicating that
	the thermal approximation can be used to calculate the
	transition rates.
	Taken from Draine \& Li (2001).
	}
\label{fig:PAH_IR_spec}
\end{figure}

\begin{figure}
\begin{center}
\includegraphics[width=.8\textwidth,angle=0]{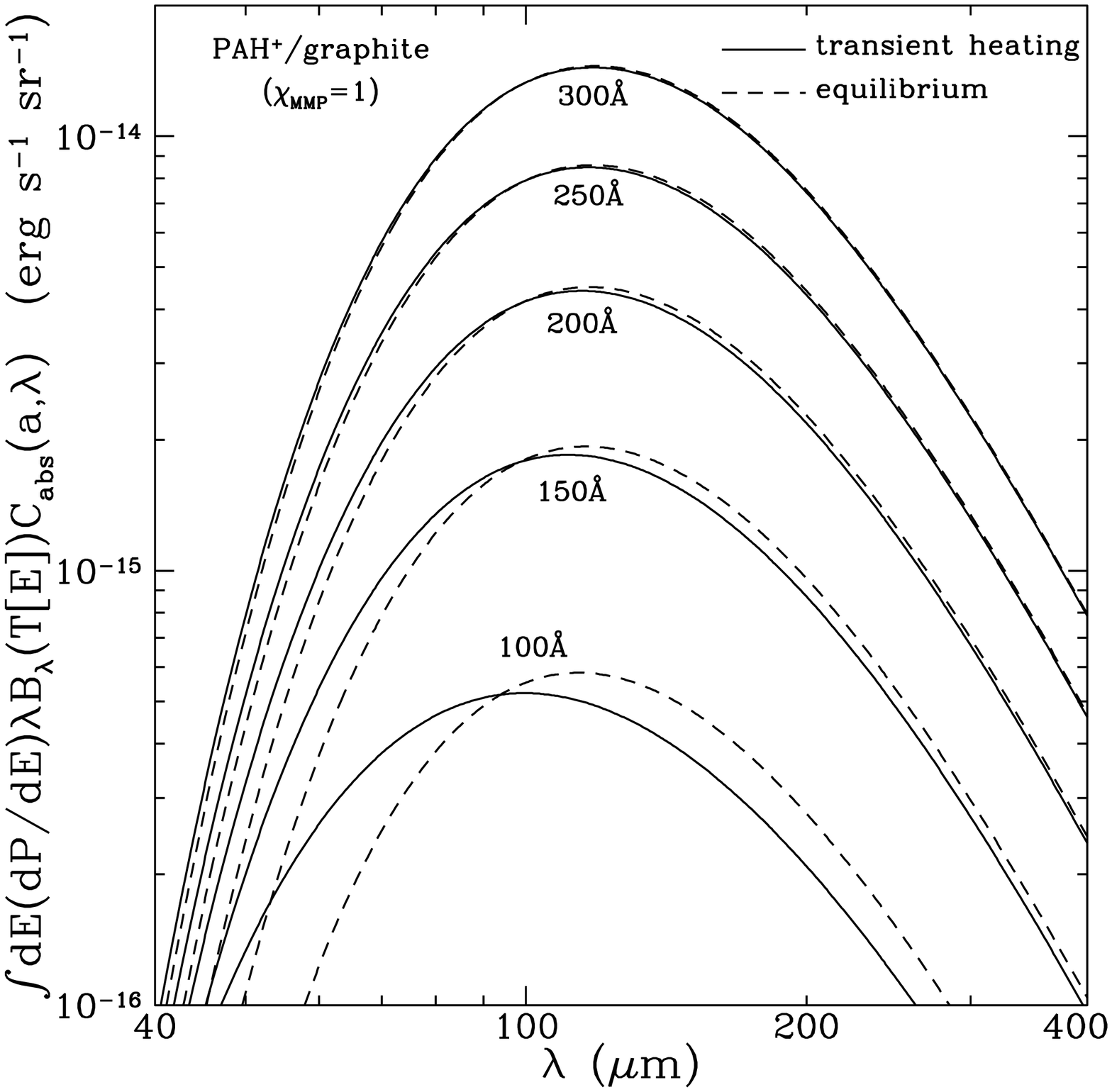}
\end{center}
\caption[]{IR emission per grain for carbonaceous grains of various
	sizes illuminated by the average ISRF.
	Solid line: full stochastic heating calculation.
	Broken line: assuming steady-state temperature $T_\stst$.
	For $a\gtsim200\Angstrom$ in the ISRF 
	we see that stochastic heating has
	little effect on the emission spectrum, but for $a\ltsim 150\Angstrom$
	it significantly modifies the emission spectrum.
	Taken from Li \& Draine (2001).
	}
\label{fig:carb_IR_spec}
\end{figure}

\subsection{IR and Far-IR Emission Spectrum}

With the energy distribution function
calculated as discussed above, we can now calculate the
time-averaged emission spectrum for a carbonaceous grain:
\beq
F_\lambda = 4\pi \int dE \frac{dP}{dE} C_\abs(\lambda) B_\lambda(T(E))
~~~.
\eeq
In Fig.\ \ref{fig:PAH_IR_spec} we show the emission spectrum of PAH$^+$
molecules of various sizes heated by the ISRF,
and in Fig.\ \ref{fig:carb_IR_spec} we compare the emission spectrum calculated
sing the energy distribution functions $dP/d\ln E$ with emission spectra
calculated for dust with steady temperature $T_\stst$.
We see that stochastic heating is important even for grains as large
as $\sim100\Angstrom$ in the ISRF.

\begin{figure}
\begin{center}
\includegraphics[width=.75\textwidth,angle=270]{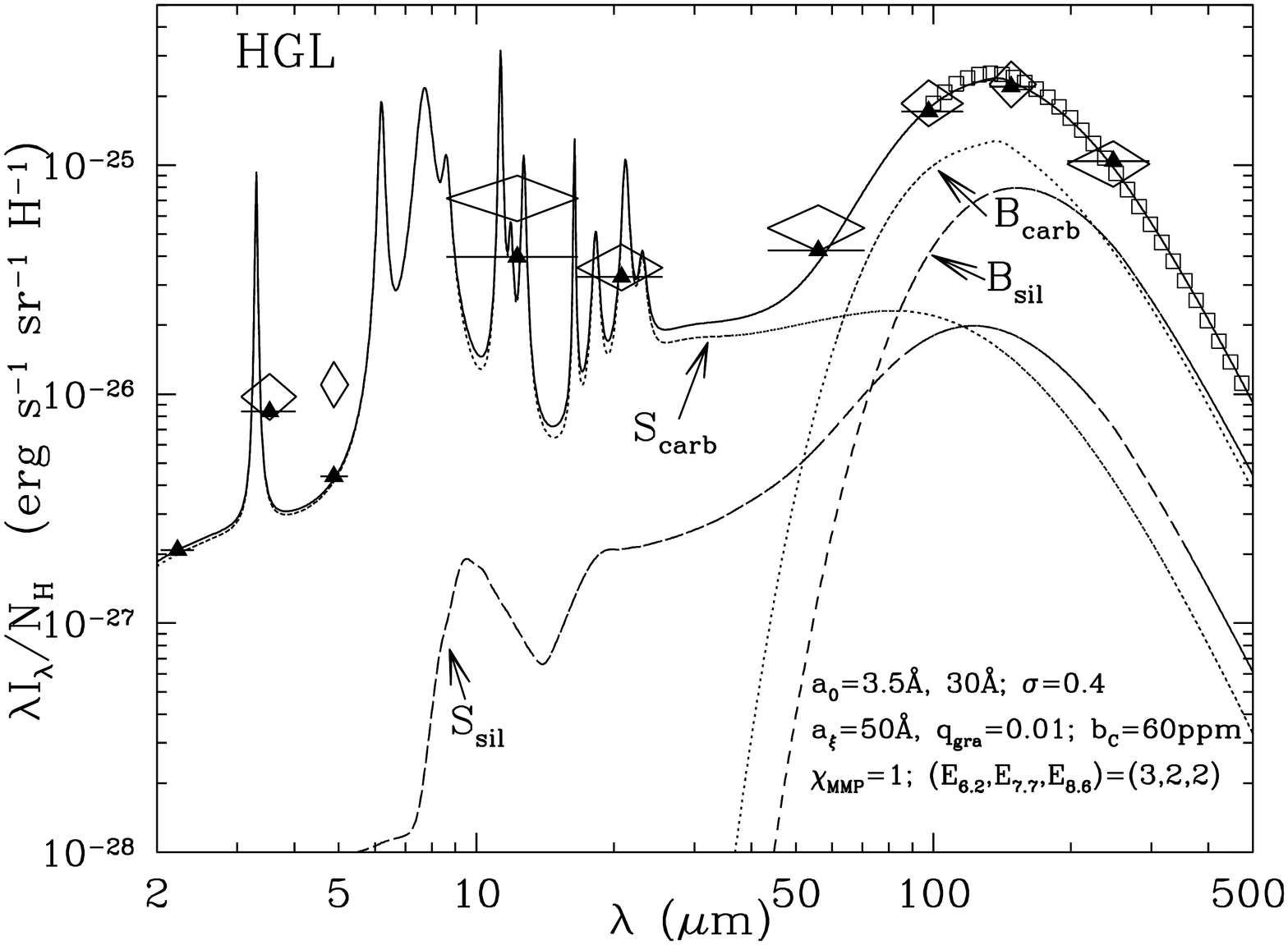}
\end{center}
\caption[]{IR emission per H for dust mixture illuminated by the ISRF.
	Diamonds: DIRBE (Arendt et al 1998).
	Squares: FIRAS (Finkbeiner et al. 1999).
	From Li \& Draine (2001).
	}
\label{fig:HGL_IR_spec}
\end{figure}

\begin{figure}
\begin{center}
\includegraphics[width=.9\textwidth,angle=0]{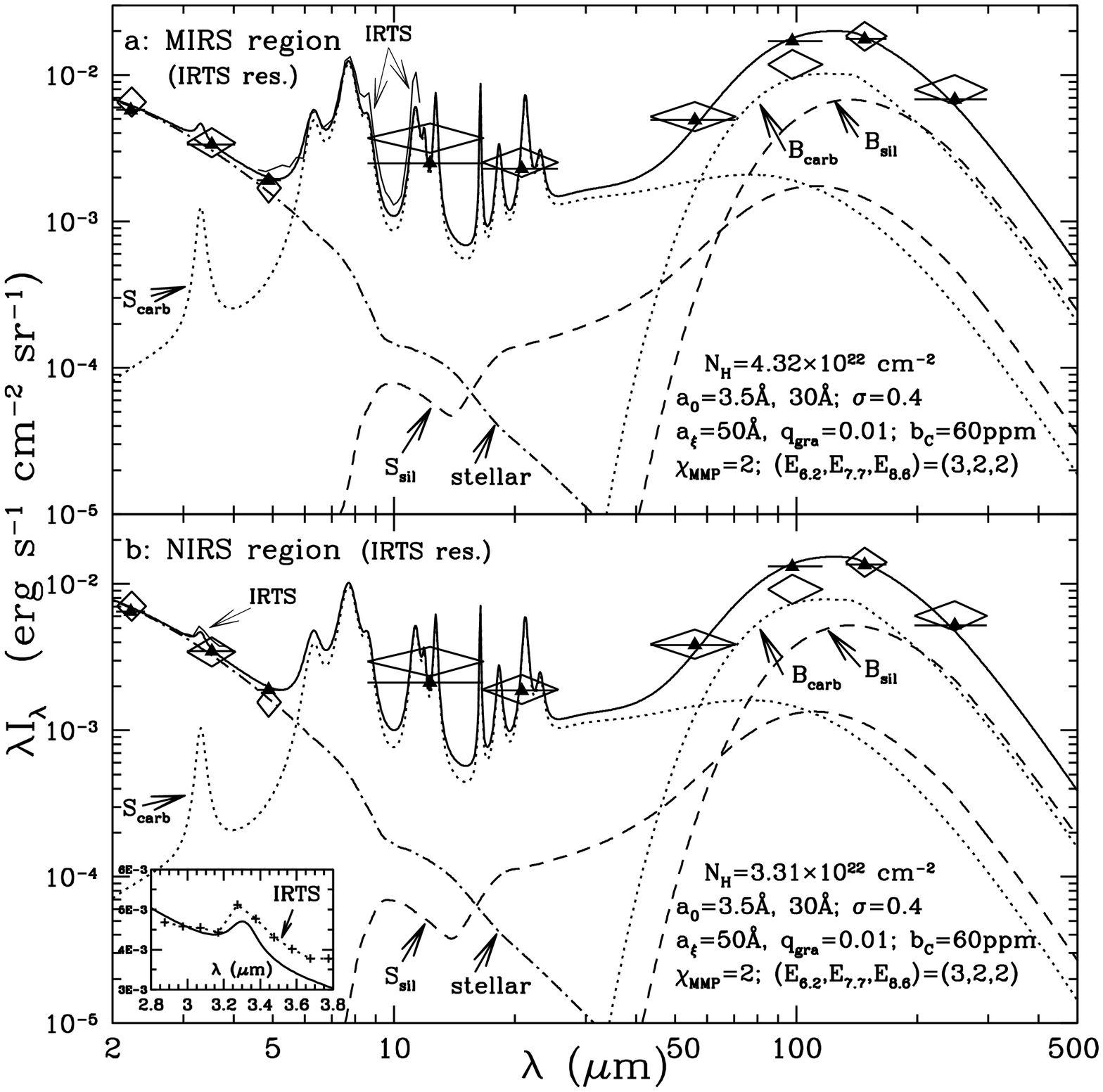}
\end{center}
\caption[]{IR surface brightness toward (a) $l=44^\circ20^\prime$,
	$b=-0^\circ20^\prime$; (b) $l=47^\circ45^\prime$.
	Diamonds: DIRBE photometry.
	Thin line, upper panel: 5-12$\micron$ spectrum measured by 
	MIRS on IRTS (Onaka et al. 1996).
	Thin line, lower panel: 2.8-3.9$\micron$ spectrum measured by
	NIRS on IRTS (Tanaka et al. 1996).
	Solid line: carbonaceous-silicate dust model 
	illuminated by $2\times$ISRF, plus reddened starlight.
	From Li \& Draine (2001).
	}
\label{fig:IRTS_IR_spec}
\end{figure}

In Fig.\ \ref{fig:HGL_IR_spec} we show the emission spectrum calculated 
for a mixture of carbonaceous and silicate grains with size distributions
consistent with the average ($R_V=3.1$) interstellar extinction curve,
illuminated by the local ISRF.
This should be the emission spectrum of the ``cirrus'' clouds.
Also shown in Fig.\ \ref{fig:HGL_IR_spec} are observations of the 
emissivity per H atom of the dust at high galactic latitudes.

Diamonds show DIRBE photometry; the small triangles show our calculated
spectrum convolved with the DIRBE filters, for comparison with the
diamonds.
The squares show the FIRAS determination of the emissivity per H nucleon.
The far-infrared emission is in excellent agreement with the predictions of
our dust model.
The model is in very good agreement with DIRBE photometry at 3.5$\micron$
and 25$\micron$; the model appears too low by about a factor 2 at
5 and 12$\micron$.  The observations are difficult; it is hoped that
SIRTF will be able to measure the spectrum of selected cirrus clouds
for comparison with our model.

In Fig.\ \ref{fig:IRTS_IR_spec} we compare our model with observations
taken on the galactic plane, where the FIR surface brightness is much higher.
In addition to DIRBE photometry, spectroscopic observations made with the
IRTS satellite are available.

These observations are
looking in the galactic plane at the ``tangent'' point of
gas in the 6 kpc ring, at a distance of 6 kpc from us.
At this distance, the 40$^\prime$ DIRBE beam subtends 70 pc, so
the observations sum the emission from dust in a region $\sim$70$\times$70 pc
on a side, and many kpc long.  We compare the observations to a model
which is obviously oversimplified: a uniform slab of dust heated
by a uniform radiation field with the spectrum of the local ISRF but whose
intensity we scale so that the dust emission best reproduces the observed
spectrum: the best fit is obtained with the starlight intensity equal to
twice the ISRF.  The DIRBE beam obviously includes emission from stars,
which we believe dominate at $\lambda < 4\micron$, so our model includes
light from stars assumed to be mixed uniformly with the dust.
This simple model is in good agreement with the observations.
In particular, we successfully reproduce the very strong emission in the
5-12$\micron$ region.  
The discrepancy at 100$\micron$ is somewhat
surprising, but may in part be due to oversimplification of our model,
and perhaps in part to uncertainties in the DIRBE absolute calibration.

\subsection{The Small Magellanic Cloud (SMC)}

\begin{figure}
\begin{center}
\includegraphics[width=.75\textwidth,angle=270]{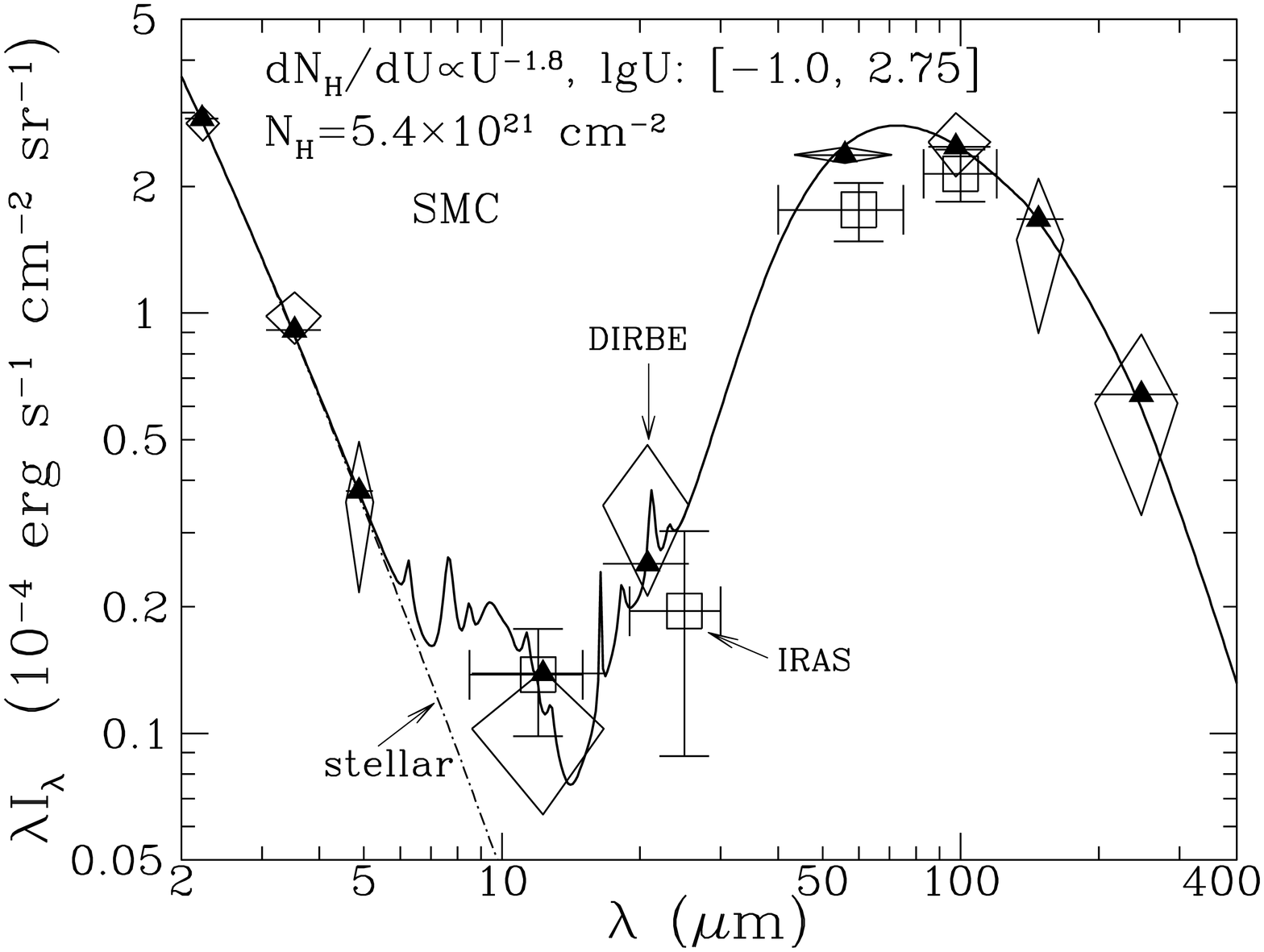}
\end{center}
\caption[]{Comparison between dust model (solid line) and the observed
	spectrum of the SMC obtained by COBE/DIRBE (diamonds) and
	IRAS (squares) averaged over a 6.25 deg$^2$ region including the
	optical bar and the Eastern Wing (Stanimirovic et al 2000).
	Model spectrum includes infrared emission from 
	dust heated by a range of starlight intensities.
	plus starlight, which dominates at $\lambda \ltsim 7\micron$.
	Triangles show the model spectrum convolved with the DIRBE
	filters.
	Taken from Li \& Draine (2002).
	}
\label{fig:SMC}
\end{figure}

How successful is this dust model in reproducing the emission observed
from other galaxies?  
The SMC, with metallicity $\sim$10\% of
solar (Dufour 1984), and dust-to-gas ratio $\sim$10\% of
the Milky Way (Bouchet et al. 1985) is a good test case.
Li \& Draine (2002) sum over both quiescent medium and regions of active
star formation in the SMC, find that to match the observed spectrum they
require a distribution of starlight intensities.
Following Dale et al. (2001)
they adopt
a power-law distribution of starlight intensities.

The dust extinction in the SMC is known to differ from Milky Way dust,
primarily in the absence of a 2175\AA\ extinction bump on most (but not
all) SMC sightlines (Gordon \& Clayton 1998).
Since we take the view that
the 2175\AA\ feature is due to PAHs, the absence of this feature implies
a low PAH abundance.
We adopt the dust mixture of Weingartner \& Draine (2001a), which
reproduces the measured SMC bar extinction curve.

With suitable choice of illuminating radiation field, the dust model
is able to reproduce the observed IR emission from the SMC.
It appears that the carbonaceous/silicate grain model
(Weingartner \& Draine 2001a; Li \& Draine 2001) can
reproduce both the observed interstellar extinction by dust and the
observed IR/FIR emission from dust.

\section{Charging of Interstellar Dust}

Why do we care about the charging of dust grains?  There are a number of
reasons:
\begin{itemize}
\item Charged dust grains are coupled to the magnetic field,
	whereas neutral grains are not.  This is important not only
	for the motions of the dust grains, but also as a mechanism
	for coupling magnetic fields to neutral gas.
\item Charged dust grains undergo stronger gas drag due to Coulomb
	interaction with ions in the gas.
\item The photoelectric charging process injects energetic photoelectrons
	into the gas, which is a major mechanism for heating interstellar
	gas.
\item Neutral and negatively-charged dust grains can play an important
	role in neutralization of ions.
\item In cold, dense regions, the dust grains may ``lock up'' a
	substantial fraction of the electrons.
\end{itemize}

\subsection{Collisions with Electrons and Ions}

Consider a spherical grain of radius $a$ and 
charge $Z_ge$, and a charged
particle of charge $ze$, kinetic energy $E$, on a trajectory
with impact parameter $b$.
If the interaction potential is taken to be a Coulomb potential, then
conservation of energy and angular momentum allow one to find the
critical impact parameter $b_\crit$ for the particle to just graze
the grain surface.  Trajectories with $b<b_\crit$ will collide with
the grain, and trajectories with $b>b_\crit$ will miss.
The collision cross section is simply
\begin{eqnarray}
C &=& \pi b_\crit^2 = \pi a^2 \left[1 - \frac{zeU}{E}\right]  
~~~{\rm for}~ E > zeU
~~~,
\\
&=& 0 ~~~~~~~~~~~~~~~~~~~~~~{\rm for}~ E \leq zeU
~~~,
\end{eqnarray}
where $U\equiv Z_g e/a$ is the electrostatic potential at the
grain surface.
If the velocity distribution is thermal, then the
rate at which the charged particles reach the grain surface is just
\beq
\frac{dN}{dt} = 
n \pi a^2 \left(\frac{8kT}{\pi m}\right)^{1/2}
\times\left\{
\begin{array}{ll}
e^{-zeU/kT}			& ~~~{\rm for} ~zeU > 0 \\
\left[1-\frac{zeU}{kT}\right]	& ~~~{\rm for} ~zeU < 0
\end{array}
\right.
~~~.
\eeq
This is a wonderfully simple result.  If the sphere is located in
a plasma of electrons and ions, and the only charging process is
collisional, then the sphere will become negatively charged since
the electrons move more rapidly than the ions.  The steady-state
charge is such that suppression of the electron arrival rate, and
enhancement of the ion arrival rate, make them equal:
\beq
n_e s_e\left(\frac{8kT}{\pi m_e}\right)^{1/2}e^{eU/kT} =
n_I s_I\left(\frac{8kT}{\pi m_I}\right)^{1/2}\left[1-\frac{eU}{kT}\right]
~~~,
\eeq
where we have assumed $Z_I=+1$.  Taking logarithms of both sides we
obtain
\beq
\frac{eU}{kT} = -\frac{1}{2}\ln\left(\frac{m_I}{m_e}\right) +
\ln\left(\frac{s_I}{s_e}\right) + 
\ln\left(1-\frac{eU}{kT}\right)
~~~,
\eeq
with the solution $eU/kT=-2.51$ for a hydrogen plasma ($m_I/m_e=1836$)
with $s_I=s_e$.

However, the Coulomb potential does not fully describe the interaction
between a charged particle and a grain.
In the case of a neutral grain, the electric field of the approaching
charged particle will polarize the grain, resulting in an attractive
potential -- the same effect that causes ion-neutral scattering rates
to be large, even at low temperatures.
\footnote{This is just the familiar ``image potential'' from 
	electrostatics for a charge near a conductor.}
Draine \& Sutin (1987) have calculated collision rates including
the image potential.
Image potential effects are important for neutral grains when
$e^2/a \gtsim kT$, or $aT\ltsim 1.67\times 10^{-3}\cm\K$.
Thus it is an important correction for small grains in cold
gas (e.g., $a\ltsim 1.67\times10^{-5}\cm$ for $T=100\K$).

The ``sticking coefficients'' for small electrons and ions are
not well known.
In the case of an ion with an ionization potential larger than
the work function for the grain material, it seems likely that
the ion, upon arrival at the grain surface, will either remain
stuck to the grain or will strip an electron
from the grain and depart; in either case the ion ``sticking coefficient''
$s_I=1$.
The case of impinging electrons is less clear.  The electron might
be elastically reflected from the grain surface, or the electron might
pass through a small grain and out the other side without energy loss,
in which case the electron would have enough energy to return to infinity.

\begin{figure}
\begin{center}
\includegraphics[width=.8\textwidth,angle=0]{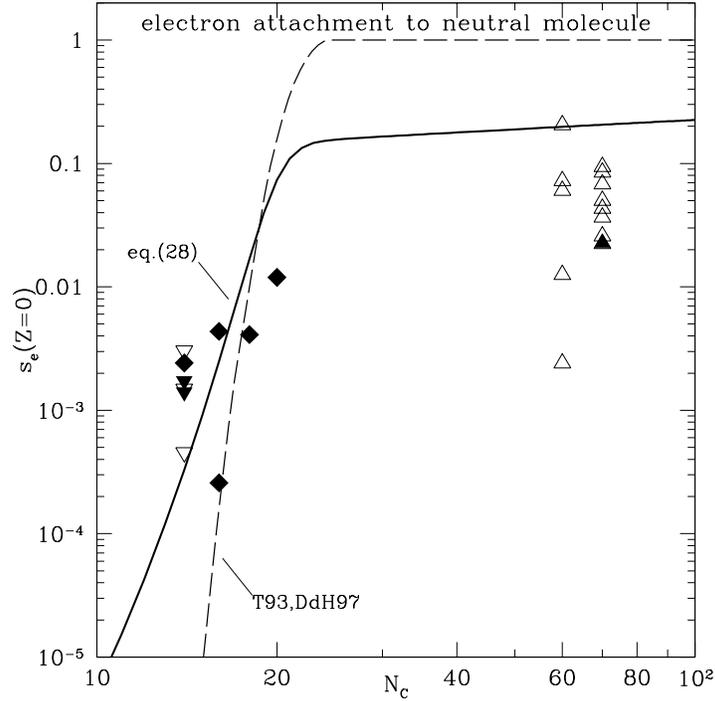}
\end{center}
\caption[]{Electron sticking coefficient for small neutral grains.
	From Weingartner \& Draine (2001c).}
\label{fig:se(Z=0)}
\end{figure}

\begin{figure}
\begin{center}
\includegraphics[width=.8\textwidth,angle=0]{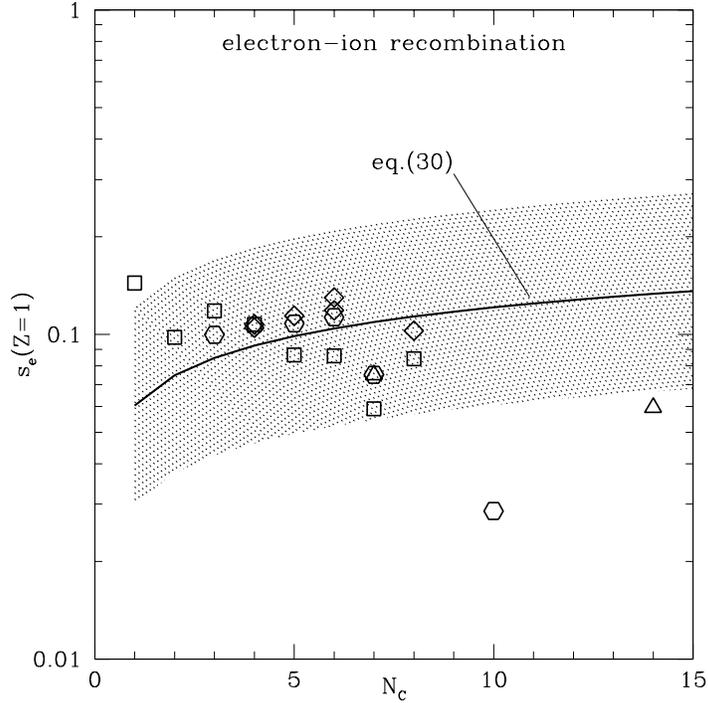}
\end{center}
\caption[]{Electron sticking coefficient for small grains with charge $Z=+1$.
	From Weingartner \& Draine (2001c).}
\label{fig:se(Z=1)}
\end{figure}

Experimental sticking coefficients for small neutral molecules or 
molecular ions can be
estimated by dividing the measured electron capture rate coefficient by
the estimated rate at which the electrons would reach the surface of
the ion.
Figs.\ \ref{fig:se(Z=0)} and \ref{fig:se(Z=1)} 
show the electron sticking coefficients for small neutral and charged 
carbonaceous clusters containing $10 < N_{\rm C} < 100$ carbon atoms.

\begin{figure}
\begin{center}
\includegraphics[width=.8\textwidth,angle=0]{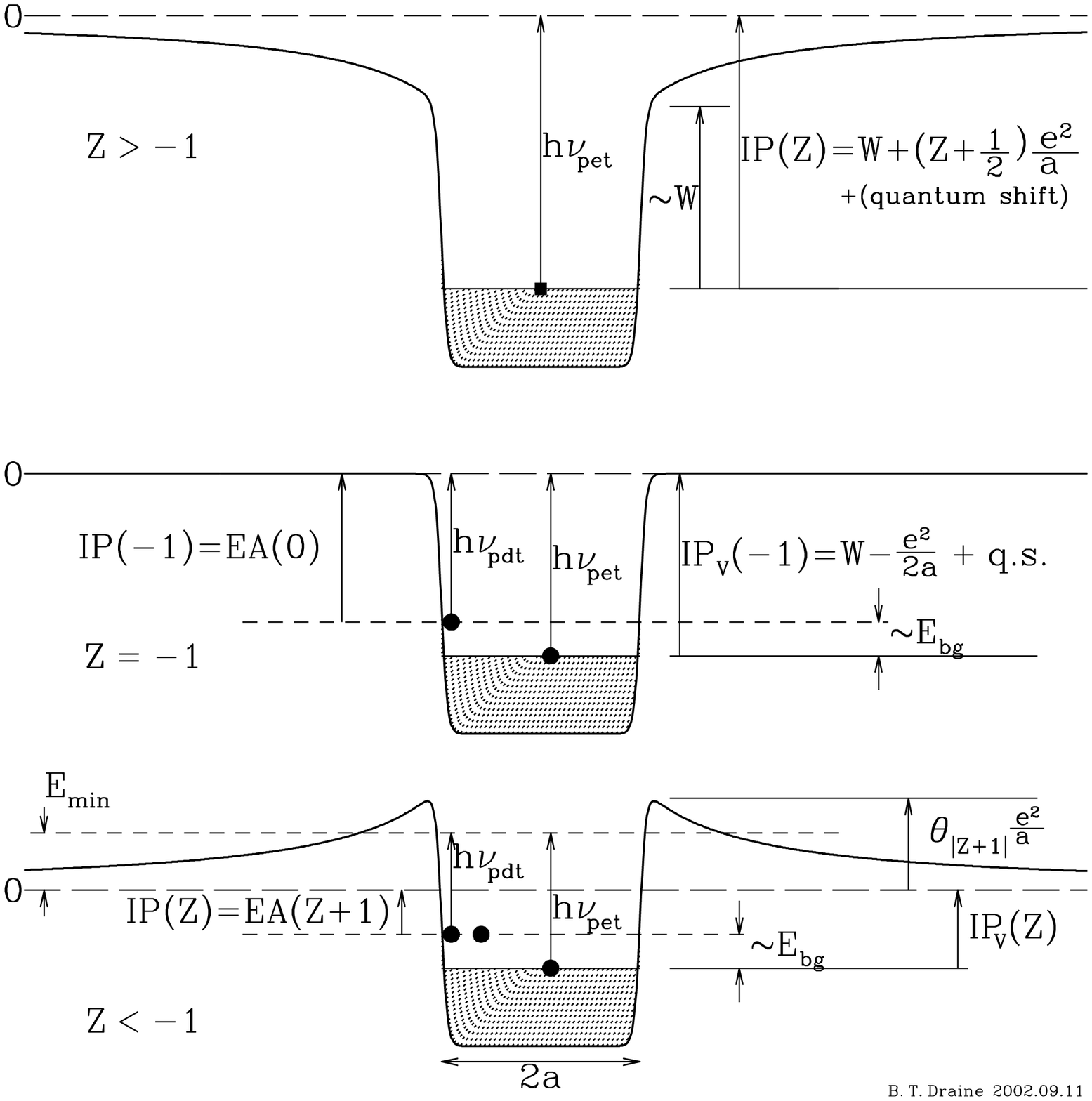}
\end{center}
\caption[]{Model for the potential confining electrons in a grain
	with charge $Ze$.  Shaded regions show occupied energy levels.
	$W$ is the work function for bulk material; $IP(Z)$ is
	the ionization potential; $EA(Z)$ is the electron affinity;
	$h\nu_{\rm pet}$ is the photoelectric threshold energy;
	$h\nu_{\rm pdt}$ is the photodetachment threshold energy (the
	minimum photon energy required to detach an electron from a negatively
	charged grain);
	$IP_V(Z)$ is the energy difference between infinity and the top
	of the valence band;
	$E_\bg$ is the bandgap energy for the material
	($E_\bg=0$ for a metal).
	Taken from Weingartner \& Draine (2001c).
	}
\label{fig:pot_wells}
\end{figure}

In hot gas, it is possible for impinging electrons 
to eject an electron from the grain.  For molecules, this is called
``ionization''; for larger objects, it is called ``secondary electron
emission''.
It is therefore possible for grains to become positively charged
by collisions in a hot ($T\gtsim 10^6\K$) plasma (Draine \& Salpeter 1979a).

\subsection{Photoelectric Emission}

The electrons in a neutral grain can be thought of a being
confined within a potential well, as shown in Fig.\ \ref{fig:pot_wells}.
The potential well is produced by a ``double layer'' of charge
at the boundary of the grain (the electron charge density extends
beyond the ion charge density, resulting in a region near the
surface with a radial electric field -- see standard texts on solid
state physics, e.g., Ashcroft \& Mermin 1976).

Three cases are shown:
\begin{itemize}
\item
For a grain with $Z\geq0$, the excited electron will be subject to
a long-range Coulomb attraction by the rest of the charge.  In this
case the electron which physically gets outside of the grain must
still have additional energy in order to reach $\infty$.
\item
For a grain with charge $Z=-1$, the excited electron sees a system
with zero net charge, so once the electron is physically outside the
double layer, it will continue to infinity.
\item
For a grain
with charge $Z<-1$, an electron which gets outside the double layer
will be accelerated away from the grain.
An excited electron can in principle penetrate the double layer by
tunneling, so that it is not actually necessary for the electron to
have an energy higher than the top of the double layer.
\end{itemize}
An insulating grain has all the ``valence band'' energy levels occupied
when it is neutral.
The difference in energy between the top of the valence band and
the first available vacant energy level (in the ``conduction band'')
is referred to as the ``band gap''.
If an insulating grain is to be negatively charged, the excess
electrons must occupy available energy levels in the conduction band.
Less energy is required to photoeject one of these ``excess'' electrons
(a process known as ``photodetachment'')
than to eject an electron from the top of the valence band.

When an ultraviolet photon is absorbed in a dust grain, it raises
an electron to an excited state.  If the electron has an energy high
enough to reach infinity, and does not lose this energy to inelastic
scattering, it will escape and be counted as a ``photoelectron''.
To calculate the rate of photoelectric charging of a grain,
we require the ``photoelectric yield'' $Y(h\nu,Z,a)$, which is
the probability that absorption of a photon $h\nu$ will produce 
a photoelectron.
To calculate the rate of photoelectric heating of the gas, we also
need to know the energy distribution function $f(E,h\nu,Z,a)$ of the
energy $E$ of the
photoelectrons leaving the grain.

The photoelectric ejection process consists of four stages:
\begin{enumerate}
\item Excitation of an electron of appropriate energy.
\item Motion of the electron from the point of excitation to the
grain surface, 
\item Penetration of the surface layer and overcoming of the image potential.
\item Once outside the grain surface, the electron must
overcome the Coulomb potential (if the grain is now positively charged)
to reach infinity.
\end{enumerate}
Weingartner \& Draine (2001c) have estimated the photoelectric yield
for small grains, both neutral and charged, taking into account the
above effects.
Allowance is made for dependence of the image potential on the
grain radius $a$, and the possibility of tunneling through the
double layer in the case of a negatively charged grain.


\begin{figure}
\begin{center}
\includegraphics[width=.9\textwidth,angle=0]{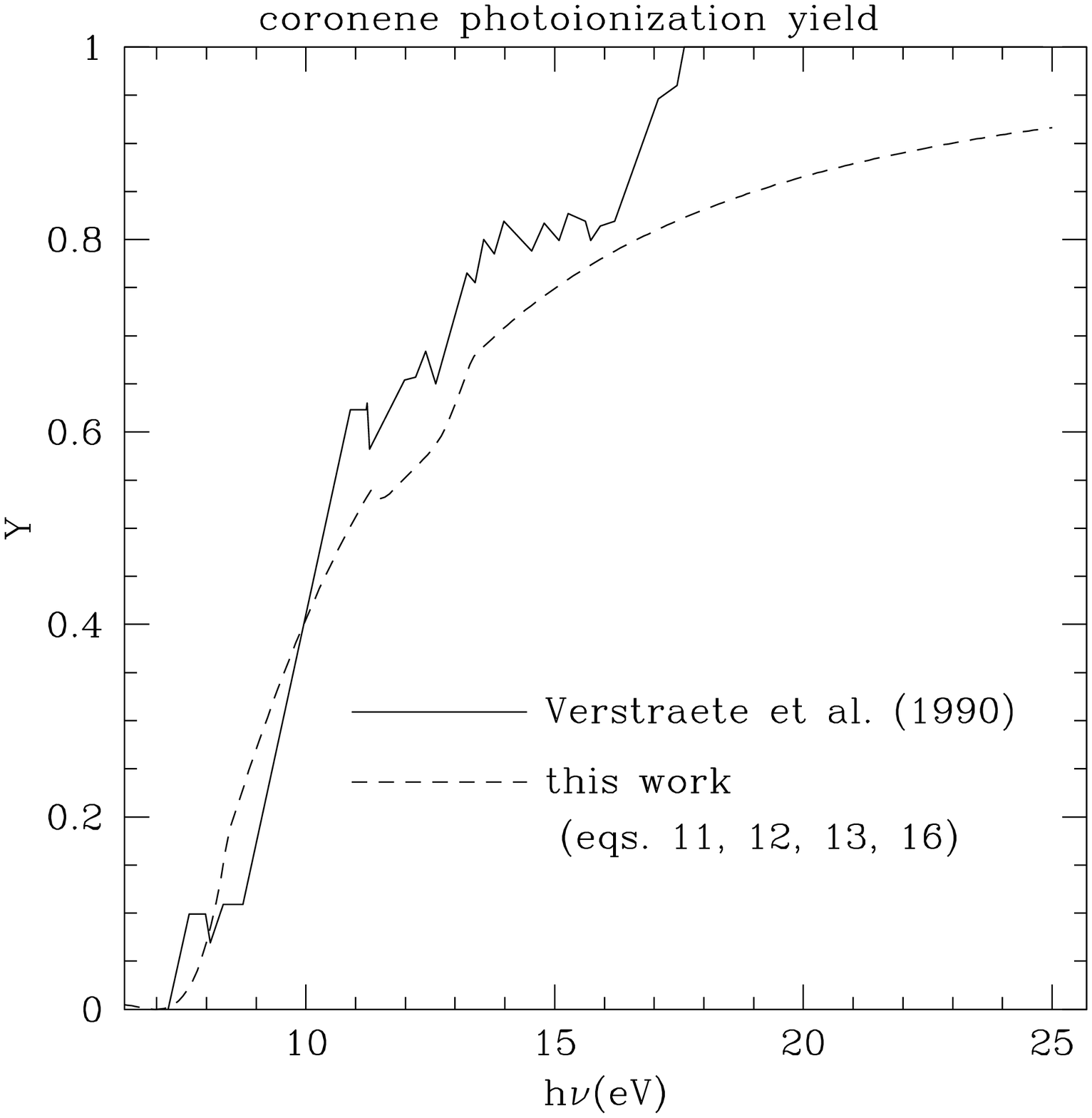}
\end{center}
\caption[]{
	Photoelectric yield for neutral coronene, as measured by
	Verstraete et al.\ (1990) and as estimated in the approach of
	Weingartner \& Draine (2001c).
	Eq. 11, 12, 13, 16 refer to equations in 
	Weingartner \& Draine (2001c), from which this figure is
	taken.
	}
\label{fig:peyield_coronene}
\end{figure}

\begin{figure}
\begin{center}
\includegraphics[width=.9\textwidth,angle=0]{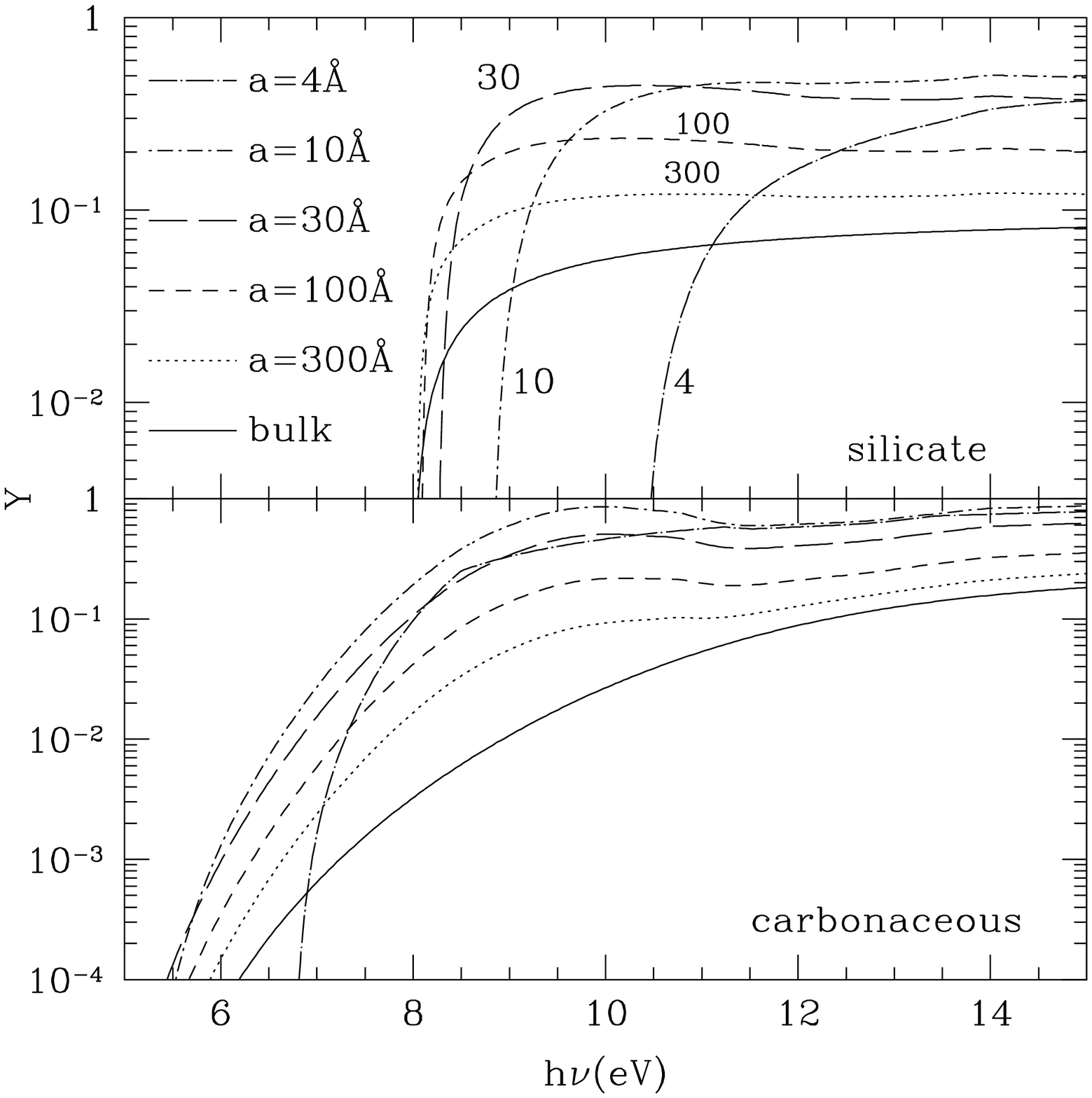}
\end{center}
\caption[]{
	Photoelectric yields (photoelectrons per absorbed photon) 
	for neutral graphite and silicate grains
	as a function of incident photon energy $h\nu$, for
	selected values of the grain radius $a$.
	From Weingartner \& Draine (2001c).
	}
\label{fig:peyield}
\end{figure}

Watson (1973) pointed out that very small particles should have
enhanced photoelectric yields because of step 2 above: in a small
particle, photoexcitations will on average be closer to the grain
surface, increasing the probability that the excited electron
will reach the grain surface without energy loss.
Draine (1978) proposed a simple function to estimate the magnitude
of this yield enhancement, which we use:
\beq
y_1 = \left(\frac{\beta}{\alpha}\right)^2
\frac{\alpha^2-2\alpha+2-2\exp(-\alpha)}
{\beta^2-2\beta+2-2\exp(-\beta)}
~~~,
\eeq
\beq
\beta = \frac{a}{l_a}   ~~~~ \alpha = \frac{a}{l_a} + \frac{a}{l_e}
~~~,
\eeq
where $l_a = \lambda/[4\pi{\rm Im}(m)]$ is the photon attenuation length
in the material, and
$l_e$, the ``electron escape length'', is essentially the mean free
path against energy loss to inelastic scattering for an excited electron.
In the limit $\alpha \ll 1$ we have
\beq
y_1 \rightarrow \frac{\alpha}{\beta}=\frac{l_a+l_e}{l_e}
~~~.
\eeq

Photon attenuation lengths $l_a\approx 300\Angstrom$ are typical in the
vacuum ultraviolet.
Martin et al.\ (1987) report $l_e\approx9\Angstrom$ for 6 eV electrons in 
thin carbon films, and
McFeely et al.\ (1990) report  $l_e\approx 6\Angstrom$
for 8 eV electrons in SiO$_2$.
Weingartner \& Draine (2001c) adopt $l_e=10\Angstrom$, independent of energy,
for both graphite and silicate grains.
Since $l_e\ll l_a$, it is clear that very large yield enhancement
factors $y_1$ are possible for very small grains.

Weingartner \& Draine write
\beq
Y(h\nu,Z,a) = \min \left[y_0\times y_1(h\nu,a)\right] \times y_2(h\nu,Z,a)
\eeq
where $y_2$ is the fraction of electrons which, having crossed the
grain surface, have sufficient energy to overcome the long-range coulomb
attraction if $Z\geq 0$ (for $Z<0$, $y_2=1$).
For PAHs and larger carbonaceous grains, $y_0$ is
chosen so that $Y=y_0y_1y_2$ approximates the photoionization yield
measured for coronene.

Fig.\ \ref{fig:peyield} shows the photoelectric yields estimated for
neutral carbonaceous and silicate grains of various radii.

\begin{figure}
\begin{center}
\includegraphics[width=.75\textwidth,angle=270]{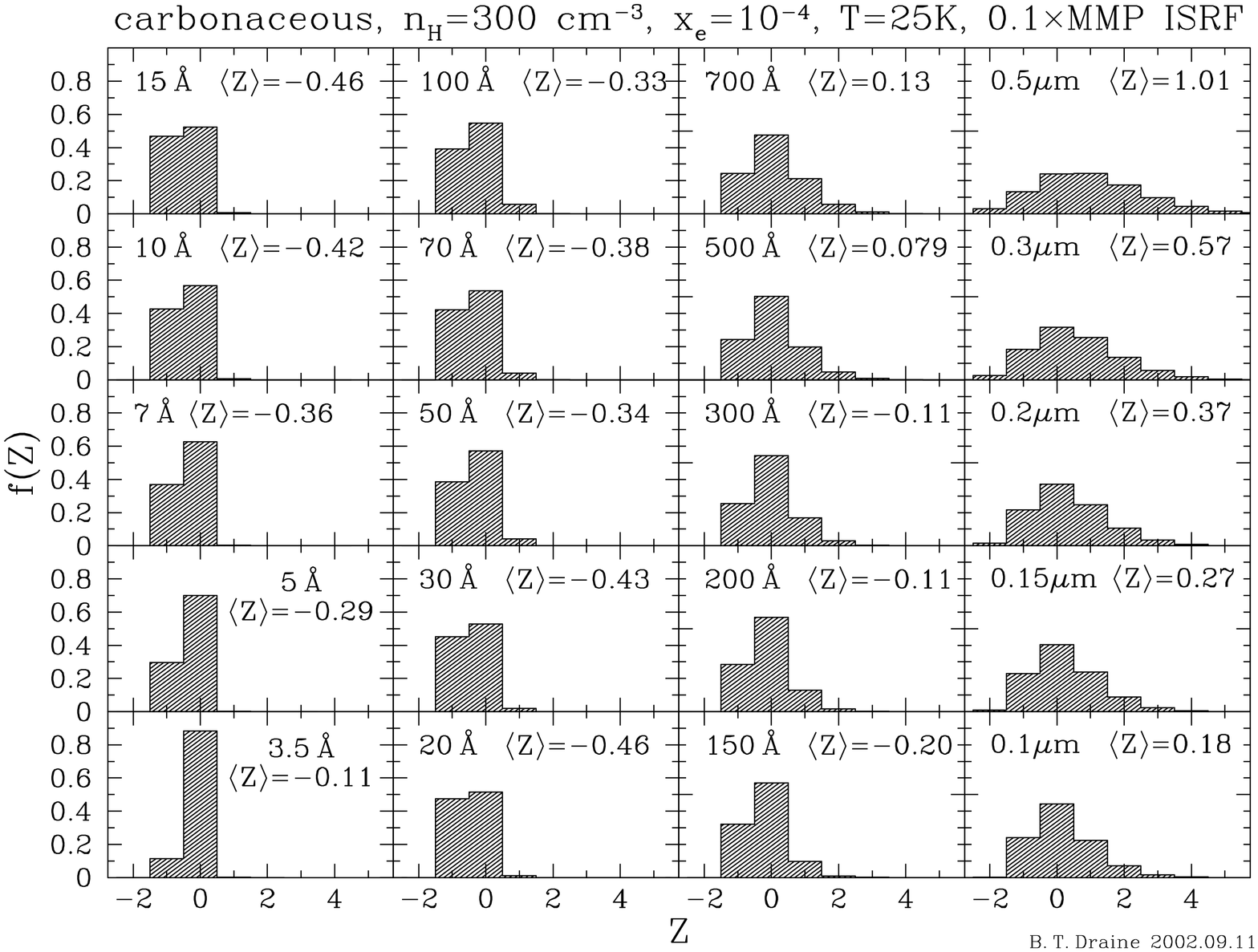}
\includegraphics[width=.75\textwidth,angle=270]{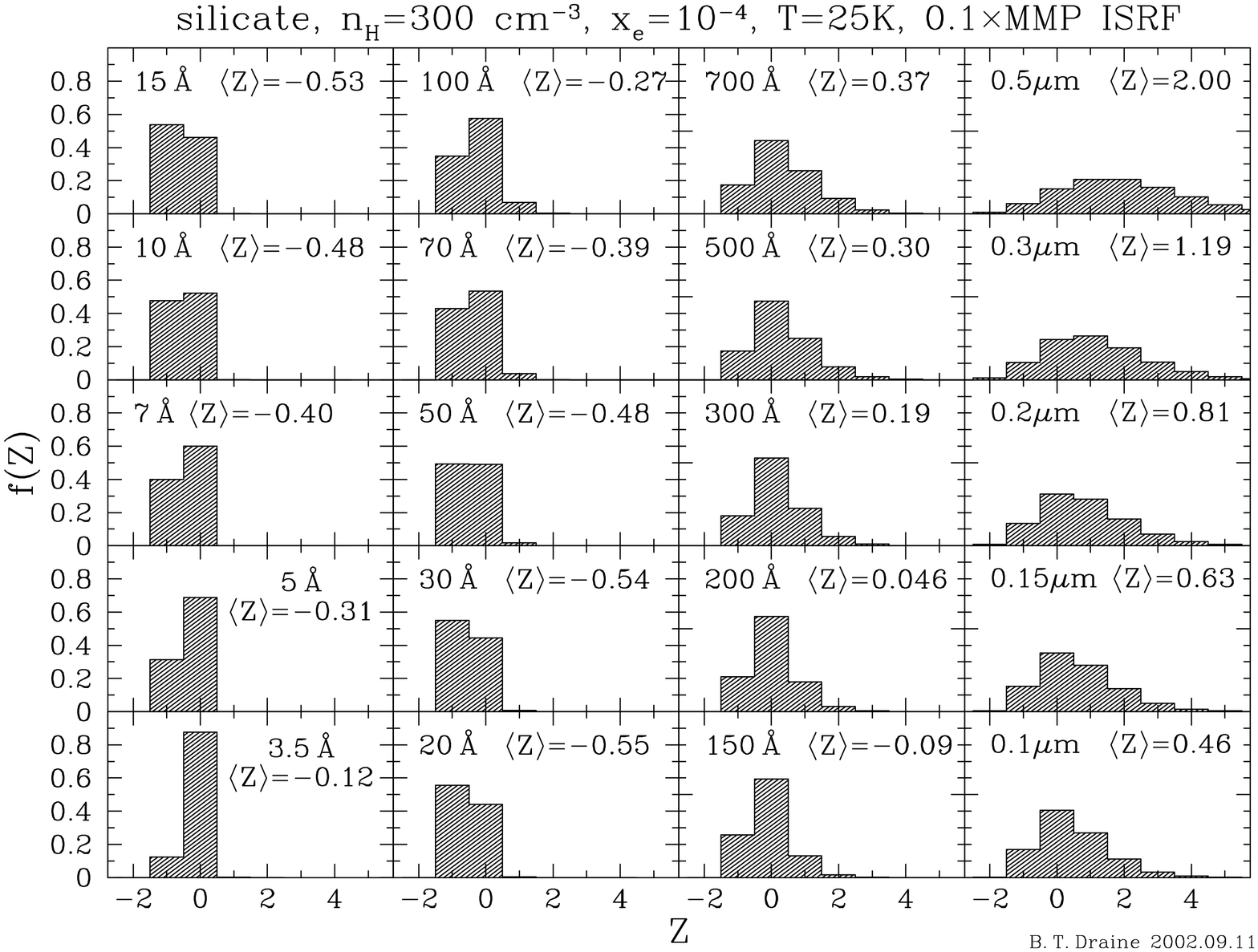}
\end{center}
\caption[]{
	Charge distribution functions for carbonaceous grains 
	and silicate grains in
	cold molecular gas.
	}
\label{fig:Zdist.mc}
\end{figure}

\begin{figure}
\begin{center}
\includegraphics[width=.75\textwidth,angle=270]{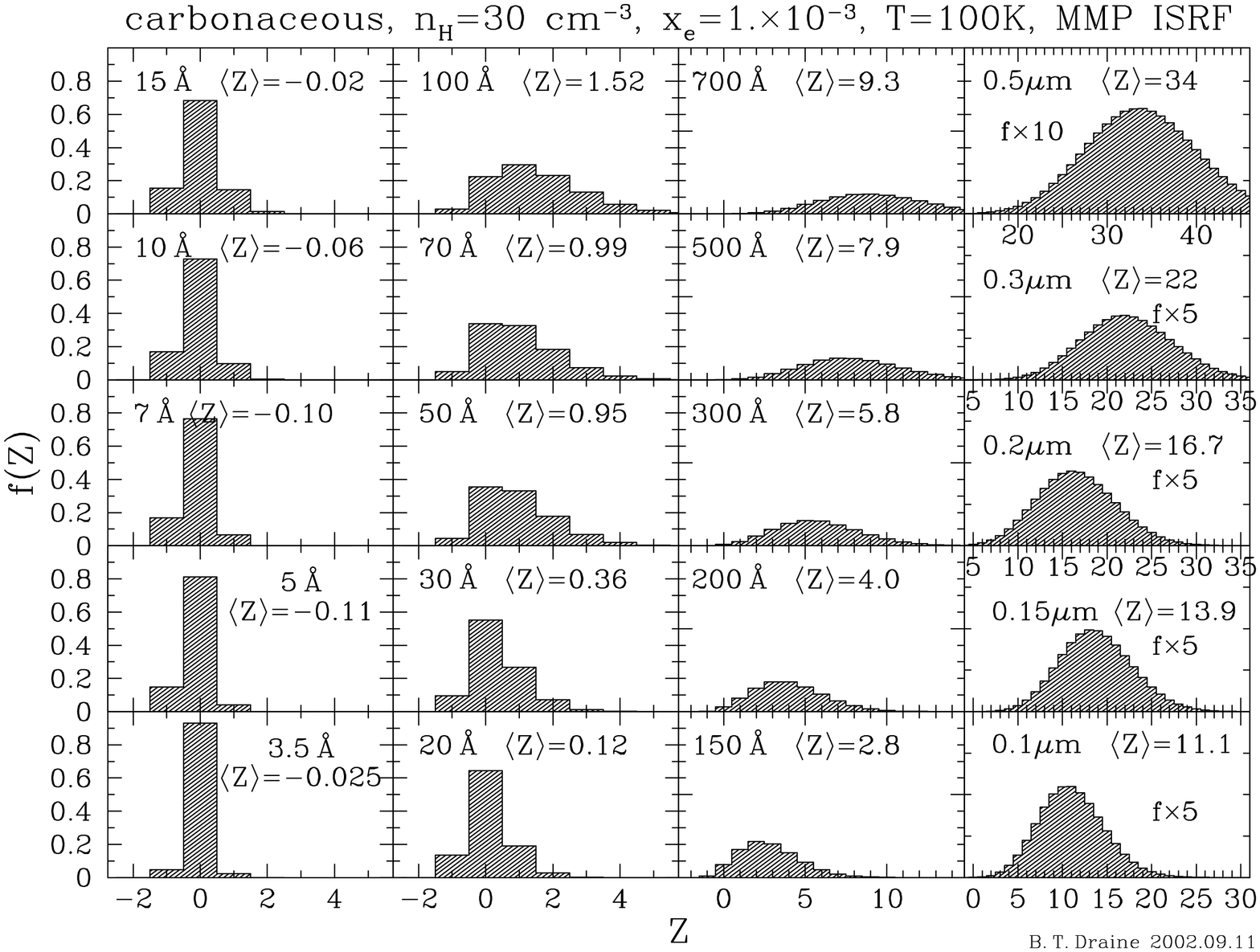}
\includegraphics[width=.75\textwidth,angle=270]{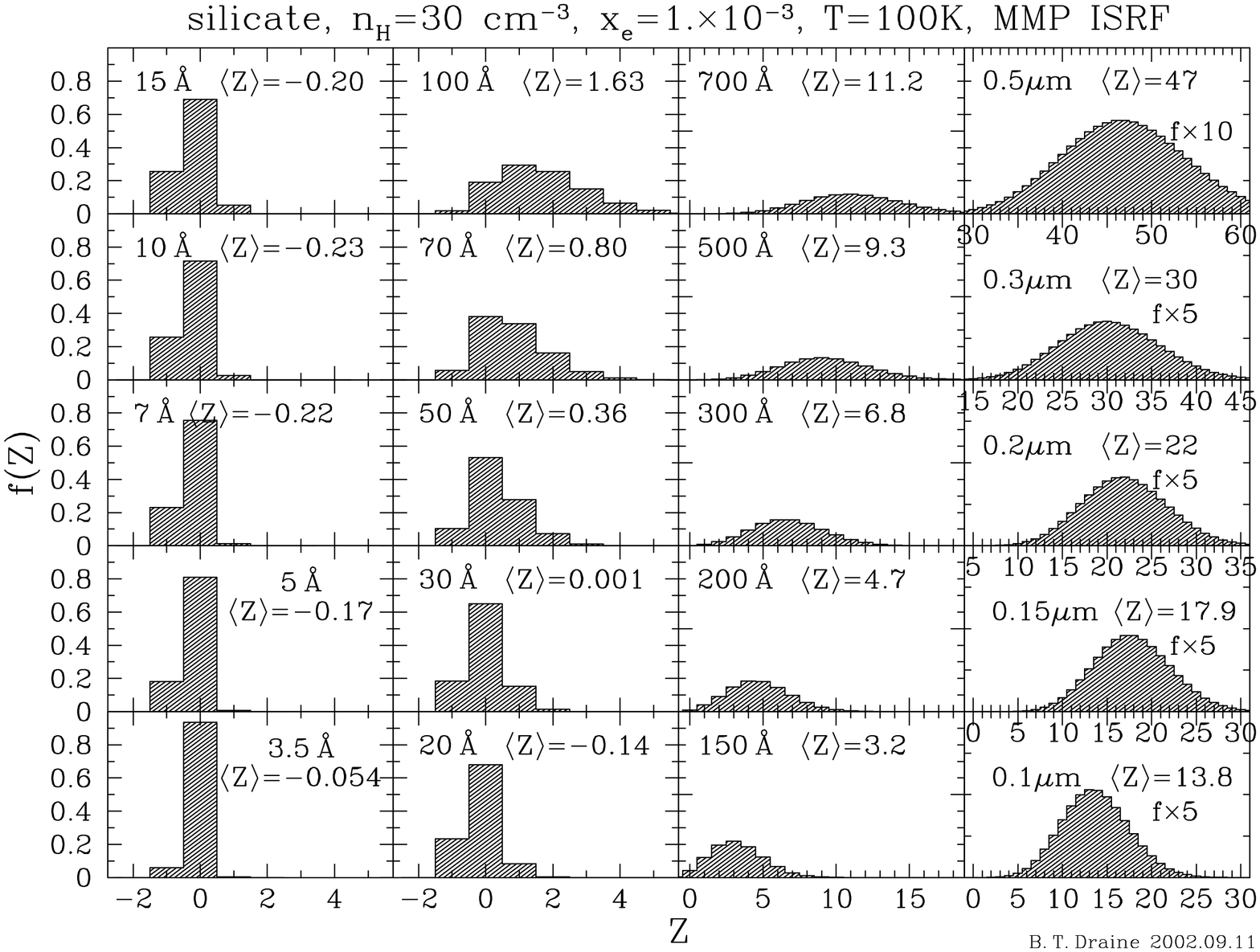}
\end{center}
\caption[]{
	Charge distribution functions for carbonaceous grains 
	and silicate grains in
	``cold neutral medium'' diffuse HI.
	}
\label{fig:Zdist.cnm}
\end{figure}

\subsection{Charge Distribution Functions}

Fig.\ \ref{fig:Zdist.mc} shows grain charge distribution functions 
calculated for dust in
a molecular cloud region.
The H nucleon density is taken to be $n_\rmH=300\cm^{-3}$, the
gas temperature $T=25\K$, and the fractional ionization
$n_e/n_\rmH = 10^{-4}$. 
The radiation field is assumed to have the spectrum of the MMP ISRF,
but with intensity a factor of 10 below the MMP value.
Under these conditions small grains ($a\ltsim 200\Angstrom$) remain
negatively charged.  For larger grains, the reduced importance of
``image charge'' effects (as well as a slight reduction in the work
function) tilts the balance in favor of photoelectric charging, but
the grain potentials remain very small.

Fig.\ \ref{fig:Zdist.cnm} shows grain charge distribution functions
calculated for a diffuse region with starlight equal to the MMP ISRF.
The electron density $n_e=0.03\cm^{-3}$ 
is taken to be the same as for the molecular cloud
region of Fig.\ \ref{fig:Zdist.mc}, but the increased
starlight intensity allows photoelectric emission to dominate the
charging for grains with $a\gtsim 25\Angstrom$.
function 
The ``CNM'' conditions are appropriate to diffuse HI at
$100\K$.  Despite the presence of ultraviolet
radiation producing photoelectric emission, note that an appreciable
fraction of the small grains can be neutral or even negatively charged.

We see that the competition between collisional charging and photoelectric
charging can go either way, depending on the grain composition,
the grain size $a$,
the electron density $n_e$ and temperature $T$,
and the spectrum and intensity of the ultraviolet background due
to starlight.

\section{Dynamics of Interstellar Dust}

What are the velocities of interstellar grains?  To answer
this question, we must understand the forces acting to accelerate
or decelerate the grains.

\subsection{Gas Drag}

If the grain is moving relative to the gas, it will experience a drag
force as momentum is transferred from the grain to the gas.
This drag force is approximately $F_\drag = M_{gr}v_{gr}/\tau_{gr}$
where $M_{gr}$ is the grain mass, $v_{gr}$ is the grain speed relative
to the gas, and $\tau_{gr}$ is the time for the grain to collide with
its own mass in gas particles.
For the case of an uncharged spherical grain in neutral gas, 
the problem can be solved exactly,
and approximated to within 1\% by
\beq
\label{eq:fdrag}
F_\drag = 2\pi a^2 kT \sum_i n_i 
\frac{8s_i}{3\sqrt{\pi}}\left[ 1+\frac{9\pi}{64}s_i^2 \right]^{1/2}
~~~~~s_i^2\equiv \frac{m_i v^2}{2kT}
~~~,
\eeq
where the sum is over the species (H, H$_2$, He) in the gas 
(Draine \& Salpeter 1979a).
If the grain is charged and ions are present, there is an additional
``Coulomb drag'' term (Draine \& Salpeter 1979a), but this is 
numerically unimportant in regions of fractional ionization $\ltsim 10^{-2}$.
Eq.\ (\ref{eq:fdrag}) is obtained for either perfectly inelastic
collision (where the gas atom is assumed to ``stick'') to the grain
or perfectly elastic specular reflections.
For subsonic motion, the drag time is
\beq
\tau_\drag = 
\frac{M_{gr}v}{F_\drag}
= \frac{\rho a \sqrt{2\pi/kT}}{\sum_i n_i \sqrt{m_i}}
\label{sec:tdrag}
~~~.
\eeq
For a cool region where the hydrogen is primarily molecular,
\beq
\tau_\drag= 1.2\times10^5\yr 
\left(\frac{a}{0.1\micron}\right)
\left(\frac{\rho}{3\g\cm^{-3}}\right)
\left(\frac{300\cm^{-3}}{n_{\rmH}}\right)
\left(\frac{25\K}{T}\right)^{1/2}
\label{eq:tdragMC}
~~~,
\eeq
while in a more diffuse region where the hydrogen is atomic,
\beq
\tau_\drag = 4.4\times10^5\yr
\left(\frac{a}{0.1\micron}\right)
\left(\frac{\rho}{3\g\cm^{-3}}\right)
\left(\frac{30\cm^{-3}}{n_{\rmH}}\right)
\left(\frac{100\K}{T}\right)^{1/2}
\label{eq:tdragCNM} ~~~.
\eeq

\subsection{Lorentz Force}

Dust grains are charged, and there are magnetic fields in clouds.
The cyclotron period is
\begin{eqnarray}
\frac{2\pi}{\omega_{c}} &=& \frac{2\pi M_{gr} c}{|\langle Z\rangle|eB}
\\ 
&=& 
\frac{5.2\times10^4 \yr}{|\langle Z\rangle|}
\left(\frac{a}{10^{-5}\cm}\right)^3
\left(\frac{\rho}{3\g\cm^{-3}}\right)
\left(\frac{3\mu{\rm G}}{B}\right)
~~~,
\end{eqnarray}
where we take $\langle Z\rangle$ because the grain
charge fluctuates on a time short compared to the cyclotron period.
Values of $\langle Z\rangle$ are indicated in Figs.\ \ref{fig:Zdist.mc} and
\ref{fig:Zdist.cnm}.
The character of the grain dynamics depends on the
ratio of the drag force to the Lorentz force, or
\begin{eqnarray}
\omega_c\tau_\drag &=&
\frac{3}{4\pi}\frac{|\langle Z\rangle| e B}{a^2 c}
\frac{\sqrt{2\pi/kT}}{\sum_i n_i \sqrt{m_i}}
\\
&=& 53 ~ 
|\langle Z\rangle| 
\left(\frac{B}{3\mu{\rm G}}\right)
\left(\frac{10^{-5}\cm}{a}\right)^2
\left(\frac{100\K}{T}\right)^{1/2}
\left(\frac{30\cm^{-3}}{n_\rmH}\right)
~,
\end{eqnarray}
where we have assumed the hydrogen to be atomic.
For grains in cold clouds, we see that we usually have 
$\omega_c\tau_\drag\gg 1$:
grains are strongly coupled to the magnetic field,
except for grain sizes where collisional charging and photoelectric
charging happen to just balance
so that $|\langle Z\rangle| \ltsim 0.02 (a/10^{-5}\cm)^2$.

\subsection{Radiation Pressure}

The interstellar radiation field is in general anisotropic.
As a result, dust grains are subject to ``radiation pressure'' forces
which can cause drift of
the grain relative to the gas.
This drift can be important as a means of transporting dust through
gas, but also because the drifting grains transfer momentum to the
magnetic field (via the Lorentz force) and to the gas (via gas drag).

Consider a unidirectional beam of photons of energy $h\nu$,
with energy density $u_\nu d\nu$ in interval $(\nu,\nu+d\nu)$.
The photons have momentum $h\nu/c$, and this
momentum can be transferred to the grain by absorption or
by scattering.  
It is easy to see that the rate of momentum transfer is
\begin{eqnarray}
F_\rad &=& \int \frac{u_\nu d\nu}{h\nu} c 
\left[ C_\abs + C_\sca(1-\langle\cos\theta\rangle)\right]
\frac{h\nu}{c}  
\\
&=& \pi a^2 \int u_\nu d\nu ~Q_\pr(\nu)~~~
Q_\pr(\nu) \equiv Q_\abs + Q_\sca(1-\langle\cos\theta\rangle)
~~~.
\end{eqnarray}
where 
$Q_\abs\equiv C_\abs/\pi a^2$,
$Q_\sca\equiv C_\sca/\pi a^2$,
and
$\langle\cos\theta\rangle$ is the mean of the cosine of the scattering angle.

We are generally interested in values of $Q_\abs$ and $Q_\pr$ averaged
over the spectrum of the radiation.

We are often content to approximate the radiation from a star by
a blackbody spectrum.
In neutral regions of the interstellar medium
we are frequently interested in averages over a blackbody spectrum
with a cutoff at 13.6 eV, and
it is useful to calculate spectrum-averaged values
\beq
\langle Q \rangle_T \equiv
\frac{\int_0^{\nu_{\max}} d\nu ~B_\nu(T) Q(\nu)}
     {\int_0^{\nu_{\max}} d\nu ~B_\nu(T)}
~~~,
\eeq
where $B_\nu(T)$ is the Planck function (\ref{eq:planck_func})
and $h\nu_{\max}=13.6\eV$.

We will also be interested in $Q$ values averaged over the 
interstellar radiation field, for which we use the estimate
of Mathis, Mezger, \& Panagia (1983).
\beq
\langle Q \rangle_\ISRF\equiv
\frac{\int_0^{\nu_{\max}} d\nu~ u_\nu^\ISRF Q(\nu)}
     {\int_0^{\nu_{\max}} d\nu~ u_\nu^\ISRF}
~~~.
\eeq

\begin{figure}
\begin{center}
\includegraphics[width=.8\textwidth,angle=0]{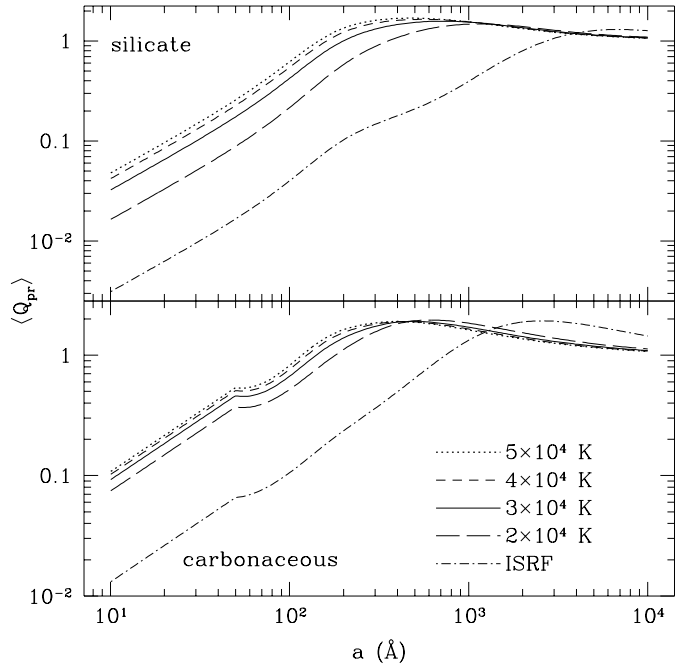}
\end{center}
\caption[]{Radiation pressure efficiency factor 
	$\langle Q_\pr\rangle$ for neutral carbonaceous
	and silicate grains, averaged over the interstellar radiation field
	(ISRF) and blackbody spectra with indicated color tempertures,
	cut off at 13.6 eV.
	Taken from Weingartner \& Draine (2001b).
	}
\label{fig:Qpr}
\end{figure}

\begin{figure}
\begin{center}
\includegraphics[width=.6\textwidth,angle=270]{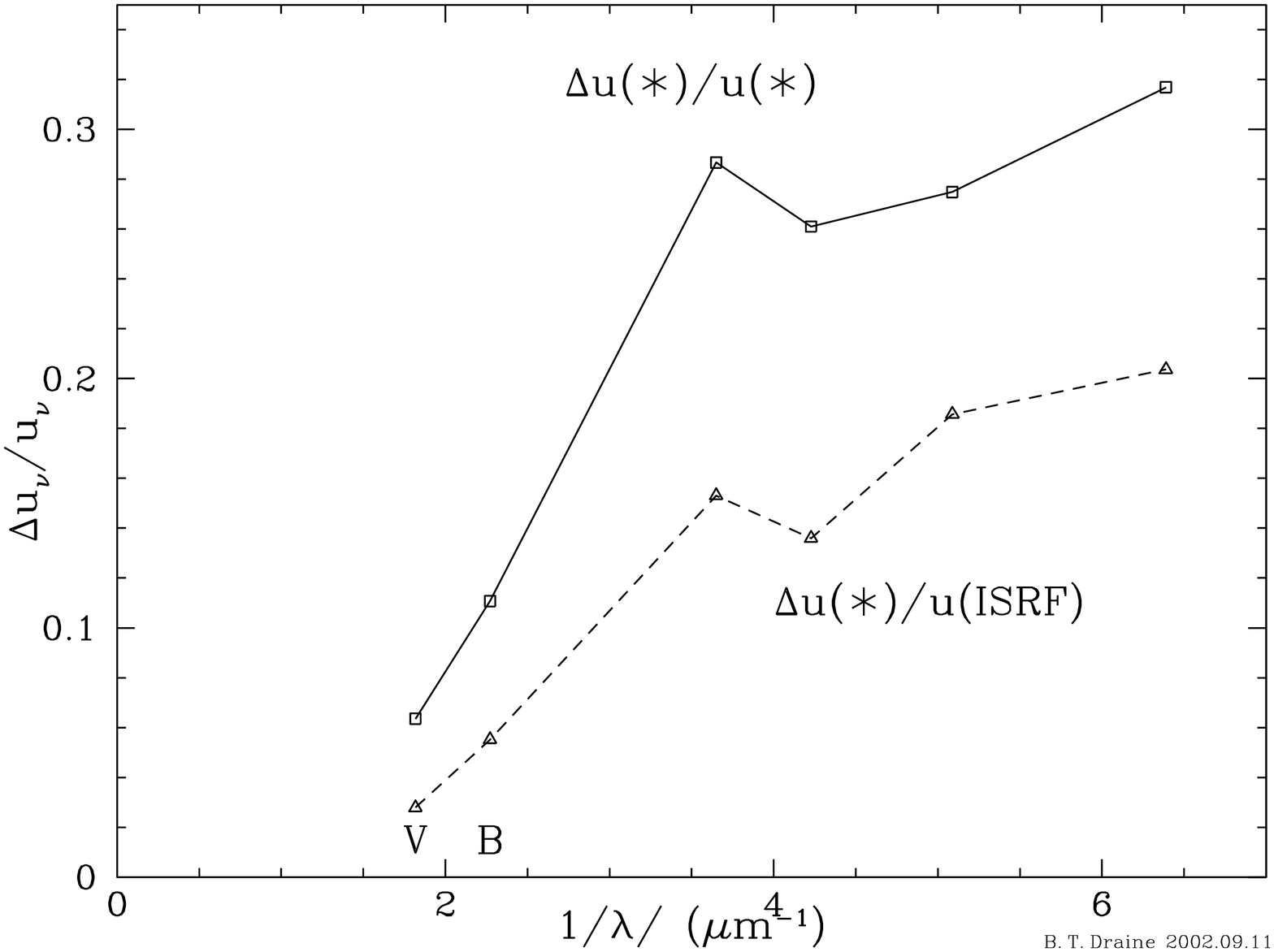}
\end{center}
\caption[]{Dipole fraction for the interstellar radiation field.
	Upper curve is the ratio of the measured dipole moment to
	the measured starlight (the measurements are incomplete).
	Lower curve is the ratio of the measured dipole moment to
	the total estimated ISRF.
	The true fraction is probably between these two curves.
	After Weingartner \& Draine (2001b).
	}
\label{fig:isrfaniso}
\end{figure}

Suppose that at some point in the interstellar medium the starlight
has a net energy flux $c\Delta u_\rad$.  By summing over the Skymap Star
Catalog (Slater \& Hashmall 1992), the Tycho Catalog (ESA 1997),
and the all-sky ultraviolet
observations of the S2/68 experiment on the TD-1A satellite,
Weingartner \& Draine (2001b) estimated the anisotropy in the
starlight background at V, B, and 4 ultraviolet bands.
In Fig.\ \ref{fig:isrfaniso} we show the net dipole moment of the
radiation field at these 6 wavelengths.

\subsection{Recoil from Photoelectric Emission and Photodesorption}

Radiation pressure is, of course, just the transfer of photon momentum
to the grain.  However, radiation can serve to energize other processes
which can exert a thrust on the grain: photoelectric emission and
photodesorption.
To see that these might be important, it is sufficient to compare the
momentum of an $h\nu\approx10\eV$ photon with the momentum 
of the photoelectron
it might produce, if the photoelectron kinetic energy $= f\times h\nu$:
\beq
\frac{p({\rm electron})}{p({\rm photon})} = 
\frac{(2m_e f h\nu)^{1/2}}{h\nu/c} =
320 f^{1/2}\left(\frac{10\eV}{h\nu}\right)^{1/2}
~~~.
\eeq
Similarly, suppose that a photon caused an $\HH$ molecule to
be desorbed with kinetic energy $f\times h\nu$:
\beq
\frac{p({\rm H_2})}{p({\rm photon})} = 
\frac{(4m_\rmH f h\nu)^{1/2}}{h\nu/c} =
1.9\times10^4 f^{1/2}
\left(\frac{10\eV}{h\nu}\right)^{1/2}
~~~.
\eeq
Therefore if an appreciable fraction of absorbed photons produce
photoelectrons or photodesorb atoms or molecules, there could be
a significant recoil on the grain.

For a spherical grain in an isotropic radiation field, the photoelectrons
would be emitted isotropically, with zero net thrust.
Suppose, however, that we have a unidirectional radiation field
(e.g., light from a single star).
In this case we might expect a higher rate of photoelectron emission from
the ``bright side'' than from the
``dark side'' of the grain.
However, very small grains do not have a ``dark side'' -- they are
effectively transparent to the incident radiation, so one must
try to estimate how this depends on grain size.
How can we estimate the bright side/dark side asymmetry?
A simple model is to assume
that
the rate of photoelectric emission from a point on the grain surface
is proportional to $|{\bf E}|^2$ just inside the surface.
Using Mie theory to calculate the
electric field intensity $\bf E$ in a spherical grain, one can then
evaluate the relative rates of photoelectric emission over the grain
surface.
Using this approach, Weingartner
\& Draine (2001b) have investigated the net thrust due to
escaping photoelectrons -- equal and opposite to
the net momentum which the escaping electrons have when they reach
infinity.
The photoelectric yield, and the energy of the photoelectrons, is of
course affected by the charge on the grain, and therefore depends on
the environmental conditions.
The analysis is complicated by
the fact that if the grain is charged, photoelectrons with nonzero
angular momentum will escape to infinity in directions different
from the direction they are travelling when they emerge from the
grain surface
(until it reaches infinity, the electric field of the photoelectron
exerts a force on the grain).

\begin{figure}
\begin{center}
\includegraphics[width=.75\textwidth,angle=0]{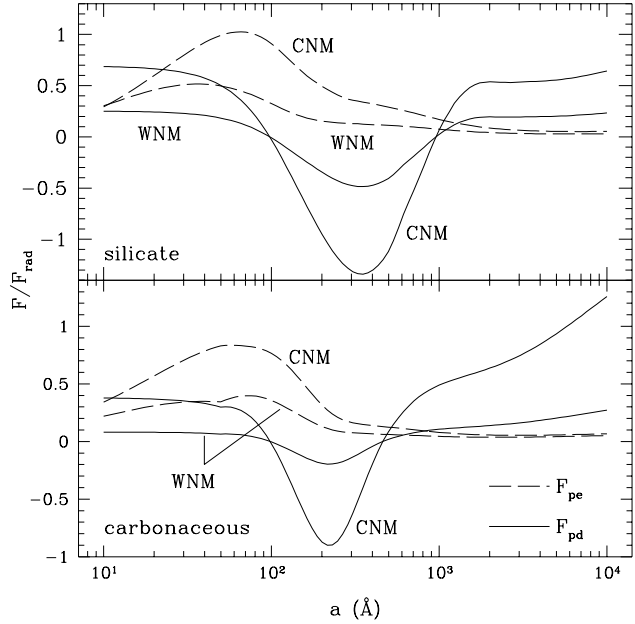}

\end{center}
\caption[]{Photoelectric thrust and photodesorption thrust
	 relative to radiation pressure
	for carbonaceous and silicate grains as a function of
	grain radius $a$.
	Curve labelled ``CNM'' is for cold diffuse clouds with
	$n_e=.03\cm^{-3}$, $T=100\K$, and the standard ISRF.
	Note that the photodesorption force can be negative for
	grains with radii $100\Angstrom\ltsim a \ltsim 500\Angstrom$.
	From Weingartner \& Draine (2001b).
	}
\label{fig:pepdthrust}
\end{figure}

Fig.\ \ref{fig:pepdthrust} shows the results of Weingartner \& Draine (2001b)
for the ratio of the photoelectric thrust to the radiation pressure force,
as a function of grain size.
We see that this is an order-unity correction for grains with
$a\approx 100\Angstrom$ in the CNM.

For photoelectric emission, it is reasonable to assume that the grain
is ``spherically symmetric'' prior to arrival of the photon -- there
are just as many electrons available for photoejection 
on the bright side as on the dark side of the grain.
For photodesorption, however, this is not a reasonable assumption.
As we will see in \S\ref{sec:rotation} below, grains are expected to
be spinning rapidly, with their principal axis of largest moment of
inertia expected to be aligned with the angular momentum.
Because of this gyroscopic stabilization, one may expect that if
photodesorption is a rapid process, then
the ``bright'' side of the grain will have fewer adsorbed molecules
ready to be photodesorbed, an effect which will depend on the
orientation of the grain's spin axis with the direction of radiation
anisotropy.

If photodesorption is rapid, the thrust will be limited by
the need to resupply the grain surface with new molecules to be
photodesorbed.
The story has a number of additional complications:
\begin{itemize}
\item The result will depend on the wavelength-dependence of the
photodesorption cross sections for the adsorbed species.
This is essentially unknown.  Weingartner \& Draine simply assume that
the photodesorption rate for a molecule on the grain surface is
proportional to $|{\bf E}|^2$ just outside the grain surface,
integrated from 6-13.6 eV.
\item Adsorbed species 
may diffuse over the grain surface, so that the ``bright side'' could
be replenished from the ``dark side''.
\item If photodesorption is rapid, 
the surface coverage of adsorbates will depend on the
orientation of the grain spin axis relative to the direction of
starlight anisotropy.
\item The spinning grain can
undergo ``polar flips'' -- the ``thermal flipping'' process described
by Lazarian \& Draine (1999a).
\item If the grain is moving, the grain motion will alter the rates
at which molecules are resupplied to the different parts of the
grain surface: the ``bright side'' may suffer a reduction in the
rate of arrival of atoms or molecules from the gas.
\item H atoms which are not photodesorbed might react on the grain
surface to form H$_2$, and could recoil from the grain with substantial
kinetic energy.  This process might happen more rapidly on the dark
side where the H concentration is higher, thus acting to oppose the
thrust due to photodesorption.
\end{itemize}
With so many uncertainties, it is not possible to reach definite
quantitative estimates for the photodesorption force, but Weingartner \&
Draine (2001b) evaluate one plausible set of assumptions; 
results are shown in Fig.\ \ref{fig:pepdthrust}
for grains in the ``warm neutral medium'' and the ``cold neutral medium''.
It is curious that there is a range of grain sizes 
(100 - 400$\Angstrom$ for carbonaceous grains, 100 - 1000$\Angstrom$ for
silicate grains) where they find that the photodesorption force is
{\it negative} -- this occurs because for this range of sizes,
interference effects cause $|{\bf E}|^2$ to be larger on what would
have been expected to be the dark side of the grain if the
grain were opaque and geometric optics were applicable.
Note that the larger grains do behave as ``expected'' for macroscopic
targets.

The ``bottom line'' is that photodesorption can make a significant
contribution to the net force on the grain, which (coincidentally) is
expected to be comparable in magnitude to the radiation pressure force
and the photoelectric thrust, but which can in principle be of the
opposite sign, depending on detailed physics of photodesorption which
are not known at this time.

\begin{figure}
\begin{center}
\includegraphics[width=.8\textwidth,angle=0]{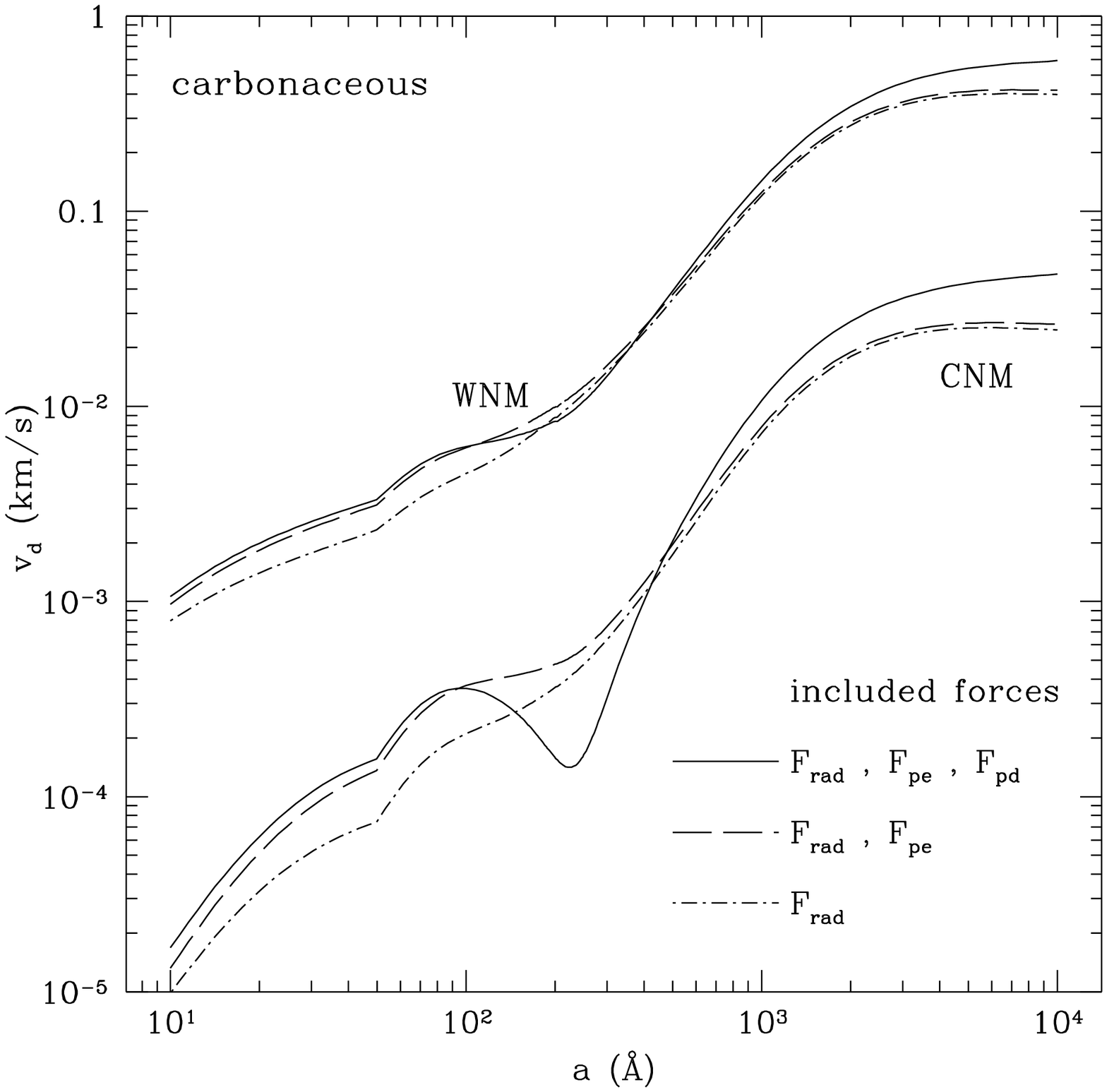}
\includegraphics[width=.8\textwidth,angle=0]{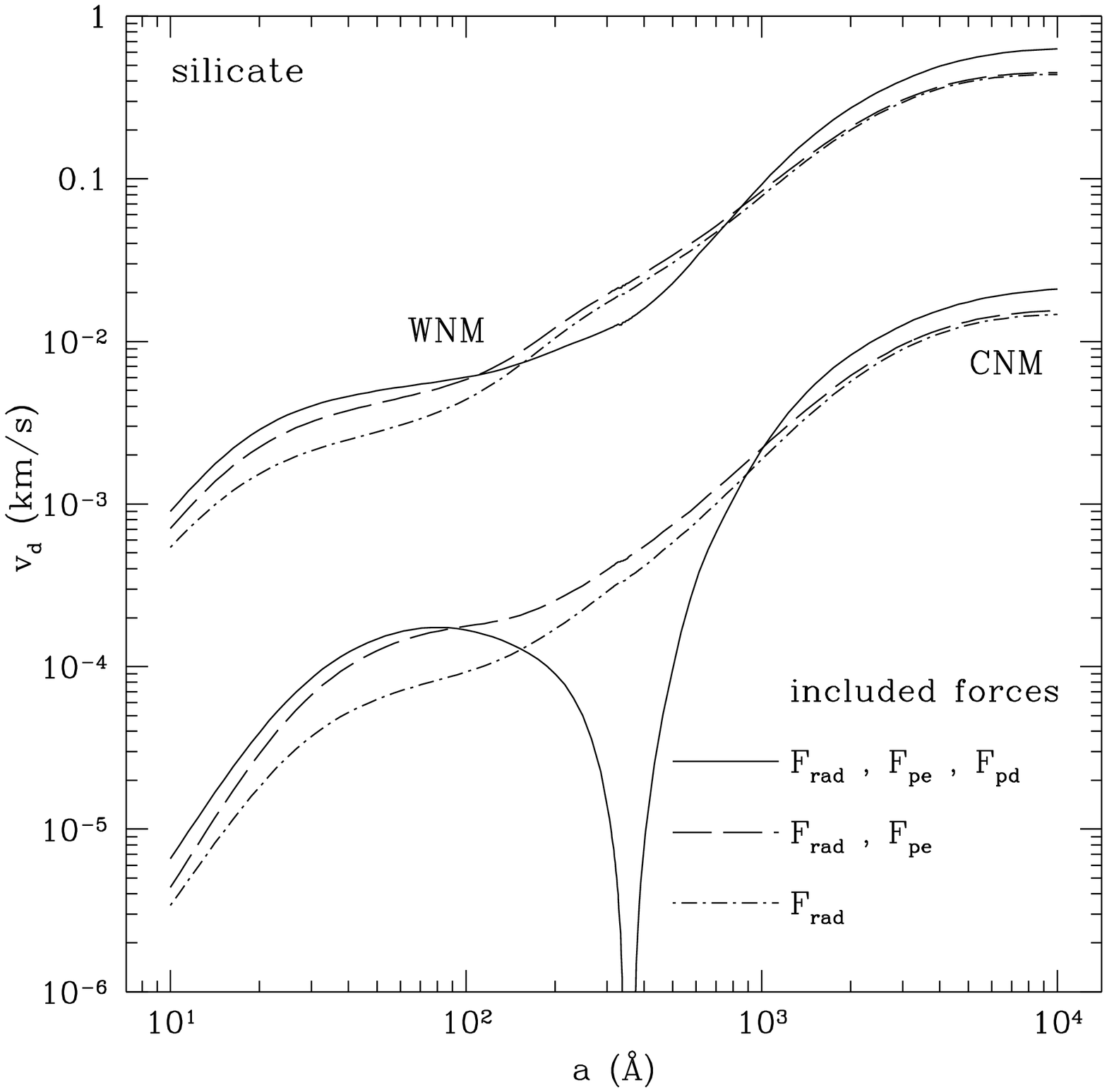}
\end{center}
\caption[]{Drift velocities for carbonaceous and silicate grains,
	as a function of grain size, in the CNM and WNM,
	for a radiation anisotropy $\Delta u_\rad/u_\ISRF=0.1$.
	Taken from Weingartner \& Draine (2001b).
	The minimum in drift velocity for $a\approx250\Angstrom$ carbonaceous
	grains or $a\approx350\Angstrom$ silicate
	grains in the CNM is due to the reversal of the photodesorption force
	for grains in this size range (see Fig.\ \ref{fig:pepdthrust}).
	}
\label{fig:vdrift}
\end{figure}

\subsection{Drift Velocities for Interstellar Grains\label{sec:vdrift}}

For the special case where the radiation anisotropy is parallel to
the magnetic field direction, we can obtain the steady-state
grain drift velocity by balancing the drag force against the
radiation-related forces:
\beq
F_\drag = F_\rad + F_\pe + F_\pd
~~~,
\eeq
where $F_\pe$ and $F_\pd$ are the forces due to photoelectric emission
and photodesorption.
If we assume subsonic motion, and neglect Coulomb drag,
this can be written
\beq
\frac{v_\drift}{\sqrt{kT/m_\rmH}}
=
\langle Q_\pr\rangle
\left(\frac{\Delta u_\rad}{n_\rmH kT}\right)
\frac{1 + F_\pe/F_\rad + F_\pd/F_\rad}
{\sum_i (n_i/n_\rmH)\sqrt{m_i/m_\rmH}}
~~~,
\eeq
so we see that the grain will drift with a ``Mach number'' which
is approximately equal to $\langle Q_\pr\rangle$ times
the ratio of the anisotropic radiation pressure to the gas pressure.
Since $Q_\pr$ reaches values of order unity (see Fig.\ \ref{fig:Qpr}),
we may expect peak Mach numbers or order 0.2 or so -- the ratio of
the anisotropic radiation energy density to the gas pressure in the
diffuse ISM.

In Figure \ref{fig:vdrift} we show drift speeds calculated by
Weingartner \& Draine for carbonaceous and silicate grains in the
WNM and CNM.  
The drift speeds are clearly a strong function of grain size, and
appear to be negligible for small $a\ltsim 100\Angstrom$ grains.
For larger grains the drift velocities are modest but potentially significant
since they could be sustained for long times:  in the CNM,
a radiation anisotropy of only $\Delta u_\rad/u_\rad=0.1$ would give
a $0.3\micron$ carbonaceous grain a drift speed of
$\sim 0.04 \kms$, so in 10 Myr it could drift 0.4 pc,
and in the WNM the drift speeds are about an order of magnitude larger.
This could in principle lead to removal of the large grains from
some gas elements and concentration in others.

\section{Rotational Dynamics of Interstellar Dust
	\label{sec:rotation}}

The rotational dynamics of interstellar grains constitutes a fascinating
story, not yet fully told.  It is intimately tied to the long-standing
problem of interstellar grain alignment, which we will discuss in 
\S\ref{sec:alignment}.

\subsection{Brownian Rotation?\label{subsec:brownian_rot}}

Early discussions of the rotational dynamics of interstellar grains
generally
assumed the grains to be rigid spheres, which appeared to capture the
essential physics, and discussed the 
scattering of impinging atoms from the grain surface.
As far as grain rotation was concerned, it seemed obvious that the grain
would simply undergo a form of Brownian motion, exchanging linear and
angular momentum with the gas.
Once statistical equilibrium was attained, a rigid grain (with purely
elastic scattering of atoms from its surface) would have an
expectation value of $kT_{\rm gas}/2$ for the kinetic energy
in each of its degrees of freedom:
three translational and three rotational.
Thus the r.m.s. translational velocity of the grain would be
\begin{eqnarray}
\langle v^2\rangle^{1/2} &=& 
\left(\frac{3kT}{M}\right)^{1/2} = 
\left(\frac{9kT}{4\pi\rho a^3}\right)^{1/2}
\\
&=& 1.82 \frac{\cm}{\s}\left(\frac{T}{100\K}\right)^{1/2}
\left(\frac{3\g\cm^{-3}}{\rho}\right)^{1/2}
\left(\frac{10^{-5}\cm}{a}\right)^{3/2}
~~~,
\label{eq:vrms}
\end{eqnarray}
which is negligible under all foreseeable circumstances.
The r.m.s. rotation rate
\begin{eqnarray}
\frac{\langle\omega^2\rangle^{1/2}}{2\pi} &=& 
\frac{1}{2\pi}\left(\frac{3kT}{2Ma^2/5}\right)^{1/2}
=\frac{1}{2\pi}\left(\frac{45kT}{8\pi\rho a^5}\right)^{1/2}
\\
&=& 4.6\times10^4 {\rm Hz}
\left(\frac{T}{100\K}\right)^{1/2}
\left(\frac{3\g\cm^{-3}}{\rho}\right)^{1/2}
\left(\frac{10^{-5}\cm}{a}\right)^{5/2}
\label{eq:omega_brownian}
\end{eqnarray}
is more impressive, though it simply corresponds to an equatorial
velocity equal to the r.m.s. translational velocity (\ref{eq:vrms}) times
a factor $\sqrt{5/2}=1.58$.

\subsection{Suprathermal Rotation\label{sec:suprathermal}}

Following the discovery of grain alignment in 1949, theoretical
studies of the rotational dynamics were carried out to try to
understand the alignment process, including
the seminal paper by Davis \& Greenstein (1951).
It was recognized that, since
interstellar grains were likely to have vibrational temperatures
differing from the gas temperature, eq.\ (\ref{eq:omega_brownian})
needed to be modified, and it was understood that for a spherical
grain the temperature $T$ appearing in eq.\ (\ref{eq:omega_brownian})
should be some weighted mean of the gas temperature and the grain 
temperature, since collisions of atoms with the grain surface will
not be perfectly elastic.

However, in 1975 -- 24 years after Davis \& Greenstein's study --
Purcell (1975) realized that interstellar grains could act as ``pinwheels''.
After all, a simple pinwheel weighing many grams can readily be put into
rotation with a rotational kinetic energy many orders of magnitude greater
than $kT$ by the air molecules which excite it.
Purcell (1979) showed, in fact, that there were 3 distinct physical processes
acting on interstellar grains which could each, acting alone, cause a grain
to rotate with a rotational energy $(1/2)I\omega^{2}\gg kT$:
\begin{enumerate}
\item The effective ``accomodation coefficient'' will probably vary
	over the grain surface, due to chemical inhomogeneities 
	and/or geometrical irregularities.  Such variations in accomodation
	coefficient on a nonspherical grain 
	will lead to a systematic torque (i.e., one which whose time-average
	is not expected to be zero) if the gas temperature and
	grain temperature differ.
\item Photoelectrons ejected from the grain may, on average, carry away
	angular momentum, resulting in a systematic torque.
\item H$_2$ formation on the grain surface, with the nascent H$_2$ molecules
	ejected with a significant kinetic energy, is expected to result
	in a large systematic torque, particularly if the
	H$_2$ formation occurs at a relatively small number
	of ``active sites'' on the grain surface.
\end{enumerate}
Purcell showed that of the three processes, the H$_2$ formation process
was likely to be the most important in regions where the H is atomic.

{\bf\medskip\noindent
\ref{sec:suprathermal}.1 Suprathermal Rotation Driven by H$_2$ Formation
\medskip}

It is instructive to examine the argument for suprathermal rotation
driven by H$_2$ formation at active sites on the grain surface.

To simplify the argument, let us consider a region where only
atomic H is present (no He), and
consider a cube of sides $2b\times2b\times2b$, with volume
$8b^3$, surface area $24b^2$, and effective radius 
$a_{\rm eff} = (6/\pi)^{1/3} b = 1.24 b$.

Choose a site at random on the grain surface, and
assume that $\HH$ molecules are formed at that site at a rate
$\dot{N}_1$.  
For simplicity, suppose that the newly-formed $\HH$ molecules leave
normal to the surface with kinetic energy $E_{\HH}$.
The expectation value for the square of
the moment arm is then just $(2/3)b^2$, and
the expectation value for the mean square
torque exerted by this site is 
\begin{eqnarray}
\Gamma_1^2 &=& \dot{N}_1^2 [m_{\HH} v_{\HH}]^2 \times \frac{2}{3} b^2
\\
&=& \frac{8}{3} \dot{N}_1^2 m_\rmH E_{\HH} b^2
~~~.
\end{eqnarray}
Now suppose that there are a total of $N_{site}$ such sites on the grain
surface, and suppose that a fraction $\gamma$ of the H atoms which
impinge on the surface are converted to $\HH$ (so there are $\gamma/2$
$\HH$ formation events per arriving H atom).
In a thermal gas, the rate per area at which atoms collide with
a convex surface is just $n (kT/2\pi m)^{1/2}$, so 
\beq
\dot{N}_1 = \frac{\gamma}{2N_{site}}
n(\rmH) \left(\frac{kT}{2\pi m_\rmH}\right)^{1/2} 24 b^2
~~~.
\eeq
Since the sites are assumed to be randomly located, the torques
add like a random walk, so that the expectation value for the
square of the total torque is just
\begin{eqnarray}
\langle\Gamma_{ex}^2\rangle &=& N_{site} \Gamma_1^2 
\\
&=& \frac{192}{\pi} \frac{1}{N_{site}} \gamma^2 n(\rmH)^2 E_{\HH} kT ~b^6
~~~.
\label{eq:Gamma_ex^2}
\end{eqnarray}
The rotating grain will experience gas drag.  For slow rotation rates
$\omega$,
it is not difficult to show that the drag torque is
\beq
\Gamma_{drag} = - \frac{80}{3} n(\rmH) 
\left(\frac{m_\rmH kT}{2\pi}\right)^{1/2}
b^4 \omega
~~~.
\eeq
To find $\langle\omega^2\rangle$, we now set 
$\langle \Gamma_{ex}^2\rangle = \langle\Gamma_{drag}^2\rangle$ to obtain
\beq
\langle \omega^2\rangle = 
\frac{27}{50} \frac{\gamma^2}{N_{site}} \frac{E_{\HH}}{m_{\rmH}}
\frac{1}{b^2}
~~~,
\eeq
or a rotational kinetic energy, relative to the thermal value,
\begin{eqnarray}
\frac{I\langle \omega^2\rangle}{3kT} &=&
\frac{3}{25} \frac{M}{m_\rmH} \frac{\gamma^2}{N_{site}} \frac{E_{\HH}}{kT}
\\
&=& 2.0\!\times\!10^{11} \frac{\gamma^2}{N_{site}}
\left(\frac{b}{10^{-5}\cm}\right)^3
\!
\left(\frac{\rho}{3\g\cm^{-3}}\right)
\!
\left(\frac{E_{\HH}}{\eV}\right)
\!
\left(\frac{100\K}{T}\right)  ~.~~~
\end{eqnarray}
where $M=8b^3\rho$ is the grain mass.
Now this is no small number!  The number of active sites is
uncertain.
If we suppose that there is a surface area $s^2$ per active site,
then $N_{site}=24b^2/s^2$, and
\beq
\frac{I\langle\omega^2\rangle}{3kT}
=
8.4\!\times\!10^5 \gamma^2 \left(\frac{s}{10 \Angstrom}\right)^2
\!\left(\frac{b}{10^{-5}\cm}\right)
\!\left(\frac{\rho}{3\g\cm^{-3}}\right)
\!\left(\frac{E_{\HH}}{\eV}\right)
\!\left(\frac{100\K}{T}\right)
.
\label{eq:H2superthermality}
\eeq
The r.m.s. rotation rate is
\beq
\frac{\langle\omega^2\rangle^{1/2}}{2\pi} = 
3.6\times10^5 {\rm Hz} ~\gamma 
\left(\frac{s}{10\Angstrom}\right)
\left(\frac{10^{-5}\cm}{b}\right)
\left(\frac{E_{\HH}}{1\eV}\right)
~~~,
\eeq
so that we might have MHz rotational frequencies for $b < 3\times 10^{-6}\cm$!
It is apparent that systematic torques can play a major role in
grain dynamics.

Each $\HH$ formation site can be thought of as being like a small
rocket thruster attached to the grain surface: the systematic torque
due to $\HH$ formation is fixed in body coordinates, so long as the
$\HH$ formation sites do not change.
The kinetics of $\HH$ formation on grain surfaces is poorly understood at
this time, and it is not certain how long-lived the active sites are
likely to be.

{\bf\medskip\noindent
\ref{sec:suprathermal}.2 Radiative Torques Due to Starlight
\medskip}

The three torques identified by Purcell do indeed appear to be individually
capable of driving grains to suprathermal rotation.  In fact, there appears
to be an additional physical process which can compete with the
$\HH$ formation torque for grains larger than $\sim 10^{-5}\cm$:
torques exerted on interstellar grains by starlight.

We discussed above the ``radiation pressure'' 
force which anisotropic starlight could exert on
a grain.
It turns out that starlight can also exert a torque on grains, and these
torques can be dynamically important!
\begin{figure}
\begin{center}
\includegraphics[width=.6\textwidth,angle=0]{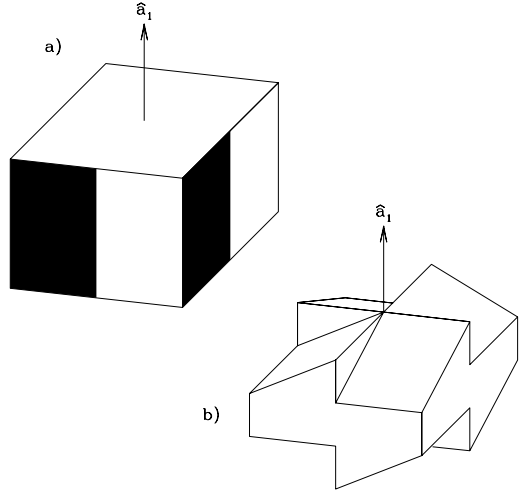}
\end{center}
\caption[]{Macroscopic examples of targets which would be subject to
	radiative torques.
	From Draine \& Weingartner (1996).
	}
\label{fig:torque_pic}
\end{figure}

How can we address this problem?  Analytic progress is difficult, because
the only cases for which we have analytic solutions are cases where
radiative torques vanish: (1) grains of complex shape but small
compared to the wavelength; (2) spheres;
(3) spheroids.
In the first case the torques vanish because only the dipole response of
the grain is important, and the dipole radiation does not carry angular
angular momentum.  In the case of spheres and spheroids, there is
no torque because of the symmetry of the target.

In order to obtain a nonzero torque we must have an asymmetric target.
Draine \& Weingartner (1996) discussed use of the 
discrete dipole approximation to calculate radiative torques.
The DDSCAT code (Draine \& Flatau 2000) 
now provides the capability to compute torques.
For an asymmetric target 
illuminated by a beam of unpolarized radiation, the direction and
magnitude of the  torque
depends on the orientation of the target relative to the beam.
It is useful to define a ``torque efficiency vector'' 
${\bf Q_\Gamma}$ defined so
that the torque on the grain is given by
\beq
{\bf\Gamma}_{rad} = \pi a_{\rm eff}^2 u_{rad} \frac{\lambda}{2\pi} 
{\bf Q_\Gamma}
~~~,
\eeq
where the effective radius 
$a_{\rm eff}$ is the radius of an equal volume sphere.
We will generally want to average $\bf Q_\Gamma$ over some spectrum;
Draine \& Weingartner (1996) calculate 
\begin{eqnarray}
\langle{\bf Q_\Gamma}\rangle &\equiv&
\frac{\int {\bf Q_\Gamma} \lambda u_\lambda d\lambda}
     {\int \lambda u_\lambda d\lambda}
~~~,
\\
\bar{\lambda} &\equiv& 
\frac{\int \lambda u_\lambda d\lambda}
{\int u_\lambda d\lambda}
~~~,
\end{eqnarray}
for $u_\lambda$ taken to be the MMP ISRF (for which
$\bar{\lambda} = 1.202\micron$).
With these definitions, the torque
\beq
{\bf \Gamma}_{rad} = \pi a_{\rm eff}^2 u_{rad} \frac{\bar{\lambda}}{2\pi}
\langle{\bf Q_\Gamma}\rangle
~~~.
\eeq

We recall from classical mechanics that an arbitrary object has three
principal axes, 
${\bf\hat{a}}_1$,
${\bf\hat{a}}_2$,
${\bf\hat{a}}_3$,
and the moment of inertia tensor is diagonal
in a coordinate
system aligned with these three axes.
Let the eigenvalues of the moment of inertia tensor be 
$I_1 \geq I_2 \geq I_3$.

For purposes of discussion, let us assume that $I_1 > I_2=I_3$.
If the grain has angular momentum $\bf J$,
then the rotational kinetic energy is
\begin{eqnarray}
E_{rot} &=& \frac{(J\cos\beta)^2}{2I_1} + \frac{(J\sin\beta)^2}{2I_2}
\\
&=& \frac{J^2}{2I_1} + J^2\frac{(I_1-I_2)}{I_1I_2}\sin^2\beta
~~~,
\label{eq:Erot}
\end{eqnarray}
with the rotational kinetic energy minized for $\beta=0$.
As we will see below, when $J$ is large we expect the grain to
remain close to this minimum energy state, with $\sin^2\beta \ll 1$.

The grain spins rapidly,  so we are concerned with
the torque after averaging over rotations of the grain around the
axis $\hat{\bf a}_1$.  This rotation-averaged torque therefore
depends only on a single angle -- the angle $\Theta$ between
$\hat{\bf a}_1$ and the direction of the incident radiation.

We will see below that we expect a spinning dust grain to precess
around the direction of the local magnetic field $\bf B$, so what
we really need is the torque averaged over grain rotations {\it and}
over precession of the axis $\hat{\bf a}_1$
around the magnetic field direction.
For a given grain and spectrum of radiation, this will now depend on
two angles: the angle $\psi$ 
between $\bf B$ and the direction of the radiation,
and the angle $\xi$ between $\hat{\bf a}_1$ and ${\bf B}$.

For purposes of spinup (or spindown), 
we are interested in the projection of the torque efficiency vector along
the spin axis, i.e., 
the axis of largest moment of inertia:
${\bf Q_\Gamma(\hat{\Theta}) \cdot \hat{a}}_1$.
Draine \& Weingartner (1996, 1997) define $H(\xi,\phi) = 
\langle {\bf Q_\Gamma\cdot \hat{\bf a}_1}\rangle$ 
averaged over rotation around $\hat{\bf a}_1$ and precession of
$\hat{\bf a}_1$ around $\bf B$.
They calculate $H(\xi,\phi)$
for three grain shapes with effective radii $a_{\rm eff}=0.2\micron$.
The numerical values of $H$ of course depend on $\xi$ and $\phi$,
but a typical value of $|H|$ would be $|H|\approx 0.005$
(see Fig.\ 5 of Draine \& Weingartner 1997), so that
the spinup torque would be
\beq
\langle {\bf\Gamma}_{rad}\cdot\hat{\bf a}_1\rangle 
= H \bar{\lambda} \Delta u_{rad}
\pi a_{\rm eff}^2
~~~.
\label{eq:radtorque}
\eeq

It is of interest to compare this to the spinup torque due to
$\HH$ formation.  We can rewrite eq.\ (\ref{eq:Gamma_ex^2}) replacing
$b$ by $(\pi/6)^{1/3}a_{\rm eff}$, and project onto a random direction
$\hat{\bf a}_1$ to obtain
\beq
\langle ({\bf\Gamma}_{ex}\cdot \hat{\bf a}_1)^2\rangle^{1/2}
=
\frac{1}{3}
\left(\frac{2}{3\pi}\right)^{1/2}
\left(\frac{6}{\pi}\right)^{1/3}
\gamma 
\left( \frac{E_{\HH}}{kT} \right)^{1/2} 
s ~ n(\rmH) kT ~ \pi a_{\rm eff}^2
~~~.
\eeq
Thus the ratio of the radiative spinup torque (\ref{eq:radtorque})
to the r.m.s. value of the $\HH$ formation torque is
\beq
\frac{\langle {\bf\Gamma}_{rad}\cdot\hat{\bf a}_1\rangle}
{\langle ({\bf\Gamma}_{ex}\cdot \hat{\bf a}_1)^2\rangle^{1/2}}
=
\frac{ H \bar{\lambda}}
{0.191 \gamma (E_{\HH}/kT)^{1/2}s} ~
\frac{\Delta u_{rad}}
{n(\rmH) k T}
\eeq
\beq
=
\frac{2.92}{\gamma} \left(\frac{H}{0.005}\right)
\left(\frac{T}{100\K}\right)^{1/2}
\left(\frac{\eV}{E_{\HH}}\right)^{1/2}
\left(\frac{10\Angstrom}{s}\right)
\frac{\Delta u_{rad}}
{n(\rmH) k T}
~~~.
\eeq
Adopting the 10\% anisotropy estimated by Weingartner \& Draine (2001b),
we have $\Delta u_{rad}\approx 8.64\times10^{-14}\erg\cm^{-3}$.
With $n({\rmH})=30\cm^{-3}$, and $T=100\K$, we have
$\Delta u_{rad}/n(\rmH)kT = 0.21$.
Therefore with an $\HH$ formation efficiency
$\gamma < 0.5$ the radiative torque would exceed the $\HH$
formation torque if the other parameters have nominal values
(in particular, $s=10\Angstrom$).
Many of the numbers are uncertain; the important point is that
for $a_{\eff}=0.2\micron$ grains the radiative torque has a magnitude
which is comparable to our estimate for the $\HH$ formation torque.

\section{Alignment of Interstellar Dust \label{sec:alignment}}

\begin{figure}
\begin{center}
\includegraphics[width=0.64\textwidth,angle=270]{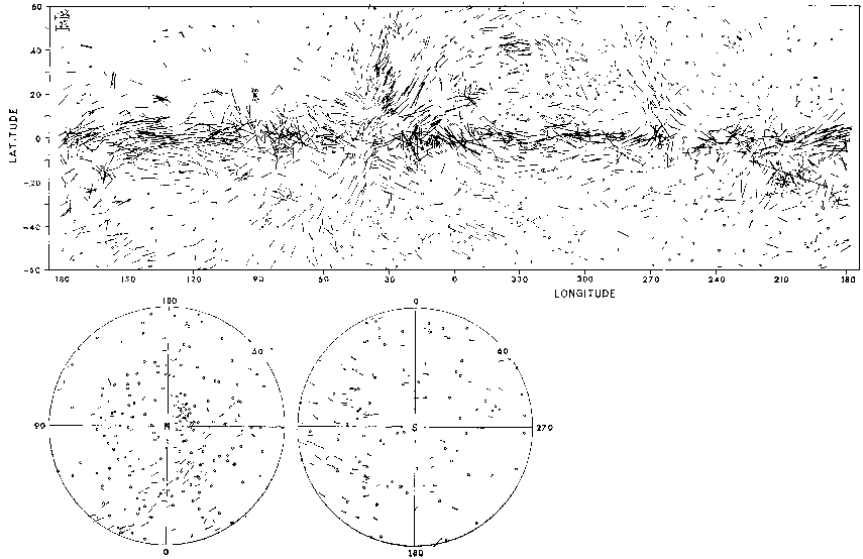}
\end{center}
\caption[]{Linear polarization of light from 1800 stars.
	From Mathewson \& Ford 1970.}
\label{fig:1800stars}
\end{figure}

Polarization of starlight was discovered in 1949.
It was
immediately deduced that the only plausible mechanism was
``linear dichroism'' of the interstellar medium due to aligned
dust grains, but understanding the mechanism responsible for this
alignment has frustrated astrophysicists for half a century.
In fact, the story of the efforts of astrophysicists to understand
grain alignment is humbling.
It is not just because it has taken us so long to solve the problem --
the humbling lesson is that several very important and relatively
simple physical effects were overlooked for decades.
It was not that these physical effects were difficult to understand, or
that they involved physics which has only recently been understood.
On the contrary -- the physics
involved would have been familiar to
well-trained physicist in 1950.
It was simply that important processes were overlooked:
\begin{itemize}
\item The Rowland effect: a charged, spinning dust grain will develop
a magnetic moment due to its circulating charge (Martin 1971).
\item The Barnett effect: a spinning dust grain with unpaired electron spins
will spontaneously magnetize (Dolginov \& Mytrophanov 1976).
\item Suprathermal rotation due to dust-gas temperature differences
	(Purcell 1975, 1979).
\item Suprathermal rotation due to photoelectric emission 
	(Purcell 1975, 1979).
\item Suprathermal rotation due to $\HH$ formation (Purcell 1975, 1979).
\item Viscoelastic dissipation of rotational kinetic energy due to
	time-varying stresses in a grain which is not rotating
	around a principal axis (Purcell 1979)
\item ``Barnett dissipation'' of rotational kinetic energy due to 
	the electron spin system (Purcell 1979).
\item Dissipation of rotational kinetic energy due to the nuclear spin
	system (Purcell 1979).
\item Suprathermal rotation due to starlight torques 
	(Draine \& Weingartner 1996).
\item Fluctuation phenomena associated with Barnett dissipation
	(Lazarian \& Roberge 1997; Lazarian \& Draine 1999a)
	and coupling to the nuclear spins (Lazarian \& Draine 1999b)
\end{itemize}
Have we thought of all the important processes?  Or are there other
important phenomena to which our eyes are still closed?

\subsection{The Motion of Spinning Grains}

As seen above, dust grains spin very rapidly.
Consider for the moment one spinning grain.
If the angular momentum $\bf J$ does not happen to be along one of the
grain's principal axes, then the grain's angular velocity $\omega$
will not be parallel to $\bf J$, and will nutate around $\bf J$, with
a nutation rate that will be smaller than (but comparable to) the
rotation rate.

Now imagine observing a single grain as it spins at kHz rotation frequencies.
Because the rotation and nutation are both rapid,
we will
quickly obtain a rotation-nutation average of the grain orientation.
During this observation, the angular momentum $\bf J$ is conserved, and
defines a preferred direction.
Now consider light propagating in some particular direction.
Because of the rotation-nutation averaging, the grain will tend to
polarize the light with electric vector 
either exactly perpendicular to $\bf J$,
or exactly parallel to $\bf J$,
depending on whether the grain on average has its ``long'' axis
perpendicular to $\bf J$, or parallel to $\bf J$.

This is for one grain.  If a population of grains shows a tendency
to polarize light, then the individual grain angular momenta cannot
be randomly-oriented -- there must be a tendency for the angular
momenta $\bf J$ to be aligned with some preferred direction in space.

The problem of grain alignment is to account for the alignment of
grain angular momenta $\bf J$.

Many alignment mechanisms have been proposed over the years.
Some directly involve the galactic magnetic field (Davis \& Greenstein 1951).
Other proposed mechanisms rely on so-called
``mechanical alignment'' due to gas (Gold 1952) 
or cosmic rays (Salpeter \& Wickramasinghe 1969) streaming
relative to the grains.
Harwit (1970) suggested that grain angular momenta might be aligned
as a result of the angular momentum absorbed from starlight photons,
if the starlight is anisotropic.
It was only in 1971 that Martin (1971) pointed out that even in the case of
nonmagnetic alignment processes, the observed alignment would be
either exactly parallel or exactly perpendicular to the magnetic field,
simply as a result of the magnetic moment on a spinning grain.

\subsection{Grain Magnetic Moments and Precession Around $\bf B$}

Grain in diffuse regions tend to be electrically charged 
(see Fig.\ \ref{fig:Zdist.cnm}).
If charged, a spinning dust grain will develop a magnetic moment due
to the ``Rowland effect'': the rotating charge on the grain constitutes
a current circulating around a loop, and there will be a resulting
magnetic field and magnetic moment interior to the loop.
Recall that a current $I$ flowing in a loop with area $A$
generates a magnetic moment $m=IA/c$.
The mean of $x^2+y^2$ over the surface of a sphere is just
$(2/3)a^2$, so the magnetic moment due to charge $Ze$ distributed uniformly
over the surface of a spinning sphere is
\beq
{\bf \mu} = \frac{{\bf\omega} Ze}{2\pi c} \pi \frac{2a^2}{3} = 
\frac{{\bf\omega} Z e a^2}{3c}
~~~.
\eeq
If $\bf\omega$ is at an angle $\theta > 0$ relative to $\bf B$, the
spinning sphere will precess with a precession frequency
\begin{eqnarray}
\omega_p &=& \frac{\mu B \sin\theta}{I\omega\sin\theta}
= \frac{5}{8\pi}\frac{ZeB}{\rho a^3c}
\\
&=& 1.01\times10^{-4}\yr^{-1} Z \left(\frac{B}{3\mu{\rm G}}\right)
\left(\frac{3\g\cm^{-3}}{\rho}\right)
\left(\frac{10^{-5}\cm}{a}\right)^3
~~~.
\end{eqnarray}
We have seen above (Fig.\ \ref{fig:Zdist.cnm}) 
that in a CNM environment, we might expect
an $a=10^{-5}\cm$ grain to have $\langle Z\rangle\approx 10$,
for a precession period $\sim 6000\yr$.
This time is short compared to the other dynamical times: the gas drag
time (\ref{eq:tdragCNM}), or (as we shall see below) the time for grain
alignment to occur.  

We can therefore conclude that regardless of
what process is responsible for alignment of interstellar grains, each
grain will have a time-averaged angular momentum distribution which 
must be symmetric around $\bf B$.
Therefore the interstellar dust mixture will tend to polarize
starlight either exactly parallel to $\bf B$ or exactly perpendicular
to $\bf B$, for any conceivable alignment mechanism!

\subsection{Barnett Effect}

Dolginov \& Mytrophanov (1976) pointed out that the ``Barnett effect''
would result in grain magnetic moments much larger than due to the
Rowland effect.  The Barnett effect can be stated very simply:
an uncharged object rotating with angular velocity $\omega$ 
tends to spontaneously magnetize, with
a magnetization $M=\chi \omega/\gamma$,
where $\chi$ is the susceptibility, and $\gamma$ is the
``gyromagnetic ratio'' for the material -- the
ratio of (magnetic moment)/(angular momentum) for the electron orbitals
or electron spins responsible for the magnetizability of the material.
For electron spins, $\gamma = \mu_B/(\hbar/2) = e/2m_e c$, 
where $\mu_B\equiv e\hbar/2m_e c$ is the Bohr magneton.

The spontaneous magnetization is
equal to that which would be produced in a stationary
sample in a (fictitous) applied magnetic field $H_{Be}=\omega/\gamma = 
2 m_e c \omega/e$ --
the so-called ``Barnett equivalent field''.

At first sight the Barnett effect seems surprising, but it is not difficult
to understand.  Consider an unmagnetized sample (equal numbers
of ``spin-up'' and ``spin-down'' electrons), with
angular momentum $J$, spinning at angular velocity $\omega$.  
At constant total angular momentum, the
angular velocity of the sample can be reduced,
with consequent reduction of the rotational
kinetic energy $I\omega^2/2$, if some of the angular momentum is put into
the electron spin system.
The electrons have magnetic moments, so
with unequal numbers of electrons in the two spin states, the sample
is now magnetized.  The reduction in rotational kinetic energy is
accompanied by heating of the grain, with an increase in entropy.
The effect was observed in the laboratory by Barnett (1915).

For normal paramagnetic materials, the magnetic susceptibility is
just
$\chi = n_s \mu_B^2/kT_g$, where $n_s$ is the number density of unpaired
electrons, and $T_g$ is the grain temperature.
The magnetic moment due to the Barnett effect is thus
\beq
\mu = \frac{4\pi a^3}{3} \frac{n_s\mu_B^2}{kT} \frac{2 m_e c}{e} \omega
~~~.
\eeq
The ratio of the ``Barnett effect moment'' to the ``Rowland effect moment''
is
\begin{eqnarray}
\frac{\mu({\rm Barnett})}{\mu({\rm Rowland})}
&=&
\frac{2\pi a n_s \hbar^2}{Z kT_g m_e}
\\
&=&
\frac{2.8\times10^5}{Z}
\left(\frac{a}{10^{-5}\cm}\right)
\left(\frac{20\K}{T_g}\right)
\left(\frac{n_s}{10^{22}\cm^{-3}}\right)
~~~.
\end{eqnarray}
An unpaired electron density $n_s\approx 10^{22}\cm^{-3}$ is typical for
normal paramagnetic materials,
so it is clear that the Barnett effect is very important for interstellar
grains, and will result in precession periods of order days or months
rather than the thousands of years for the Rowland effect.

\subsection{Alignment by Paramagnetic Dissipation: Davis-Greenstein Mechanism}

Davis and Greenstein (1951) described a mechanism
which could produce grain alignment.
Consider a grain spinning with angular velocity $\bf \omega$
in a static magnetic field ${\bf B}_0$.
Let $\theta$ be the angle between ${\bf B}_0$ and $\bf\omega$.

The grain material has a magnetic susceptibility $\chi$, and it will
try to magnetize in response to an applied magnetic field.
In grain body coordinates, the applied field has a static component
parallel to the grain rotation axis, with magnitude 
$B_\parallel=B_0\cos\theta$,
and a component of magnitude $B_\perp=B_0\sin\theta$ which appears to
be rotating, with angular velocity $\omega$ relative to the grain.
The rotating field ${\bf B}_\perp$ is perpendicular to $\bf\omega$.

It seems clear that the grain material will develop a magnetization
$\bf M$
which will have a static component 
${\bf M}_\parallel = \chi_0 B_\parallel$ parallel to $\bf\omega$, and
a component ${\bf M}_\perp$ which rotates in body coordinates
(but is stationary in inertial coordinates),
with magnitude $M_\perp = |\chi(\omega)|B_\perp$,
where $\chi(\omega)$ is the (complex) magnetic susceptibility for
a magnetic field rotating at frequency $\omega$.
In body coordinates, the rotating magnetization ${\bf M}_\perp$ 
will ``lag'' behind the rotating
field ${\bf B}_\perp$ -- there will be an ``in-phase'' component
$\chi_1(\omega)B_\perp$ and an ``out-of-phase'' component
$\chi_2(\omega)B_\perp$.

The galactic magnetic field ${\bf B}_0$ will exert a torque on the
grain magnetic moment $V{\bf M}$.
The torque is obviously $\perp$ to ${\bf B}_0$.
A bit of calculation shows that the out-of-phase component of the
magnetization results in a torque of magnitude
$V \chi_2 B_\perp B_0 = V \chi_2 B_0^2 \sin\theta$ which acts to
leave the component of the angular momentum $\parallel$ to $B_0$
unchanged (the torque, after all, must be $\perp {\bf B}_0$) but
acts to reduce the component 
$J_\perp=J\sin\theta$ perpendicular to ${\bf B}_0$:
\beq
\frac{d}{dt} J_\perp = - V \chi_2 B_0^2\sin\theta
~~~.
\eeq
At low frequencies the imaginary component of the magnetic
susceptibility varies linearly with frequency:
$\chi_2 = K\omega$.  With $\omega = J/I = J_\parallel/I\cos\theta$,
noting that $J_\parallel=const$,
we obtain
\beq
\frac{d}{dt}J_\perp =
J_\parallel \frac{d}{dt} \tan\theta = 
- J_\parallel \frac{VKB_0^2}{I} \tan\theta
~~~,
\eeq
with solution
\beq
\tan\theta = \tan(\theta_0) e^{-t/\tau_{\rm DG}}
~~~,
\eeq
\beq
\tau_{\rm DG} \equiv \frac{I}{VK B_0^2}
~~~.
\eeq
The magnetic susceptibilities of paramagnetic materials are quite well
understood at the frequencies of interest (see, e.g., Morrish 1965), with
\beq
K \approx 1.25\times 10^{-13} \s \left(\frac{20\K}{T}\right)
~~~.
\eeq
With this value, and $I\approx (2/5)\rho V a^2$, 
the Davis-Greenstein alignment time
\begin{eqnarray}
\tau_{\rm DG} &=& \frac{2\rho a^2}{5K B_0^2}
\\
&=&
1.2\times10^6\yr
\left(\frac{a}{10^{-5}\cm}\right)^2
\left(\frac{\rho}{3\g\cm^{-3}}\right)
\left(\frac{T}{20\K}\right)
\left(\frac{5\mu{\rm G}}{B_0}\right)^2
~~~.
\end{eqnarray}
Thermally-rotating grains are disaligned by random collisions with
gas atoms on a time scale $\tau_{\rm drag}$.
The ratio of the alignment time to the disalignment time is
\beq
\frac{\tau_{\rm DG}}{\tau_{\rm drag}} =
2.8 
\left(\frac{n_\rmH}{30\cm^{-3}}\right)
\left(\frac{T_g}{20\K}\right)
\left(\frac{5\mu{\rm G}}{B_0}\right)^2
\left(\frac{T}{100\K}\right)^{1/2}
\left(\frac{a}{10^{-5}\cm}\right)
~~~,
\eeq
where we have assumed the hydrogen to be atomic.
Thus we see that even for $a=10^{-5}\cm$ grains and a relatively
high value of the magnetic field strength, alignment is slow compared
to disalignment, and the ratio is linear in the grain radius $a$.
The Davis-Greenstein alignment mechanism favors alignment of {\it smaller}
grains, but this is a problem, since small grains are observed to NOT
be aligned, while large grains are (Kim \& Martin 1995).
As we will see below, some important physics has been overlooked.

\subsection{Effect of Suprathermal Rotation}

Consider now a grain rotating suprathermally.
The Davis-Greenstein alignment time is independent of the rotation
rate, since the torque is proportional to the rate of rotation.
On the other hand,
the time scale for random collisions to change the direction of
the angular momentum is approximately $(J/J_{th})^2 \tau_{\rm drag}$,
where $J_{th}$ is the angular momentum expected for Brownian
rotation.
So if the grain rotation is suprathermal, with $J/J_{th}\gg 1$,
we can virtually ignore the disaligning effects of random collisions,
and Davis-Greenstein paramagnetic dissipation will bring the
grain into alignment with ${\bf B}_0$ in a few million years.
As we have seen in \S\ref{sec:suprathermal}, suprathermal rotation
appears to be expected.  Could it for some reason be suppressed in
small grains?

\subsection{Thermal Flipping vs. Suprathermal Rotation}

Why aren't small grains aligned?  From eq.\ (\ref{eq:H2superthermality})
it certainly appears that $\HH$ formation should be able to drive
grains as small as $10^{-6}\cm$ into suprathermal rotation, and yet
there is strong evidence that
grains smaller than $10^{-5}\cm$ are not aligned (Kim \& Martin 1995).

Lazarian \& Draine (1999a) have proposed an explanation for this -- the
process of ``thermal flipping'' may prevent small grains from achieving
suprathermal rotation.

The key issue is that the grain does NOT behave like a ``rigid body''
subject only to external torques --
the grain has internal vibrational degrees of freedom.
The grain rotational kinetic energy depends on the angle $\beta$ between
principal axis $\hat{\bf a}_1$ and $\bf J$ through eq.\ (\ref{eq:Erot}):
\beq
E_{rot} = \frac{J^2}{2I_1} + J^2 \frac{(I_1-I_2)}{I_1I_2} \sin^2\beta
~~~.
\eeq
In the absence of external torques (and ignoring the small amount of
angular momentum in the spin system), $J$ is fixed, but the rotational
kinetic energy can change if energy is exchanged with the vibrational
degrees of freedom of the grain.
Thus we expect the angle $\beta$ to fluctuate.
Suppose, for example, that we start at $\beta=0$.
Thermal fluctuations will cause $\beta$ to ``explore'' the region
$0 \leq \beta < \pi/2$.
If
\beq
J^2 \frac{(I_1-I_2)}{I_1I_2} \ltsim kT
~~~,
\eeq
then thermal fluctuations could allow $\beta$ to reach $\pi/2$: the
maximum possible rotational kinetic energy.
At this point $\beta$ can go either way, with equal probabilities,
so the grain may ``flip over'' to the state $\pi/2 < \beta \leq \pi$.

Now consider the effects of $\HH$ formation torques.
These torques are {\it fixed} in body coordinates.
If the grain ``flips'' from
$\beta=0$ to $\beta=\pi$, the torques change sign in inertial coordinates!
As a result, if the grain flips frequently, the $\HH$ torques can be
averaged out to zero, and have no significant effect on the grain
dynamics, so that the grain rotation remains essentially thermal!
Lazarian \& Draine (1999a) refer to this as ``thermal trapping''.

The question now comes down to estimating the rate of flipping, and how
the rate depends on grain size.
Lazarian \& Draine (1999a) discussed the role which the Barnett effect
(i.e., the electron spin system) can play in coupling the grain
rotation to the vibrational modes, and concluded that flipping would
be rapid enough to result in thermal trapping 
for grain sizes $a\ltsim 0.1\micron$.
This seemed to successfully account for the absence of
grain alignment for $a\ltsim 0.1\micron$.

However, further analysis of grain dissipational processes showed
that the {\it nuclear} spin system could be important at the lower
rotational rates appropriate to larger grains
(Lazarian \& Draine 1999b).
It was found that the coupling of grain rotation to the nuclear
spin system would lead to rapid thermal flipping even for grains as
large as $\sim 1\micron$!

Thermal flipping therefore now appears able to prevent $\HH$ formation
torques from being able to achieve superthermal rotation even for
the $a\approx 0.2\micron$ grains which are observed to be aligned!
How is grain alignment to be achieved?

\subsection{Effect of Radiative Torques}

So long as the grain surface is not modified, 
the $\HH$ formation torques are fixed in body coordinates, and
thermal flipping will cause these torques to average to zero if
the grain spends equal amounts of time in the two ``flip states'' 
(i.e., $\beta < \pi/2$
and $\beta > \pi/2$).
However, the grain is also subject to other systematic torques,
the most important of which is the radiative torque due to anisotropic
starlight.
This torque is {\it not} fixed in body coordinates, so that even
if the grain spends equal amounts of time in the two flip states, there
is a secular change in angular momentum due to the radiative torques.
Radiative torques are unimportant for small $a\ltsim 0.05\micron$ grains,
but very important for $a\gtsim 0.1\micron$ grains.
It therefore appears that ``thermal trapping'' will prevent small
($a\ltsim 0.05\micron$) grains from achieving suprathermal rotation,
but that larger ($a\gtsim 0.1\micron$) grains {\it can} achieve
suprathermal rotation, following which they can be gradually aligned
by the Davis-Greenstein mechanism.

But life turns out to be more complicated.  
Draine \& Weingartner (1997) found that radiative torques
act not only to spin up grains, but also to change their alignment.
The characteristic time scale for radiative torques to change the
grain alignment turns out to be just the drag time $\tau_{\rm drag}$, which
is shorter than the Davis-Greenstein alignment time, so that the
radiative torques appear able to dominate over the systematic effects
of the Davis-Greenstein alignment torque.

Study of three different grain shapes and different angles between
the magnetic field direction and the starlight anisotropy direction
showed that under some conditions the starlight torques brought about
alignment (the aligned state was an ``attractor''), 
while under other conditions the grain did not settle down in any
stationary state.  There appeared to be an overall tendency for
grain alignment to take place, but the sample of 3 grain shapes was
insufficient to reach general conclusions.

Further study of this problem is underway, and it is hoped that in a year
or two we will have definite conclusions on the role of radiative torques
in the grain alignment problem.  But for the moment 
the situation appears very promising -- we seem to finally understand
the absence of alignment of small grains, and it appears that the
physics we now understand will explain the alignment of the larger ones,
with radiative torques due to starlight playing a critical role.

\begin{figure}
\begin{center}
\includegraphics[width=.8\textwidth,angle=0]{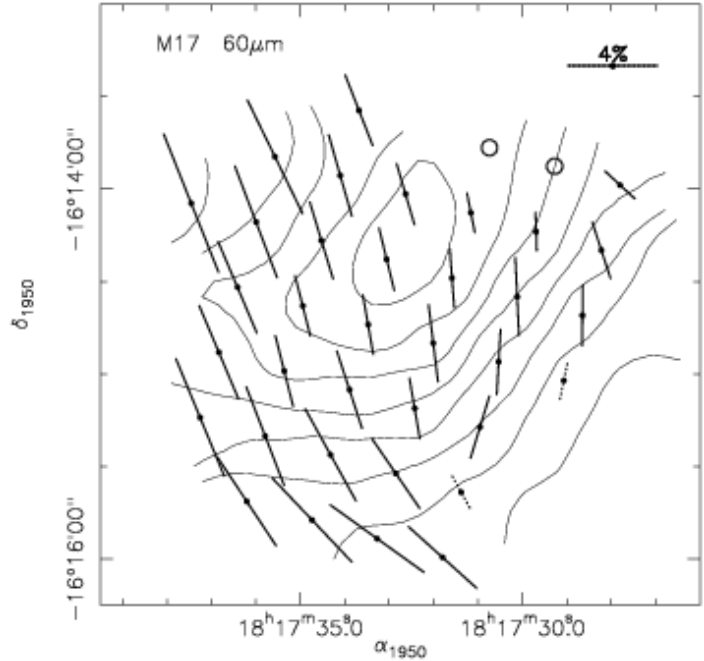}
\end{center}
\caption[]{60$\micron$ linear polarization toward M17.
	Flux contours are at 20\%, 30\%, ...,90\% of the peak flux.
	From Dotson et al (2000).
	}
\label{fig:M17}
\end{figure}

\subsection{Alignment and Disalignment in Different Regions}

There is evidence that the degree of grain alignment varies from one
region to another.
Observations of polarization of starlight indicate that 
the $a\gtsim 0.1\micron$ grains in diffuse clouds are
substantially aligned, but JHK observations of quiescent dark clouds
(Goodman et al. 1995) indicate that the dust grains in the dark interiors
are not aligned.  How can we understand this?

There are several possible plausible answers:
\begin{itemize}
\item The inner dark regions have insufficient starlight
for the radiative torques to achieve grain alignment.
\item The inner dark regions are molecular, so that $\HH$ formation torques
cannot drive suprathermal rotation.
\item Photoelectric emission is suppressed in the inner dark regions.
\end{itemize}
It is also possible, of course, that either the magnetic field structure
is completely tangled in the interior, or the grains in the central regions
are for some reason more spherical, but these do not seem plausible.

It is also interesting to note that in some dark clouds the grains are
reasonably well aligned: for example, the M17 molecular cloud 
(Fig.\ \ref{fig:M17}) shows 
linear polarization (averaged over the 22$^{\prime\prime}$ beam) 
as large as 5\%.
Note, though, that although this cloud is ``dark'' (i.e., we cannot
see into it at optical wavelengths) it is full of stars -- the
infrared spectrum indicates a grain temperature $T\approx 45 \K$, and
the grains must be heated by starlight: the mean starlight intensity
must be $\sim (45/20)^6 \approx 10^2$ times greater than the ISRF in order
to heat the grains to this temperature!
It therefore appears likely that starlight torques could drive the grains
in this cloud to suprathermal rotation rates.

Finally, a recent study by Padoan et al.\ (2001) concluded that 
submm polarization maps of quiescent cores were consistent with a model
where the grains in the outer parts of the cloud with $A_V\ltsim 3$~mag
were aligned, while grains in regions with $A_V> 3$~mag were randomly
oriented.

All of the above appear consistent with the idea that starlight torques
are an essential part of the grain alignment process.
Grains deep in quiescent dark clouds are unable to align because
starlight torques are weak or absent, whereas starlight torques
are present are do produce alignment within star-forming clouds.

\subsection{Summary of the Grain Alignment Story}

The grain alignment story has been long and complicated, with a number
of false turns.  The discussion above has skipped some of the important
physics involving coupling of the grain rotation to the electron and
nuclear spin systems (see Lazarian \& Draine 1999a,b) 
as well as the important question of
the grain dynamics during ``crossovers'', when reversals of the
systematic torques on the grain cause the grain angular momentum to
become small -- during this time the grain is very vulnerable to
disalignment (see, e.g., Lazarian \& Draine 1997).

To summarize: 
the important elements of the grain alignment story now appear to be
as follows:
\begin{itemize}
\item Grains are subject to 2 important systematic torques: 
	\begin{itemize}
	\item $\HH$ formation torques,
	\item radiative torques due to starlight.
	\end{itemize}
\item For grains with rotational kinetic energies less than a few times
	$kT_d$, thermal fluctuations allow the grain to sample different
	orientations of the principal axes relative to the instantaneous
	angular momentum $\bf J$, leading to ``thermal flipping'' of the
	grain between the two ``flip states''.
\item If the thermal flipping is sufficiently rapid, the grain will spend 
	50\% of the time in each flip state, in which case the $\HH$
	formation torque, which is fixed in body coordinates, will 
	integrate to zero in inertial coordinates: the $\HH$ formation
	torque will be unable to drive the grain to suprathermal rotation.
\item Starlight torques are not fixed in body coordinates, and hence
	do not integrate to zero even if the grain spends equal amounts
	of time in the two flip states.
\item For grains with $a\gtsim 0.1\micron$ in regions with sufficiently
	strong starlight, the systematic torque due to starlight
	will be able to drive the grain to suprathermal rotation.
\item Once a grain achieves suprathermal rotation, thermal flipping will
	cease, the grain will settle into a state where the principal
	axis $\hat{\bf a}_1$ is either parallel or antiparallel to ${\bf J}$,
	and all systematic torques will operate simultaneously.
\item Grains in suprathermal rotation are essentially immune from disalignment
	by random collisions.  Alignment will result from
	the combined effects of Davis-Greenstein alignment and 
	starlight torques.
\item Under some circumstances, starlight torques do not result in
	strong alignment.  However, numerical experiments suggest that,
	averaged over irregular grain shapes, starlight torques lead
	to overall alignment.
\item The fact that starlight torques are important only for grains
	with $a\gtsim 0.1\micron$ appears to account for the observation
	that $a\ltsim0.05\micron$ grains in diffuse clouds are not aligned,
	while $a\gtsim 0.1\micron$ grains appear to be well-aligned.
\end{itemize}

\section{Evolution of the Grain Population}

The interstellar grain population in a galaxy is the result of a rich and
complicated mix of processes.

We would like to know the composition of interstellar dust, how and
where it forms, and the extent to which individual grains are
either homogeneous or conglomerate.  We would like to know what the
size distribution of the different types of grains is, how and
where the size distribution is fashioned, and the manner in which
the size distribution varies from one region to another.
If we understood the balance of processes acting in our part of the
Milky Way, perhaps we could deduce how grains might differ in the inner
or outer parts of the Milky Way, and in the LMC, SMC, and other galaxies.

Unfortunately, the overall grain evolution problem seems to me to be beyond
our grasp at this time -- I don't think we understand the parts well 
enough to justify trying to put them together as a whole.
At this time, we have to try to clarify the individual elements
of the story.

\subsection{Dust Formation in Stellar Outflows}

There is abundant evidence that dust grains form in some stellar
outflows: we see infrared emission from 
dust around red giants, around carbon stars,
and in planetary nebulae.  
In all these cases the dust condensed out
of material which at an earlier time was completely gaseous.
Observations show that different types of stellar outflows form different
types of dust.

The dust around red giants with O/C $>$ 1 shows a 10$\micron$ emission feature
which is due to the Si-O stretching mode in amorphous silicate material;
around some OH/IR stars the 10$\micron$ feature appears in absorption
(e.g., Demyk et al. 2000), along with emission features revealing 
the presence of crystalline silicates.

The dust around carbon stars, on the other hand, does not show the
10$\micron$ feature, which is consistent with the fact that we do not
expect silicates to form in an atmosphere with O/C $<$ 1.
Some carbon stars (e.g. IRC+10216) instead show a 11.3$\micron$ emission
feature which is attributed to SiC (e.g., Blanco et al. 1998).

It should also be mentioned that we find presolar grains in meteorites
whose isotopic composition clearly indicates formation in outflows from
evolved stars (Hoppe \& Zinner 2000); 
these include nanodiamonds, SiC, graphite, and
Al$_2$O$_3$.  Some of these grains may have formed in supernova ejecta.

So there is no question that stellar outflows contribute dust to the
interstellar medium.  However, this does {\it not} 
necessarily imply that the bulk of
the dust in the interstellar medium was formed in stellar outflows.

The physics and chemistry of dust formation in stellar outflows is complex:
the material is generally far from thermodynamic equilibrium,
and the outflows themselves are often (perhaps always?) hydrodynamically
complex -- neither steady nor spherically symmetric.  We can not yet
reliably describe the details of the grain formation process.

\subsection{Grain Growth in the ISM\label{sec:growth}}

{\bf\medskip\noindent\ref{sec:growth}.1 Accretion in the ISM\medskip}

Dust grains in the interstellar medium can grow, since atoms from the
gas can stick to them.
We can calculate the lifetime against accretion 
for an atom or ion in the gas phase:
\beq
\tau^{-1} = A_i^{-1/2} s_i \left(\frac{8kT}{\pi m_\rmH}\right)^{1/2}
\int da \pi a^2 \frac{dn_d}{da} D(a)
~~~,
\eeq
where $A_i$ is the mass number of the ion, $s_i$ is the sticking coefficient,
and $n_d(a)$ is the number density of grains with radii less than $a$.
The enhancement factor $D(a)$ gives the collision rate relative to
what it would be in the absence of electrostatic effects:
\beq
D(a) = \sum_{Z_d} f(Z_d,a) B(Z_d,a)
~~~,
\eeq
where
\beq
B(Z_d,a) = \left\{
\begin{array}{ll}
\exp\left(-Z_d Z_i e^2/akT\right)	& {\rm for}~ Z_gZ_i > 0	
\\
\left(1-\frac{Z_dZ_ie^2}{akT}\right)	& {\rm for}~ Z_gZ_i < 0
\\
1+\left(\frac{\pi Z_i^2 e^2}{2akT}\right)^{1/2}
					& {\rm for}~ Z_g = 0
\end{array}
\right.
~~~.
\eeq
\begin{figure}
\begin{center}
\includegraphics[width=.8\textwidth,angle=0]{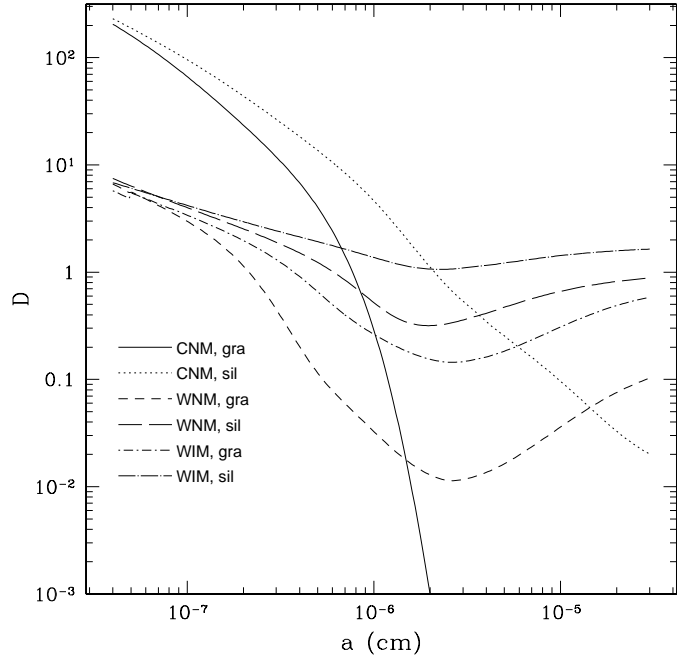}
\end{center}
\caption[]{Collision enhancement factor for positively charged ion (see
	text).
	From Weingartner \& Draine (1999).}
\label{fig:D(a)}
\end{figure}
Fig.\ \ref{fig:D(a)} shows the collisional enhancement factor $D(a)$
calculated by Weingartner \& Draine (1999)
for carbonaceous and silicate grains in various environments.
Enhancement factors $D\gtsim\sim10^2$ are found for $a\ltsim 10^{-7}\cm$
grains in the CNM.
As a result, an ion (e.g., SiII, TiII, FeII) would have a lifetime
against accretion of $\sim 1\times10^{5}\yr$ in the CNM if the
grain size distribution includes the ultrasmall grain population of
Fig.\ \ref{fig:sizedist}.
Therefore we expect that those metals which can ``stick'' to the
small grains will be strongly depleted from the gas phase;
presumably those elements which are not strongly depleted
(e.g., Na, K, S) for some reason do not remain on the ultrasmall
grains after they collide.

{\bf\medskip\noindent\ref{sec:growth}.2 Mantle Formation\medskip}

As discussed in \S\ref{sec:spec_ice}, grains in dark molecular clouds
are apparently coated with ice ``mantles''.  The overall mantle
composition is dominated by H$_2$O ice.
It is not known whether the H$_2$O molecules arrived from the
gas phase and then froze onto the grain, or whether the H$_2$O formed
by surface chemistry.
In most clouds the gas phase H$_2$O abundance is very low, and
the observed mantles would not be able to form by accretion, suggesting
that O and H atoms arriving at the grain react there to form H$_2$O
(Jones \& Williams 1984).
However, it has recently been proposed that the H$_2$O could be
produced in the gas phase during occasional passage of a shock wave, and then
quickly accreted onto the grain surfaces (Bergin et al.\ 1999).
Other molecules (e.g., CO) may be mixed with the H$_2$O, or in some
cases condensed separately, as the absorption profile in some cases
requires that the CO not be mixed with polar molecules.
This could result if the H$_2$O was deposited while the grain was
too warm to retain CO, with the CO later freezing out when the conditions
changed and the grain 
became colder.

{\bf\medskip\noindent\ref{sec:growth}.3 Coagulation\medskip}

The grain population in normal diffuse clouds has a very large population
of very small grains, required to explain the rapid rise in extinction
at short wavelengths.  In dense regions, however, the extinction at
short wavelengths appears to be reduced.  This can only happen if
some of the small grains are either destroyed altogether, or simply
coagulated onto big grains.  The latter seems more likely.

What is the time scale for such coagulation?  This depends on the
velocity differences between different grains.  Grain-grain velocities
can be the result of differential drift velocities
(see \S\ref{sec:vdrift}),
hydrodynamic turbulence (see, e.g., Draine 1985),
or MHD turbulence (Lazarian \& Yan 2002).

Grains larger than $\sim 10^{-5}\cm$ contribute a geometric cross section
per H $\Sigma \approx 0.5\times10^{-21}\cm^2$.
If the characteristic grain-grain velocity difference is $\Delta v_{d}$,
then the grain-grain coagulation time would be
\beq
t_{coag} = \left[n_{\rm H} \Sigma \Delta v_{d}\right]^{-1}
= 2 \times 10^8\yr 
\left(\frac{30\cm^{-3}}{n_{\rm H}}\right)
\left(\frac{0.1\kms}{\Delta v_d}\right)
~~~,
\eeq
so that grain coagulation will not be very important in diffuse clouds,
but could proceed in dense regions.

\subsection{Photolysis}

When grains with ice mantles are returned to diffuse regions and exposed
to ultraviolet radiation, it appears that most of the material is
driven off, but the ultraviolet radiation photolyzes the ice to
produce an organic refractory residue (e.g., Greenberg et al.\ 2000).
This residue may be responsible for the interstellar 3.4$\micron$
``aliphatic C-H'' absorption band.

\subsection{Photodesorption\label{sec:photodesorption}}

The process of photodesorption is potentially very important, but is
unfortunately not well understood because of limited laboratory studies.
Photodesorption is a quantum process whereby absorption of a single
photon can lead to ejection of an atom or molecule from a solid.
For example, an absorbed photon might cause an atom or molecule on a
substrate to be excited
to a repulsive electronic state which would then be ejected from the substrate.

Photodesorption was discussed by Draine \& Salpeter (1979b), who argued that
a molecule in a surface monolayer might have a photodesorption cross
section as large as $\sim10^{-18}\cm^2$.

If an atom or molecule adsorbed on a grain surface has a cross section 
$\sigma_{pd}=10^{-18}\sigma_{-18} \cm^2$ for $8 < h\nu < 13.6\eV$ photons,
the photodesorption rate for an adsorbed atom or molecule would
be $\sim 6\times10^{-11}\sigma_{18}\s^{-1}$ in the diffuse ISM,
since the density of $8 - 13.6\eV$ photons is $\sim2.0\times 10^{-3}\cm^{-3}$
in the MMP ISRF.
Laboratory studies show that
H$_2$O ice has a photodesorption cross section
$\sigma_{pd}\approx8\times10^{-18}\cm^2$ at
$\lambda=1215\Angstrom$
(Westley et al. 1995a, 1995b), so that it seems likely that at least
some species will have $\sigma_{-18}\gtsim 1$, with photodesorption rate
$\gtsim 6\times10^{-11}\s^{-1}$.  Note that this estimated photodesorption
rate is comparable to photodissociation rates for 
diatomic molecules
in the interstellar radiation field.\footnote{%
	Roberge et al. (1991) estimate photodissociation rates
	$8.6\times 10^{-10}\s^{-1}$ and $3.5\times10^{-10}\s^{-1}$
	for CH and OH, respectively.}

O is the most abundant species after H and He.
The probability per unit time that
an O atom will arrive at a single surface
site (of area $\sim 10^{-15}\cm^2$) is $\sim2\times 10^{-13}\s^{-1}$ in
the CNM.
It is therefore apparent that any species with $\sigma_{18}\gtsim .01$
will be photodesorbed before accretion of anything chemically interesting
(other than H)
can take place on top of it.

We conclude that photodesorption most likely plays a dominant role
in determining what kind of accretional growth takes place on
a grain.  It is no surprise that the noble gases (He, Ne, Ar) do not deplete,
but the fact that some 
chemically-reactive elements
(e.g., Na, K, S) appear to undergo minimal depletion, at
least in the diffuse ISM, may be due to photodesorption.\footnote{%
	The only
	other depletion-suppressing process 
	would be chemisputtering by H atoms: an impinging H atom might
	react with the adsorbed atom, and the reaction energy might
	eject the resulting hydride.}
Laboratory studies of photodesorption from carbonaceous or
silicate substrates would be of great value to clarify why some
elements deplete and others do not in the diffuse ISM.

\subsection{Grain Destruction in Shock Waves\label{sec:shocks}}

From time to time a shock wave will pass through an element of interstellar
gas, compressing, heating, and accelerating the gas (Draine \& McKee 1993),
and creating conditions under which grain destruction can occur
(Draine 1995).
Shock waves can be driven into cold cloud material ahead of an
ionization/dissociation front (Bertoldi \& Draine 1996), as the result of
energetic outflows from newly-formed stars, or as the result of
the explosion of a nearby supernova.

It is easiest to study the shock in the ``shock frame'' where the shock front
is stationary: in the upstream direction the grains and gas move together
toward the shock
with a speed equal to the shock speed $v_s$.  The fluid is then
decelerated at the shock front, with the postshock gas velocity
$v_g\approx v_s/4$ for a strong shock.
The postshock gas is initially hot.
If the shock is strong, and we can neglect the effects of a possible
magnetic precursor, the gas temperature will rise to
\beq
T_s \approx \frac{3}{16k}\mu v_s^2 = 
2300\times\K 
\left(\frac{\mu}{m_\rmH}\right)
\left(\frac{v_s}{10\kms}\right)^2
~~~,
\eeq
where $\mu$ is the mass per free particle ($\mu/m_\rmH \approx 1.4/1.1 = 1.27$ 
for cold atomic clouds, and $\mu/m_\rmH = 1.4/0.6 = 2.33$ for molecular gas,
where we assume He/H=0.1).

The grains move across the shock front
into the postshock fluid, where they initially have velocity 
$v_s-v_g\approx (3/4)v_s$
relative to the gas.
The grains now begin to decelerate by gas drag, but are also acted
on by the magnetic field ${\bf B}$ and electric field 
${\bf E}=-{\bf v}_g\times{\bf B}/c$ (evaluated in the shock frame).
As a result, the grain population develops grain-grain velocity differences
which can be of order the shock speed $v_s$, or even larger if
conditions are favorable for ``betatron'' acceleration of the dust
as the density and magnetic field increase in
cooling postshock gas (Spitzer 1976; Draine \& Salpeter 1979b).

{\bf\medskip\noindent\ref{sec:shocks}.1 Grain-Grain Collisions \medskip}

In the absence of betatron acceleration, a grain will slow down
by a factor $1/e$ when it has collided with
its own mass of gas.  If all grains were identical, then the chance
that a grain would run into another grain before appreciable slowing-down
would be of order the dust-to-gas mass ratio of $\sim 0.01$, and we can
therefore expect that a typical grain (one representative of the
bulk of the grain mass) will have a probability of $\sim 0.01$ of collision
with another comparable grain with a grain-grain velocity difference
$\sim v_s$.

The physics of grain-grain collisions has been discussed by Tielens
et al. (1994).  For refractory grains one expects virtually complete
destruction if the energy in the center-of-mass system is sufficient
to vaporize both grains.  For identical grains, mean atomic mass
20 amu, and binding energy per atom of 5.7 eV (these numbers are
appropriate to MgFeSiO$_4$ silicate) 
complete vaporization could occur in a head-on
collision between identical
grains with a velocity difference of $15~ \kms$, so we may expect
vaporization of perhaps of $\sim1\%$ of the grains in a hydrodynamic
shock with shock speed $v_s\gtsim 20\kms$.
At lower grain-grain collision speeds, shattering may occur.
Thus we should expect that several percent of the large dust grains
will be shattered in shocks with $v_s\gtsim 10 \kms$.
Grain-grain collisions may be responsible for maintaining the population
of small grains in the interstellar medium.

{\bf\medskip\noindent\ref{sec:shocks}.2 Sputtering\medskip}

\begin{figure}[h]
\begin{center}
\includegraphics[width=0.65\textwidth,angle=270]{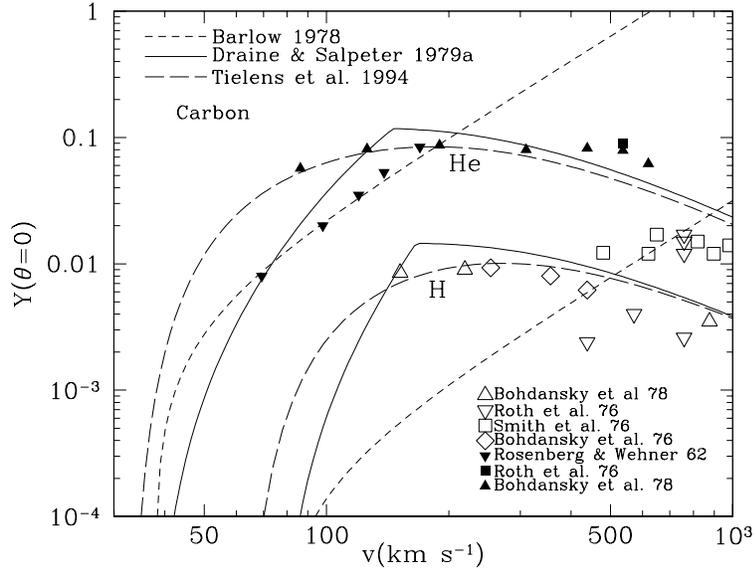}
\end{center}
\caption{Sputtering yields for H and He on carbon estimated by
	Barlow (1978), Draine \& Salpeter (1979a), and Tielens et al.\ (1994),
	and compared with laboratory data.
	}
\label{fig:C_yields}
\end{figure}
\begin{figure}
\begin{center}
\includegraphics[width=0.6\textwidth,angle=270]{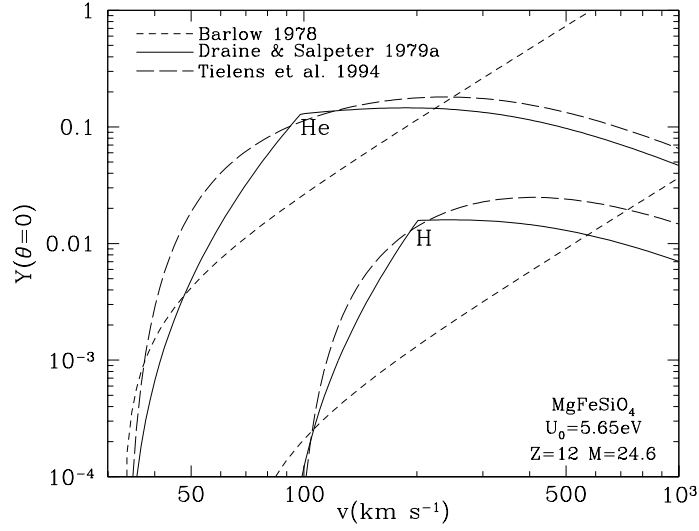}
\end{center}
\caption{Sputtering yields for H and He on silicate material estimated
	by Barlow (1978), Draine \& Salpeter (1979a), 
	and Tielens et al.\ (1994).
	}
\label{fig:sil_yields}
\end{figure}

\begin{figure}[t]
\begin{center}
\includegraphics[width=1.0\textwidth,angle=0]{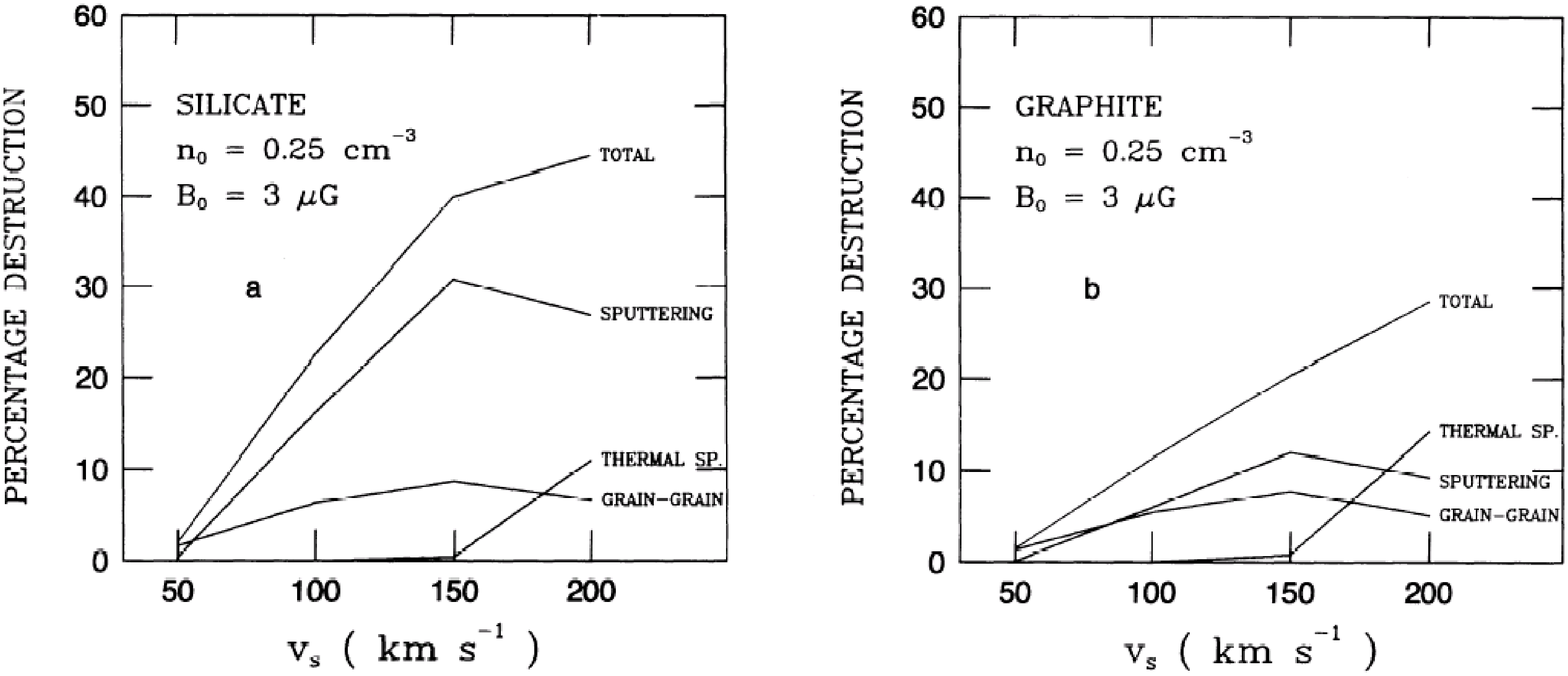}
\end{center}
\caption{Fractional grain destruction calculated for graphite and
	silicate grains in interstellar shock waves in low density
	gas.
	Figure from Jones et al. (1994).
	}
\label{fig:Jones_etal_1994}
\end{figure}
\begin{figure}
\begin{center}
\includegraphics[width=0.8\textwidth,angle=0]{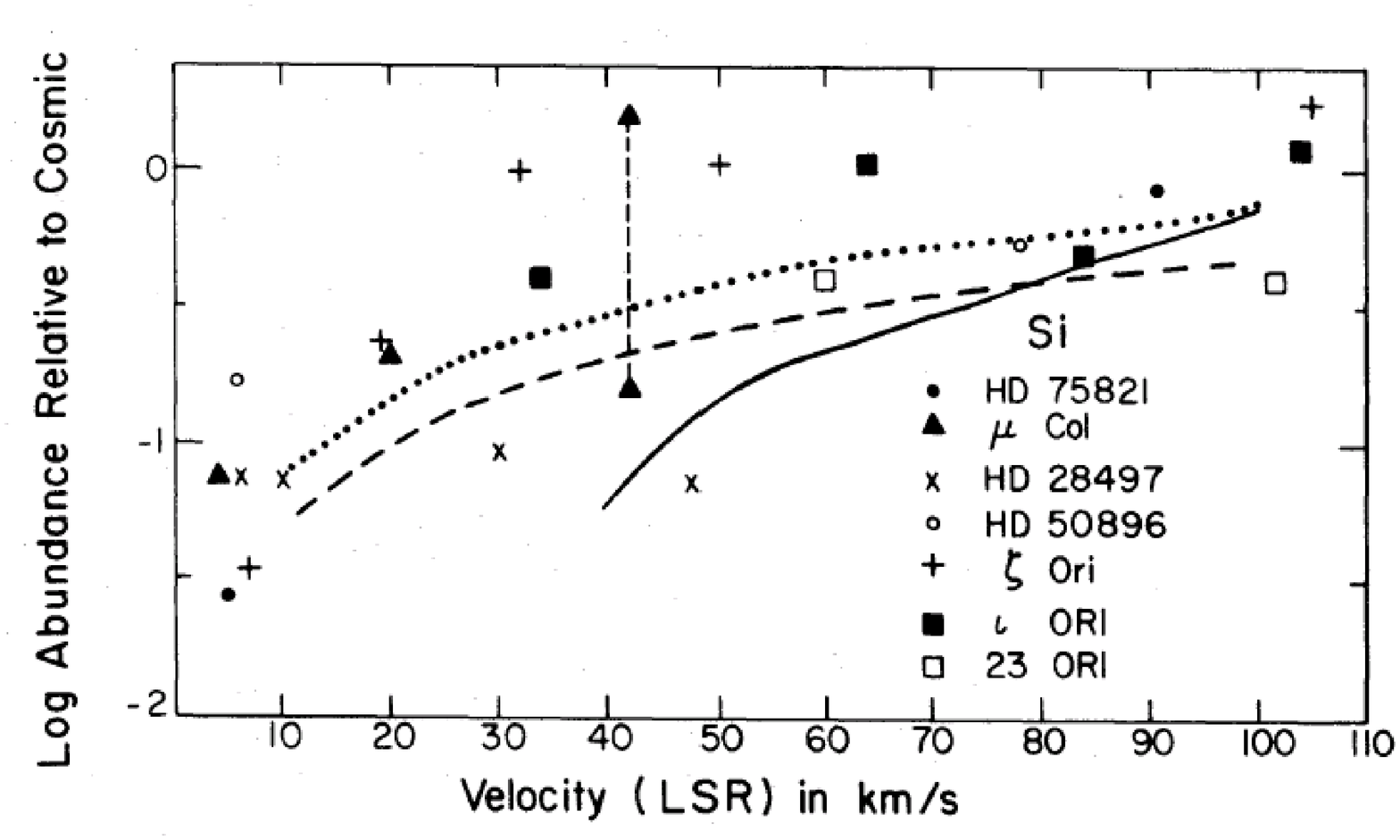}
\end{center}
\caption{Observed gas-phase Si abundances: evidence for refractory grain
	destruction in high-velocity shocks.
	Solid, dashed, and broken lines are theoretical models
	discussed by Cowie (1978).  Figure from Cowie (1978).
	}
\label{fig:Cowie_1978}
\end{figure}

In a fast shock, the grain material can be eroded by the process
of ``sputtering'', where individual atoms or ions from the gas
collide with the grain, occasionally ejecting one of the grain atoms.
In interstellar gas, the sputtering rate depends mainly on the
sputtering ``yields'' $Y(E)$ for H and He projectiles incident on silicate or
carbonaceous target materials.
In ionized gas the sputtering rates will be affected by the grain
charge.
An expression for the rate of sputtering of a moving grain, and
estimates for sputtering yields $Y$, are given by Draine \& Salpeter (1979a).

In the initial postshock region where the gas is hot, the sputtering is
due to a combination of thermal energy and the motion of the grain
through the gas with velocity $\sim3v_s/4$.
If the gas cools before the grain slows down, then subsequent sputtering
is entirely due to the rapid motion of the grain through the gas,
with He atoms having kinetic energies 4 times larger than H atoms.
The sputtering continues
until the grain slows to the point where the kinetic energy of
the impinging He atoms is too low for appreciable sputtering.

Detailed models of grain destruction in shock waves are in general
agreement that shock speeds $v_s\gtsim 100\kms$ result in substantial
grain destruction, with essentially complete grain destruction
for $v_s > 200\kms$ in the case of a radiative shock -- see
Fig.\ \ref{fig:Jones_etal_1994}
(Draine \& Salpeter 1979b; Jones et al. 1994).

These estimates for grain destruction sppear to be in agreement with the
observation that high-velocity gas (which has been shocked) tends to
have enhanced gas phase abundances of elements, such as Si, which are normally
heavily depleted in interstellar gas, an effect first noted for Ca
by Routly \& Spitzer (1952).
Fig.\ \ref{fig:Cowie_1978} shows this effect for Si.

\subsection{Thermal Sublimation}

Ice mantles on grains can be removed if the grains become too
warm.  What is the critical grain temperature for ice mantle
removal?

The probability per unit time for a surface molecule to desorb 
is approximately
\beq
\frac{dP}{dt} = \nu_0 e^{-U_0/kT}
~~~,
\eeq
where $U_0$ is the binding energy of the molecule, and
$\nu_0\approx 10^{13}\s^{-1}$ is a characteristic vibration
frequency.  If there are $N_{\rm mono}$ monolayers of ice on the
grain and we want it to survive a time $\Delta t$, the
grain temperature should not exceed
\beq
T_{\rm sub} = \frac{U_0/k}{\ln(\nu_0 \Delta t/N_{\rm mono})}
\eeq
Some heating phenomena are brief (e.g., a stellar flare or
a supernova) but we are usually interested in longer time
scales.
For example, a massive star in or near the cloud might have
a lifetime $\Delta t\approx 5\times10^6\yr$.
We can write
\beq
T_{\rm sub} = \frac{U_0/54k}{1+0.018\ln[(\Delta t/10^6\yr)(10^3/N_{\rm mono})]}
\eeq
Thus, a H$_2$O ice mantle ($U_0=0.5\eV$) $10^3$ monolayers
thick will survive $10^6\yr$ provided $T < 108\K$.
Therefore we should not be surprised to find H$_2$O ice in 
clouds where the grain temperature is 80 K (as in the molecular
cloud OMC-1), but it would be a great surprise to find H$_2$O
ice if the grain temperature were, say, 120K.

\subsection{Gamma Ray Bursts}

Gamma ray bursts may destroy dust grains out to a substantial
distance.  While this is probably uninteresting in terms of
evolution of the overall grain population in galaxies, it may
nevertheless be of observational interests because it affects
the observable properties of gamma-ray bursts.
This is of considerable interest because there is evidence that
gamma-ray bursts are associated with star-forming regions in
galaxies, and yet the optical afterglows from a number of bursts
show little evidence for reddening.

Grain destruction near GRBs has been discussed by
Waxman \& Draine (2000), 
Fruchter, Krolik \& Rhoads (2001), and
Draine \& Hao (2002).
There are two separate mechanisms which could be important.
Thermal sublimation (as discussed above, but now with larger
binding energies $U_0/k\approx 7\times10^4\K$ and shorter times
$\Delta t\approx 10 \s$ for the optical transient, so the
grain temperature can rise to $\sim 3000\K$.
Draine \& Hao find that for a reasonable (but very uncertain)
estimate of the GRB peak optical luminosity, dust grains can
be destroyed out to distances of $\sim$5 pc.

A more exotic grain destruction mechanism may operate; if effective,
it could destroy grains out to greater differences.
This is the mechanism of ``Coulomb explosion'' (Waxman \& Draine 2000).
The grains will be quite highly charged due to the intense
X-ray irradiation from the GRB.  If the grains are sufficiently
weak, the electrostatic stresses on the grain might conceivably
tear it apart.  Fruchter et al give arguments in favor of this
outcome, in which case GRBs could destroy grains out to quite
large distances.

\subsection{Time Scales}

When all grain destruction processes are considered, it is found
that sputtering in shock waves dominates the overall destruction
of refractory grains. 
The frequency with which grains are overtaken by such shocks depends
on the frequency of the supernovae which drive these blastwaves,
and on the structure of the interstellar medium, since material
in dense regions is ``sheltered'' from the effects of a passing blastwave.
The problem is complex, and therefore there are many places
where different assumptions could alter the quantitative conclusion,
but Draine \& Salpeter (1979b) estimated a lifetime 
$\sim1.3\times10^8\yr$ for silicate material in a diffuse cloud,
while Jones et al. (1994) find a lifetime $2.2\times10^8\yr$
for silicate grains.

Draine \& Salpeter (1979b) stress that this relatively short
lifetime was impossible to reconcile with the idea that
``interstellar grains are stardust'':
nearly all of the Si in the interstellar medium is in solid
form  (e.g., see Fig.\ \ref{fig:depletions}, showing that only
2.5\% of the Si is in the gas phase in the cloud toward $\zeta$Oph).
The ``turnover'' time of the interstellar medium is equal
to the ratio of (mass in gas)/(rate of mass going into stars),
or $\sim 5\times10^9M_\odot/5 M_\odot\yr^{-1} \approx 1\times10^9\yr$.
Suppose that all the Si leaving stars entered the ISM in solid form,
but there was no grain growth in the ISM.
A grain destruction time $\sim2\times10^8\yr$ -- only $\sim$20\% of the
turnover time -- would imply that $\sim80\%$ of the Si would be found
in the gas phase.
However, observations of diffuse clouds typically show $<10\%$ of the
Si in the gas gas phase (see, e.g., Fig.\ \ref{fig:depletions}).

The ineluctable conclusion is that {\it most} of the Si atoms in grains
must have joined those grains in the ISM through some process of
grain growth.  The kinetics of accretion implies that only in
denser regions can accretion rates be large enough to remove
Si from the gas fast enough to maintain the low observed
gas-phase abundances.
This in turn requires that there must be fairly rapid exchange
of mass between the dense and less dense phases of the ISM.
A more quantitative discussion of the required mass exchange
rates  may be found in
Draine (1990) and Weingartner \& Draine (1999).

\section{Effects of Dust on Interstellar Gas}

\subsection{Photoelectric Heating}

Photoelectrons emitted from dust grains depart with nonzero kinetic energy.
This is often the dominant mechanism for heating interstellar gas,
whether in diffuse clouds or in photodissociation regions.

Photoelectric heating rates have been reestimated recently by
Bakes \& Tielens (1994) and Weingartner \& Draine (2001c).
For starlight with the spectrum of the ISRF, Weingartner \& Draine find
a heating rate per H nucleon
$\Gamma/n_{\rm H} \approx 6\times10^{-26}\erg\s^{-1}$ for conditions
characteristic of cool H~I clouds.  This heating rate is much larger
than other heating mechanisms, such as ionization by cosmic rays or
X-rays.

\subsection{H$_2$ Formation and Other Chemistry}

The H$_2$ molecule is central to interstellar chemistry.
The rovibrational lines can be
an important coolant for gas at temperatures
$100\ltsim T \ltsim 2000\K$.

The principal gas-phase formation processes are via H$^-$
\beq
\label{eq:Hminus}
{\rm H} + e^- \rightarrow {\rm H}^- + h\nu ~~;~~ 
{\rm H}^- + {\rm H} \rightarrow {\rm H}_2 + e^-
\eeq
or 3-body collisions:
\beq
{\rm H} + {\rm H} + X \rightarrow  {\rm H}_2 + X + \Delta E
\label{eq:3body}
\eeq
where the third body $X$ could be H, He, or another H$_2$.
The H$^-$ formation pathway (\ref{eq:Hminus})
is slow because of the generally low
abundance of H$^-$ in diffuse gas, and the 3 body process
(\ref{eq:3body}) requires very high densities to become important.
In the interstellar medium of the Milky Way, the dust abundance is
such that H$_2$ formation is overwhelmingly due to grain surface catalysis.
The grain catalysis process was first proposed by Gould \& Salpeter (1963),
and the basic picture has not changed:
\begin{enumerate}
\item An H atom arrives at a grain surface and ``sticks''.
\item The H atom diffuses over the grain surface until it
	becomes trapped by either chemisorption or physisorption
\item A second H atom arrives at the surface, sticks, diffuses over
	the surface, and encounters the trapped H atom.
\item The two H atoms react to form H$_2$, releasing $\sim4.5\eV$ of
	energy, and ejecting the H$_2$ from the grain surface into the gas.
\end{enumerate}
Part of the 4.5eV goes into overcoming the binding of the two H atoms
to the surface, part goes into vibrational excitation of the grain
lattice (``heat''), part appears as kinetic energy of the H$_2$, and 
part appears in vibrational and rotational energy
of the newly-formed H$_2$ molecule.

In the above picture, the H$_2$ formation rate is
\beq
\frac{dn({\rm H}_2)}{dt} = R n_{\rm H} n({\rm H})
\eeq
\begin{eqnarray}
R &=& \frac{1}{2} \left(\frac{8kT}{\pi m_{\rm H}}\right)^{1/2} 
\int da \frac{1}{n_{\rm H}}\frac{dn_d}{da} \pi a^2 \gamma(a)
\\
&=& 7.3\times10^{-17}\cm^3\s^{-1}
\left(\frac{T}{100\K}\right)^{1/2} 
\frac{A \langle \gamma \rangle}{10^{-21}\cm^2{\rm H}^{-1}}
\end{eqnarray}
\beq
A \equiv \int da (1/n_{\rm H})(dn_d/da)\pi a^2
\eeq
\beq
\langle\gamma\rangle \equiv 
\frac{1}{A}\int da (1/n_{\rm H})(dn_d/da)\pi a^2 \gamma(a)
\eeq
where $\gamma(a)$ is the fraction of H atoms colliding with a grain of radius
$a$ which are converted to H$_2$; $\gamma < 1$ since not all impinging H
atoms will ``stick'', and some of those which stick may later be removed
from the surface by some process other than H$_2$ formation.
Gould \& Salpeter argued for $\gamma$ of order unity under interstellar
conditions.

Dust models which reproduce the observed extinction of starlight have
\beq
A\equiv \int da \frac{1}{n_{\rm H}}\frac{dn_d}{da} \pi a^2 \gtsim 10^{-21}\cm^2
\eeq
For example, the dust model of Weingartner \& Draine (2001a) has
$A=6.7\times10^{-21}\cm^2/{\rm H}$.
The H$_2$ formation ``rate coefficient'' $R$ can be determined from
ultraviolet observations of H$_2$ in diffuse regions; Jura (1975) found
$R\approx 3\times10^{-17}\cm^3\s^{-1}$, which implies 
$\langle \gamma\rangle\approx 0.06$.
It appears, then, that the very small grains which dominate 
the grain surface area (in particular, the PAHs) must not be efficient
catalysts for H$_2$ formation.

How might H$_2$ formation on the smallest grains be suppressed?
The time scale between photon absorption events is
\beq
\tau_\abs \approx 10^6\s (10\Angstrom/a)^3
~~~,
\eeq
while the time scale
between arrival of H atoms is
\beq
\tau_{\rm H} \approx 8 \times 10^6\s \left(\frac{10\Angstrom}{a}\right)^2 
\left(\frac{30\cm^{-3}}{n_{\rmH}}\right)
\left(\frac{100\K}{T}\right)^{1/2}
\eeq
so it is possible that an H atom physisorbed on a very small grain
may be removed when the grain is heated by a photon absorption event,
so that only rarely would an arriving H atom find another H atom
with which to recombine.  The efficacy of this process depends on
the binding energy of the H atom.

\subsection{Grains as Sources of Complex Molecules}

Observational studies of star-forming regions sometimes find molecular
abundances in the gas phase which are difficult to understand in the
context of pure gas-phase chemistry.
An extreme example of this is the observation  of
D$_2$CO/H$_2$CO=0.003 toward Orion (Turner 1990), or 
D$_2$CO/H$_2$CO=0.03-0.16 toward
the low-mass protostar IRAS 16293-2422 (Ceccarelli et al 2001).
Such extreme deuteration seems impossible to envision in the gas phase,
but could occur by chemistry on grain mantles during a precollapse phase,
with the molecules put into the gas phase when the dust is warmed by
energy from the protostar.
There seems little doubt that at least some regions have molecular abundances
which are heavily influenced by grain surface chemistry.

\subsection{Ion Recombination on Dust Grains}

In diffuse clouds, 
the ion/neutral fraction
elements with ionization potentials $< 13.6\eV$ is indicative of
the relative rates for photoionization of starlight vs. recombination
of ions with electrons. 
It is often assumed that the only channel for neutralization of
metal ions X$^+$ is radiative recombination,
${\rm X}^+ + e^- \rightarrow X + h\nu$.
However,
collisions with neutral or
negatively charged grains can be more important than radiative recombination
for neutralization of metal ions for conditions typical of the
diffuse interstellar medium
(Weingartner \& Draine 2001d).
Weingartner \& Draine provide estimates for the effective
``rate coefficient'' for ion neutralization via collisions with
dust grains.
Recombination on dust grains is particularly 
effective for protons in H~I regions, 
and therefore is involved in regulating the free electron density.

\subsection{Coupling Neutral Gas to Magnetic Fields}

In gas of low fractional ionization, a significant fraction of the
``free'' charge present may be located on dust grains.
The fraction of the free charge residing on the grains 
depends very much on the
numbers of very small dust grains, but appears likely to become
significant in regions of low fractional ionization
$n_e/n_{\rm H}\ltsim 10^{-6}$ (Draine \& Sutin 1987).  

Coupling of the magnetic field to the gas is due to the force exerted
on moving charges; in regions of low fractional ionization, the neutral
gas will be unaffected by the magnetic field unless the charged species
are collisionally coupled to the neutral atoms and molecules.
Because dust grains have relatively large physical cross sections,
the charge trapped on dust grains can be important at coupling the
magnetic field to the neutral gas in MHD shocks in gas with fractional
ionization
$n_e/n_{\rm H}\ltsim 10^{-7}$ (Draine 1980; Draine, Roberge \& Dalgarno 1983).

\subsection{Dust Grains as Magnetometers?}

Magnetic fields are dynamically important in the ISM, but difficult to 
observe remotely.
Nature has been kind enough to strew microscopic magnetometers -- 
dust grains -- throughout the interstellar medium.
Unfortunately
we have not yet figured out how these magnetometers work, so that even
if we could measure their degree of alignment, we would not be able
to determine the magnetic field strength.
But we {\it can} use dust grains as ``compasses'' to indicate the
{\it direction} of the magnetic field.

As discussed above, dust grains have their angular momenta $\bf J$
systematically aligned with the magnetic field ${\bf B}_0$, at least
in regions where sufficient starlight is present.
Here we review the main features of this alignment:
\begin{itemize}
\item It is virtually certain that the sense of alignment is to have
${\bf J}\parallel {\bf B_0}$ (to within a sign -- parallel and antiparallel
are equally favored).
\item Suppose the grain has 3 principal axes $\hat{\bf a}_1$, 
$\hat{\bf a}_2$, and $\hat{\bf a}_3$, with moments of inertia
$I_1 \ge I_2 \ge I_3$.
If the grain has rotational kinetic energy
$E_{\rm rot}> (3/2)kT_{\rm gr}$, then the grain will tend to have
its shortest axis $\hat{\bf a}_1 \parallel {\bf J}$ so as to minimize
its rotational kinetic energy at constant ${\bf J}$.
In regions where the grains are cooler than the gas 
(the most common situation), this is expected to
be true for both large grains (which rotate suprathermally) and small
grains (which rotate thermally).
\item Thus we expect the grains to have their ``long axes''
($\hat{\bf a}_2$, $\hat{\bf a}_3$) $\perp {\bf B_0}$.
\item Starlight extinction will be greatest for starlight with
${\bf E} \parallel$
the long axes, and therefore we expect the transmitted starlight
to have ${\bf E} \parallel {\bf B}_0$.
\item Far-infrared and submm emission will be polarized $\parallel$ to the
grains' long axes, and hence we expect the FIR/submm emission to be
polarized with ${\bf E}\perp {\bf B}_0$.
\end{itemize}
Therefore we can obtain a map of the magnetic field by either
measuring the polarization of starlight for many stars,
or by mapping the polarized far-infrared or submm emission from the
cloud.
CCDs and infrared detectors make possible extensive measurements of
starlight polarization, even in dense regions,
and FIR/submm arrays (e.g., SCUBA on the JCMT) make possible
measurements of polarized submm emission.

\begin{figure}
\begin{center}
\includegraphics[width=.7\textwidth,angle=0]{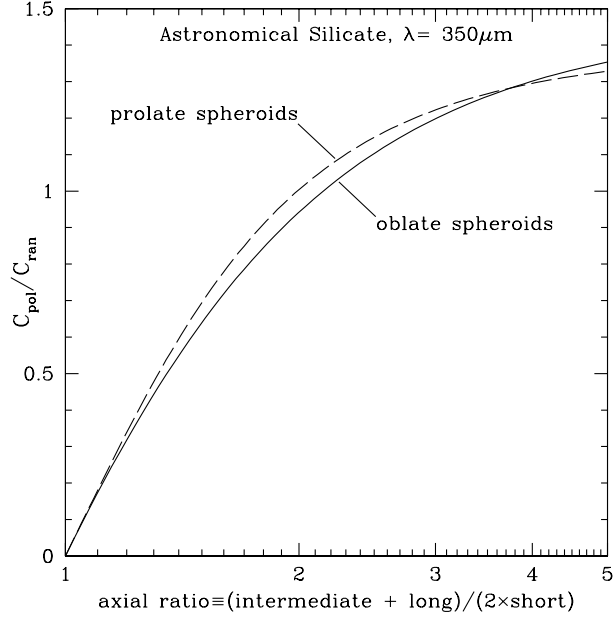}
\end{center}
\caption[]{Polarization efficiency $C_{\rm pol}/C_{\rm ran}$
	for spheroidal dust grains
	(oblate and prolate) as a function of the axial ratio.
	From Padoan et al. (2001).
	}
\label{fig:pol_eff}
\end{figure}

Grains are likely to have a fairly high polarizing efficiency in the FIR/submm.
Define
\begin{eqnarray}
C_{\rm pol} \equiv \frac{1}{2}
\left[C({\bf E}\parallel \hat{\bf a}_2)+
	C({\bf E}\parallel \hat{\bf a}_3)\right]
-C({\bf E}\parallel \hat{\bf a}_1)
\\
C_{\rm ran} \equiv 
\frac{1}{3}
\left[C({\bf E}\parallel \hat{\bf a}_1)+
	C({\bf E}\parallel \hat{\bf a}_2)+
	C({\bf E}\parallel \hat{\bf a}_3)
\right]
\end{eqnarray}
If the grains were spherical, there would of course be no polarization
in extinction or emission.
Fig.\ \ref{fig:pol_eff} shows how the polarizing efficacy
depends on the axial ratio of the dust grains.
The grains are assumed to be either oblate or prolate spheroids,
and the dielectric function of the grains is taken to be that of
``astronomical silicate'' at 350$\micron$.

For perfect spinning alignment, the maximum polarization is obtained
for a sightline $\perp {\bf B}_0$:
\beq
P_{\rm max} = \frac{C_2+C_3-2C_1}{C_2+C_3+2C_1} = 
\frac{C_{\rm pol}/C_{\rm ran}}{1-\frac{1}{6}\frac{C_{\rm pol}}{C_{\rm ran}}}
\eeq
Suppose the grain shape has (intermediate+long)/(2$\times$short) =
1.4 (e.g., axial ratios 1:1.2:1.6).
Then $C_{\rm pol}/C_{\rm ran}\approx 0.5$ for astronomical silicate.
Perfect alignment would give polarization in emission
$P=0.5/(1-.08)=0.55$.
However, the largest observed polarizations are $\sim10\%$,
implying either
\begin{itemize}
\item imperfect alignment
\item less elongation
\item disorder in the magnetic field on short length scales.
\end{itemize}
Unfortunately, it will be difficult to determine the relative importance
of these three effects.
For perfectly aligned grains in a uniform magnetic field perpendicular to
the line-of-sight, $P=0.1$ would be produced by grains with
axial ratio of only $\sim 1.05$ -- very minimal elongation.

\section{Concluding Remarks}

There are many unanswered questions in the astrophysics of interstellar
dust grains in cold clouds.  The questions are interesting in their own
right, but one is also driven to answer them by the need to understand
dust grains well enough to use them as diagnostics of interstellar
conditions, and to understand the effects which dust grains have on
the dynamics and evolution of interstellar gas.
With the advent of powerful infrared and submillimeter observing facilities --
including {\it SIRTF, SOFIA, ALMA, Herschel,} and {\it Planck} -- 
we can anticipate
that there will be progress on the astrophysics of dust, 
driven by the challenge
of understanding the new data.
An exciting decade lies before us!

I thank Robert Lupton for making available the SM software package.
This work was supported in part by NSF grant AST 9988126.


%

\end{document}